\documentclass[preprint,aps,nofootinbib]{revtex4}
\usepackage{dcolumn}
\usepackage{bm}
\usepackage{epsfig}
\usepackage{graphicx}
\usepackage{axodraw}
\usepackage{amsmath}
\usepackage{amssymb}
\usepackage{mathrsfs}
\usepackage{dsfont}
\usepackage{float}
\usepackage{color}
\usepackage[usenames,dvipsnames]{xcolor}
\usepackage{hyperref}
\usepackage[caption=false]{subfig}

\def\fun#1#2{\lower3.6pt\vbox{\baselineskip0pt\lineskip.9pt
  \ialign{$\mathsurround=0pt#1\hfil##\hfil$\crcr#2\crcr\sim\crcr}}}

\def\simlt{\stackrel{<}{{}_\sim}}
\def\simgt{\stackrel{>}{{}_\sim}}

\input epsf

\newenvironment{Eqnarray}%
         {\arraycolsep 0.14em\begin{eqnarray}}{\end{eqnarray}}
\newcommand{\be}{\begin{equation}}
\newcommand{\ee}{\end{equation}}
\newcommand{\bea}{\begin{Eqnarray}}
\newcommand{\eea}{\end{Eqnarray}}

\def\nn{\nonumber}
\def\half{\tfrac{1}{2}}
\def\phm{\phantom{-}}

\def\cosb{c_\beta}
\def\sinb{s_\beta}
\def\tanb{\tan\beta}

\def\lsim{\mathrel{\raise.3ex\hbox{$<$\kern-.75em\lower1ex\hbox{$\sim$}}}}
\def\gsim{\mathrel{\raise.3ex\hbox{$>$\kern-.75em\lower1ex\hbox{$\sim$}}}}
\def\lsub#1{_{\lower 1.5pt\hbox{$\scriptstyle#1$}}}
\def\lsup#1{^{\lower 2pt\hbox{$\scriptstyle#1$}}}

\def\eq#1{Eq.~(\ref{#1})}
\def\eqs#1#2{Eqs.~(\ref{#1}) and (\ref{#2})}

\def\eqst#1#2{Eqs.~(\ref{#1})--(\ref{#2})}

\def\Eq#1{Eq.~(\ref{#1})}

\def\Eqs#1#2{Eqs.~(\ref{#1}) and (\ref{#2})}

\begin{document}

\begin{flushright}
\vspace*{-1.cm}
\vspace{-0.8cm} EFI-15-32\\ 
FERMILAB-PUB-15-407-T \\ MCTP-15-15\\ SCIPP 15/12 \\ WSU-HEP-1505\\
\end{flushright}

\vspace*{1.2cm}

\title{On the Alignment Limit of the NMSSM Higgs Sector}

\author{
\vspace{0.5cm}
\mbox{\bf Marcela Carena$^{\,a,b,c}$, Howard E.~Haber$^{\,d}$, Ian Low$^{\,e,f}$,} \\
\mbox{\bf Nausheen R.~Shah$\,^{g,h}$, and Carlos E.~M.~Wagner$^{\,b,c,e}$}
 }
\affiliation{
\vspace*{.1cm}
$^a$\mbox{\footnotesize{Fermi National Accelerator Laboratory, P.O. Box 500, Batavia, IL 60510}}\\
$^b$\mbox{\footnotesize{Enrico Fermi Institute, University of Chicago, Chicago, IL 60637}}\\
$^c$\mbox{\footnotesize{Kavli Institute for Cosmological Physics, University of Chicago, Chicago, IL 60637}}\\
$^d$\mbox{\footnotesize{Santa Cruz Institute for Particle Physics, University of California, Santa Cruz,  CA 95064}} \\
$^e$\mbox{\footnotesize{High Energy Physics Division, Argonne National Laboratory, Argonne, IL 60439}}\\
$^f$\mbox{\footnotesize{Department of Physics and Astronomy, Northwestern University, Evanston, IL 60208}} \\
$^g$\mbox{\footnotesize{Michigan Center for Theoretical Physics,
University of Michigan, Ann Arbor, MI 48109}}\\
$^h$\mbox{\footnotesize{Department of Physics and Astronomy, Wayne State University, Detroit, Michigan 48201}}\\
}

\vskip -0.5in

\begin{abstract}
The Next-to-Minimal Supersymmetric extension of the Standard Model~(NMSSM) with a Higgs boson of mass 125 GeV can be compatible with stop masses of  order of the electroweak scale, thereby  reducing the degree of fine-tuning necessary to achieve electroweak symmetry breaking.
Moreover, in an attractive region of the NMSSM parameter space, corresponding to  the ``alignment limit'' in which one of the neutral Higgs fields lies approximately in the same  direction in field space
as the doublet Higgs vacuum expectation value, the observed Higgs boson is predicted to have Standard-Model-like properties.  We derive  analytical expressions for the alignment conditions  and show that they point toward a more natural region of  parameter space for electroweak symmetry breaking, while allowing for perturbativity of the theory up to the Planck scale. Moreover, the alignment limit in the NMSSM leads to a well defined spectrum in the Higgs and Higgsino sectors, and yields  a rich and interesting Higgs boson  phenomenology that can be tested at the LHC.   We discuss the most promising channels for discovery  and present several benchmark points for further study. 
\end{abstract}
\thispagestyle{empty}

\maketitle

 \section{Introduction}

The recent discovery of a scalar resonance \cite{Aad:2012tfa,Chatrchyan:2012ufa}, with properties similar to the Higgs boson of the Standard Model~(SM) motivates the study of  models of electroweak symmetry breaking which are weakly coupled at the weak scale.
Low energy supersymmetric theories with flavor independent mass parameters are  particularly well motivated models of this class, in which electroweak symmetry breaking
is triggered by radiative corrections of the Higgs mass parameters induced by supersymmetry breaking effects in the top-quark sector.  

The Higgs sector of the Minimal Supersymmetric extension of the Standard Model (MSSM)  is a two-Higgs-doublet model (2HDM) in which the tree-level mass of the CP-even Higgs boson associated with electroweak symmetry breaking is bounded from above by the $Z$ boson mass, $m_Z$.  Consistency with the observed Higgs mass may be obtained by means of large radiative corrections, which depend logarithmically on the scalar-top quark (stop) masses, and on the stop mixing mass parameters in a quadratic and quartic fashion~\cite{mssmhiggsradcorr}--\cite{Lee:2015uza}.  The large values of the stop mass parameters and mixings necessary to obtain the proper Higgs mass also lead to large negative corrections to the Higgs mass parameter that in general must be canceled by an appropriate choice of the supersymmetric Higgsino mass parameter $\mu$ in order to obtain the proper electroweak symmetry breaking scale.  For large stop masses, such a cancelation is
unnatural in the absence of specific correlations among the supersymmetry breaking parameters (whose origins are presently unknown).

The Next-to-Minimal Supersymmetric extension of the Standard Model (NMSSM)~\cite{Ellwanger:2009dp} shares many properties with the MSSM, but the Higgs sector is extended by the addition of a singlet superfield, leading to two additional neutral Higgs bosons. The tree-level Higgs mass receives additional contributions proportional to the square of the superpotential coupling $\lambda$ between the singlet and the doublet Higgs sectors and thus is no longer bounded from above by $m_Z$. Such contributions become negligible for large values of $\tan\beta$,  the ratio of the two Higgs doublet vacuum expectation values (VEVs).  Therefore, for sizable values of $\lambda$ and values of $\tan\beta$ of order one, an observationally consistent Higgs mass may be obtained without the need for large radiative corrections, enabling a more natural breaking of the electroweak symmetry than in the MSSM. 

The SM-like properties of the 125 GeV Higgs boson in both the MSSM and the NMSSM may be ensured via the decoupling limit, where all the Higgs bosons (excluding the observed Higgs boson with a mass of 125 GeV) are much heavier than the electroweak scale. However, the decoupling limit is not the only way to achieve a SM-like Higgs boson, as observed in Ref.~\cite{Gunion:2002zf} and rediscovered recently in Refs.~\cite{Delgado:2013zfa, Craig:2013hca, whitepaper, Carena:2013ooa, Haber:2013mia}: a SM-like Higgs can be obtained by way of the ``alignment limit," where one of the neutral Higgs mass eigenstates is approximately aligned in field space with the Higgs doublet VEV. Subsequent studies  have continued to focus on the alignment limit in 2HDMs~\cite{Carena:2014nza, Dev:2014yca}. In particular, approximate alignment may be obtained in the MSSM for moderate to large values of $\tan\beta$ and for large values of the ratio $\mu A_t/M_S^2$, where $A_t$ is the stop mixing mass parameter, $\mu$ is the supersymmetric Higgsino mass parameter, and $M_S^2$ is the average of the two stop squared-masses \cite{Carena:2013ooa}. Moreover, there is an interesting complementarity between precision measurements of the SM-like Higgs properties and direct searches for non-standard Higgs bosons in the MSSM~\cite{Carena:2014nza}.

In this work we  extend the study of alignment without decoupling beyond the 2HDM to the NMSSM where there is an additional singlet scalar as well as the two doublet scalars. In fact, it will become clear that our analysis is quite general and can be applied even beyond the NMSSM. 
We demonstrate that the alignment conditions in the NMSSM Higgs sector are fulfilled in regions of parameters space consistent with a natural breaking of the electroweak symmetry, where the stop mass parameters are of the order of the electroweak scale. Moreover, under the assumption that all couplings remain perturbative up to the Planck scale, we show that the requirements of natural electroweak symmetry breaking and the alignment limit in the Higgs sector lead to well defined spectra for Higgs bosons and Higgsinos that may be tested experimentally in the near future at the LHC.

 There have been several recent works analyzing similar questions in the 
NMSSM after the discovery of the Higgs boson (for example, see Refs.~\cite{Ellwanger:2011aa}--\cite{King:2014xwa}).  
In particular, in  Ref.~\cite{King:2014xwa} a numerical scan over the NMSSM parameter space was employed to
determine the regions of the NMSSM parameter space that are consistent with present Higgs boson precision measurements and searches for other Higgs boson states and supersymmetric particles.  These parameter regions include those that are consistent with the alignment conditions examined in this paper.   Consequently, the benchmark scenarios presented in  Ref.~\cite{King:2014xwa} exhibit similar features to the ones presented in Appendix~\ref{Benchmarks} of this work.  In contrast to previous studies, in this paper we develop an analytic understanding of the alignment conditions that lead to consistency with the observed Higgs physics, and we present a detailed phenomenological study of the non-SM-like Higgs boson couplings to fermions and gauge bosons.     

This paper is organized as follows. In section~\ref{alignment}, we analyze the alignment conditions in extensions of the Higgs sector with two doublets and one singlet, and discuss the NMSSM as a particular example.  In section \ref{Higgspheno}, we examine the associated Higgs phenomenology.  In section \ref{HiggsDecays}, we study the Higgs production and decay modes relevant for run~2 of the LHC, and we present our conclusions in section \ref{Conclusions}.  In Appendix A, details of the scalar potential in the Higgs basis for the two doublets/one singlet model are given, along with the corresponding expressions for the NMSSM Higgs sector.   Explicit expressions for the rotation matrix elements relating the Higgs basis and mass eigenbasis are provided in Appendix~B.  In Appendix C, we exhibit the trilinear scalar self-couplings and the couplings of the neutral scalars to the $Z$ boson.  Finally, in Appendix D we present several NMSSM benchmark scenarios that illustrate features of the Higgs phenomenology considered in this paper.

\section{NMSSM Alignment Conditions}
\label{alignment}

\subsection{Generalities}
\label{sect:general}

The scalar sector of the NMSSM consists of two electroweak doublets and one electroweak singlet. We first present some general considerations of the ``alignment limit" in the Higgs sector that can be applied broadly to any Higgs sector made up of two doublets and one singlet. Similar to the case of the 2HDM, the discussion is most transparent when one adopts the {\em Higgs basis} \cite{higgsbasis,Branco:1999fs}, in which only the neutral component of one of the two doublet scalars acquires a non-zero VEV.\footnote{Here, we are implicitly assuming that no charge-breaking minima exist; that is, all charged scalar VEVs are zero.  As shown in Ref.~\cite{Ellis:1988er}, the condition for a local charge-conserving minimum in the NMSSM is equivalent to the requirement that the physical charged Higgs bosons of the model have positive squared-masses.} 

In the paradigm of spontaneous electroweak symmetry breaking, a tree-level scalar coupling to massive electroweak gauge bosons is directly proportional to the strength of the VEV residing in any scalar with non-trivial SU(2)$_L\times$U(1)$_Y$ quantum numbers.  Thus, in the Higgs basis, if the scalar doublet Higgs field with the nonzero VEV coincides with one of the scalar mass eigenstates (the so-called \textit{alignment limit}), then this state couples to $W$ and $Z$ bosons with full SM strength and is the natural candidate to be the SM-like 125 GeV Higgs boson. Non-zero couplings of the other mass eigenstates to the massive gauge bosons arise only away from the alignment limit. 

In the Higgs basis, we define the hypercharge-one doublet fields $H_1$ and $H_2$ such that the VEVs of the corresponding neutral components are given by
\be \label{hbasisvev}
\langle H^0_1\rangle=\frac{v}{\sqrt{2}}\,,\qquad\quad
\langle H^0_2\rangle=0\,,
\ee
where $v\simeq 246$~GeV. The singlet scalar field $S$ also possesses a non-zero VEV, 
\be \label{svev}
\langle S\rangle\equiv v_s\,.   
\ee
We shall make the simplifying assumption that the scalar potential preserves CP, which is \textit{not} spontaneously broken in the vacuum.  Thus, the phases of the Higgs fields can be chosen such that $v_s$ is real.  We then define the following neutral scalar fields,
\bea
\label{eq:HSMdef}
H^{\rm SM}& \equiv &\sqrt{2}\,{\rm Re}~H_1^0-v\,,\qquad\quad H^{\rm NSM}\equiv \sqrt{2}\,{\rm Re}~H_2^0\,,\qquad H^{\rm S}\equiv \sqrt{2}\,({\rm Re}~S-v_s)\,,\\
\label{eq:ASMdef}
A^{\rm SM}& \equiv &\sqrt{2}\,{\rm Im}~H_1^0\,,\qquad\qquad\quad A^{\rm NSM}\equiv \sqrt{2}\,{\rm Im}~H_2^0\,,\qquad A^{\rm S}\equiv \sqrt{2}\,{\rm Im}~S\,,
\eea
where $A^{\rm SM}$ is the Goldstone field that is absorbed by the $Z$ and provides its longitudinal degree of freedom.   Under the assumption of CP conservation, the scalar fields $H^{\rm SM}$, $H^{\rm NSM}$ and $H^{\rm S}$ mix to yield three neutral CP-even scalar mass eigenstates of the following real symmetric squared-mass matrix,
\be
\label{eq:higgsbasism2}
\mathcal{M}_S^2 =
\left(
\begin{array}{ccc}
\mathcal{M}^2_{11} & \mathcal{M}^2_{12}  & \mathcal{M}^2_{13}\\  
\mathcal{M}^2_{12} & \mathcal{M}^2_{22}  & \mathcal{M}^2_{23}\\  
\mathcal{M}^2_{13} & \mathcal{M}^2_{23}  & \mathcal{M}^2_{33}
\end{array} 
\right) \ ,
\ee
The exact alignment limit is realized when the following two conditions are satisfied
\be
 \mathcal{M}^2_{12}= 0 \ , \qquad  \mathcal{M}^2_{13} = 0 \ .
\label{eq:alt}
\ee
In this case, $H^{\rm SM}$ is a CP-even mass-eigenstate scalar with squared mass $\mathcal{M}^2_{11}$, and its couplings to massive gauge bosons and fermions are precisely those of the SM Higgs boson. In practice, we only need to require that the observed 125 GeV scalar (henceforth denoted by~$h$) is SM-{\em like}, which implies that the alignment limit is approximately realized.  In this case, \eq{eq:alt} is replaced by the following conditions:
\be \label{approxalign}
 \mathcal{M}^2_{12} \ll \mathcal{O}(v^2) \ , \qquad\quad  \mathcal{M}^2_{13} \ll \mathcal{O}(v^2) \ ,
\ee
which imply that
\be \label{approxmh}
m_h^2\simeq \mathcal{M}^2_{11}=(125~{\rm GeV})^2\,.
\ee
Corrections to \eq{approxmh} appear only at second order in the perturbative expansion and thus are proportional to the squares of $\mathcal{M}^2_{12}$ or $\mathcal{M}^2_{13}$, respectively.

We shall denote the CP-even Higgs mass-eigenstate fields by $h$, $H$, and $h_S$,  where $h$ is identified with the observed SM-like Higgs boson, $H$ is a dominantly doublet scalar field and $h_S$ is a dominantly singlet scalar field.\footnote{The special case where the mass eigenstate is evenly split by the doublet and the singlet fields constitutes a region of parameter space of measure zero and will be ignored in this work.} The mass-eigenstate fields $\{h,H,h_S\}$ are related to $\{H^{\rm SM},H^{\rm NSM},H^{\rm S}\}$ by a real orthogonal matrix $\mathcal{R}$,
\be \label{mixmat}
\begin{pmatrix}  h \\ H \\  h_S\end{pmatrix}=\mathcal{R}
  \begin{pmatrix} \phm H^{\rm SM} \\ H^{\rm NSM} \\ \phm H^{\rm
      S}\end{pmatrix}\,,
  \ee
where\footnote{$R_{23}^\prime$ is an improper rotation matrix, resulting in $\det\mathcal{R}=-1$.  The reason for this choice is addressed below.}
 \bea \label{rmatrix}
\mathcal{R}=R_{23}^\prime R_{13}R_{12} &=& \left( \begin{array}{ccc}
1\quad &\phm 0 &\phm 0\\
0\quad &\,\,\,\, -c_{23} \phm & -s_{23}\\
0\quad &-s_{23} &\phm c_{23}\end{array}\right)
\left( \begin{array}{ccc}
\phm c_{13}\quad &\phm 0 & \phm s_{13}\\
\phm 0\quad & \phm 1&\phm 0\\
-s_{13}\quad &\phm 0 &\phm c_{13}\end{array}\right)
\left( \begin{array}{ccc}
\phm c_{12}\,\, &\phm s_{12}\quad &\phm 0\\
-s_{12}\,\, &\phm c_{12}\quad &\phm 0\\
\phm 0\,\, &\phm 0\quad &\phm 1\end{array}\right)
\nonumber \\[10pt]
&=&
\left( \begin{array}{ccc}
  c_{13}c_{12} & \quad c_{13}s_{12} &\quad  s_{13}\\[6pt]
    c_{23}s_{12}+c_{12}s_{13}s_{23} & \quad
    -c_{12}c_{23}+s_{12}s_{13}s_{23}
    & \quad -c_{13} s_{23} \\[6pt]
     -c_{12}c_{23}s_{13}+s_{12}s_{23} & \quad  
  -c_{23}s_{12}s_{13}-c_{12}s_{23} & \quad 
   \phm c_{13}c_{23}\end{array}\right)\,,
\eea
where $c_{ij}\equiv \cos\theta_{ij}$ and $s_{ij}\equiv\sin\theta_{ij}$. The mixing angles $\theta_{ij}$ are defined modulo $\pi$.  It is convenient to choose $|\theta_{ij}|\leq\half\pi$, in which case $c_{ij}\geq 0$.  The mixing angles $\theta_{ij}$ are determined by the diagonalization equation,
\be \label{diag}
\mathcal{R}~ \mathcal{M}^2_S~\mathcal{R}^T={\rm
  diag}(m_h^2,m_H^2,m_{h_S}^2)\,.
\ee
We can use \eqs{rmatrix}{diag} to obtain exact expressions for the mixing angles in terms of $m^2_h$ and the elements of $\mathcal{M}^2_S$ as follows.  Multiply \eq{diag} on the left by $\mathcal{R}^T$ and consider the first column of the resulting matrix equation.  This yields three equations, which can be rearranged into the following form,
\bea
x\mathcal{M}^2_{12}+y \mathcal{M}^2_{13}&=& m_h^2-\mathcal{M}^2_{11}\,,\label{e1}\\
x\mathcal{M}^2_{23}+\mathcal{M}^2_{13}&=& y\bigl(m_h^2-\mathcal{M}^2_{33}\bigr)\,,\label{e2}\\
y\mathcal{M}^2_{23}+\mathcal{M}^2_{12}&=& x\bigl(m_h^2-\mathcal{M}^2_{22}\bigr)\,,\label{e3}
\eea
where $x\equiv s_{12}/c_{12}$ and $y\equiv s_{13}/(c_{12}c_{13})$ and correspond to the ratios of the NSM and S components to the SM component of $h$, respectively. Eliminating $y$ from \eqs{e2}{e3} yields an expression for $x$.  To obtain the corresponding expression for $y$, it is more convenient to return to \eqs{e2}{e3} and eliminate $x$.  The resulting expressions are:   
\bea
x\equiv\frac{s_{12}}{c_{12}}&=&\frac{\mathcal{M}^2_{13}\mathcal{M}^2_{23}-\mathcal{M}^2_{12}\bigl(\mathcal{M}^2_{33}-m_h^2\bigr)}{\bigl(\mathcal{M}^2_{22}-m_h^2\bigr)\bigl(\mathcal{M}^2_{33}-m_h^2\bigr)-\mathcal{M}^4_{23}}\,,\label{x}\\[10pt]
y\equiv\frac{s_{13}}{c_{12}c_{13}}&=&\frac{\mathcal{M}^2_{12}\mathcal{M}^2_{23}-\mathcal{M}^2_{13}\bigl(\mathcal{M}^2_{22}-m_h^2\bigr)}{\bigl(\mathcal{M}^2_{22}-m_h^2\bigr)\bigl(\mathcal{M}^2_{33}-m_h^2\bigr)-\mathcal{M}^4_{23}}\,,\label{y}
\eea
which are equivalent to \eqs{h1}{h2} of Appendix~\ref{app:eigenstates}. Inserting the above results for $x$ and $y$ back into \eq{e1} then yields a cubic polynomial equation for $m_h^2$, which we recognize as the characteristic equation obtained by solving the eigenvalue problem for $\mathcal{M}_S^2$. The approximate alignment conditions given in \eq{approxalign} imply that $|s_{12}|\ll 1$ and $|s_{13}|\ll 1$, in which case one can approximate $m_h^2\simeq \mathcal{M}^2_{11}$ in \eqs{x}{y} to very good accuracy.

Likewise, repeating the above exercise for $H$,  we can obtain the ratio of the S component to the NSM component of $H$ [cf.~\eq{H2}],
\be \label{z}
\frac{c_{13}s_{23}}{c_{12}c_{23}-s_{12}s_{13}s_{23}}=\frac{\mathcal{M}_{23}^2(\mathcal{M}_{11}^2-m_H^2)-\mathcal{M}_{12}^2\mathcal{M}_{13}^2}{\mathcal{M}_{23}^4+(\mathcal{M}_{11}^2-m_H^2)(m_H^2-\mathcal{M}_{33}^2)}\,.
\ee
In the exact alignment limit where $\mathcal{M}_{12}^2=\mathcal{M}_{13}^2=0$ (and hence $s_{12}=s_{13}=0$), and when $\mathcal{M}_{23}^2 \ll m_H^2$, \eq{z} reduces to
\be
\frac{s_{23}}{c_{23}}=\frac{\mathcal{M}_{23}^2}{m_H^2-\mathcal{M}_{33}^2}\,.
\ee

Our choice of $\det\mathcal{R}=-1$ in \eq{mixmat} requires an explanation.   In the limit where there is no mixing of $H^{\rm S}$ with the Higgs doublet fields $H^{\rm SM}$ and $H^{\rm NSM}$, we have $s_{13}=s_{23}=0$ (and $c_{13}=c_{23}=1$ by convention), which
yields~\footnote{In the original 2HDM literature, the CP-even Higgs mixing angle was defined by a transformation that rotated $\{H^{\rm SM},H^{\rm NSM}\}$ into $\{H,h\}$. With this ordering of the mass eigenstates, the determinant of the corresponding transformation matrix is $+1$.} 
\be \label{transf}
h=c_{12}H^{\rm SM}+s_{12}H^{\rm NSM}\,,\qquad
H=s_{12}H^{\rm SM}-c_{12}H^{\rm NSM}\,.
\ee
The transformation from $\{H^{\rm SM},H^{\rm NSM}\}$ to $\{h,H\}$ given in \eq{transf} employs a $2\times 2$ orthogonal matrix of determinant $-1$. Indeed, in the standard conventions of the 2HDM literature (see Refs.~\cite{Gunion:1989we,Branco:2011iw} for reviews), we identify $c_{12}=\sin(\beta-\alpha)$ and $s_{12}=\cos(\beta-\alpha)$.

We next turn to the Higgs couplings to vector bosons and fermions. The interaction of a neutral Higgs field with a pair of massive gauge bosons $VV=W^+ W^-$ or $ZZ$ arises from scalar field kinetic energy terms after replacing an ordinary derivative with a covariant derivative when acting on the electroweak doublet scalars.  After spontaneous symmetry breaking, only the interaction term $H^{\rm SM}VV$ is generated.\footnote{In the Higgs basis there is no $H^{\rm NSM} VV$ and $H^S VV$ interactions since $\langle H_2^0\rangle=0$ and $H^S$ is an electroweak singlet.}
Using \eq{mixmat},
\be \label{hsmeq}
H^{\rm SM}=\mathcal{R}_{11} h+\mathcal{R}_{21}H+\mathcal{R}_{31}h_S\,,
\ee
which then yields the following couplings normalized to the corresponding SM values,
\bea
g_{hVV}&=&\mathcal{R}_{11}=c_{13}c_{12}\,,\label{hV}\\
g_{HVV}&=&\mathcal{R}_{21}=c_{23}s_{12}+c_{12}s_{13}s_{23}\,,\label{HV}\\
g_{h_S VV}&=&\mathcal{R}_{31}=-c_{12}c_{23}s_{13}+s_{12}s_{23}\,.\label{SV}
\eea
Note that in the limit where there is no mixing of $H^{\rm S}$ with the Higgs doublet fields $H^{\rm SM}$ and $H^{\rm NSM}$,  we recover the standard 2HDM expressions $g_{hVV}=\sin(\beta-\alpha)$ and $g_{HVV}=\cos(\beta-\alpha)$.

For the Higgs interactions with the fermions, we employ the so-called Type-II Higgs-fermion Yukawa couplings~\cite{Hall:1981bc} as mandated by the holomorphic superpotential~\cite{mssmhiggs,Gunion:1984yn},\footnote{Here, we neglect the full generation structure of the Yukawa couplings and focus on the couplings of the Higgs bosons to the third generation quarks.}
\be \label{HUHD}
-\mathscr{L}_{\rm Yuk}=\epsilon_{ij}\bigl[h_b \overline{b}_R H_d^i Q^j_L+h_t\overline{t}_R Q^i_L H_u^j\bigr]+{\rm h.c.}\,,
\ee
where $Q_L =(u,d)$. The scalar doublet fields $H_d$ and $H_u$ have hypercharges $-1$ and $+1$, respectively, and define the SUSY basis.  In the SUSY basis, the corresponding neutral VEVs are denoted by\footnote{Here, we deviate from the conventions of Ref.~\cite{Ellwanger:2009dp}, where all VEVs are defined without the $\sqrt{2}$\, factor.  In this latter convention (not used in this paper), $v=174$~GeV.}
\be \label{hvev}
\langle H_d^0\rangle\equiv \frac{v_d}{\sqrt{2}}\,,\qquad\quad
\langle H_u^0\rangle\equiv \frac{v_u}{\sqrt{2}}\,,
\ee
where
$
v^2\equiv |v_d|^2+|v_u|^2=(246~{\rm GeV})^2\,,
$
is fixed by the relation $m_W\equiv \half gv$.
Without loss of generality, the phases of the Higgs fields can be chosen such that both $v_u$ and $v_d$ are non-negative.  The ratio of the VEVs defines the parameter
\be \label{tanbdef}
\tan\beta\equiv\frac{v_u}{v_d}\,,
\ee
where the angle $\beta$ represents the orientation of the SUSY basis with respect to the Higgs basis.   To relate the doublet fields $H_d$ and $H_u$ to the hypercharge-one, doublet Higgs basis fields $H_1$ and $H_2$ defined above, we first define two hypercharge one, doublet scalar fields, $\Phi_d$ and $\Phi_u$ following the notation of \cite{Gunion:1984yn},
\be
\Phi_d^j\equiv\epsilon_{ij}H_d^{*\,i}\,,\qquad\quad \Phi_u^j=H_u^j\,.
\ee
Then, the Higgs basis fields are defined by
\be \label{cphiggsbasisfields}
H_1=\begin{pmatrix}H_1^+\\ H_1^0\end{pmatrix}\equiv \frac{v_d \Phi_d+v_u\Phi_u}{v}\,,
\qquad\quad H_2=\begin{pmatrix} H_2^+\\ H_2^0\end{pmatrix}\equiv\frac{-v_u \Phi_d+v_d\Phi_u}{v}
 \,.
\ee

In terms of the Higgs basis fields, the neutral CP-even Higgs interactions given in \eq{HUHD} can be rewritten as
\be \label{lyuk}
\mathscr{L}_{\rm Yuk}=\frac{m_t}{v}\overline t_L t_R\left(H^{\rm SM}+\cot\beta H^{\rm NSM}\right)
+\frac{m_b}{v}\overline b_L b_R\left(H^{\rm SM}-\tan\beta H^{\rm NSM}\right)+{\rm h.c.}\,,
\ee
after identifying $h_t=\sqrt{2}m_t/v_u$ and $h_b=\sqrt{2}m_b/v_d$.  Using \eq{mixmat},
\be \label{hnsmeq}
H^{\rm NSM}=\mathcal{R}_{12} h+\mathcal{R}_{22}H+\mathcal{R}_{32}h_S\,,
\ee
along with \eq{hsmeq}, we can rewrite \eq{lyuk} as
\bea
\!\!\!\!\!\!\!\!
\mathscr{L}_{\rm Yuk}&=&\frac{m_t}{v}\overline t_L t_R\biggl\{(\mathcal{R}_{11}+\mathcal{R}_{12}\cot\beta)h
+(\mathcal{R}_{21}+\mathcal{R}_{22}\cot\beta)H+(\mathcal{R}_{31}+\mathcal{R}_{32}\cot\beta)h_S\biggr\}\nonumber \\[8pt]
&+&\frac{m_b}{v}\overline b_L b_R\biggl\{(\mathcal{R}_{11}-\mathcal{R}_{12}\tan\beta)h
+(\mathcal{R}_{21}-\mathcal{R}_{22}\tan\beta)H+(\mathcal{R}_{31}-\mathcal{R}_{32}\tan\beta)h_S\biggr\}\,. \label{YUK}
\eea 
In the limit where there is no mixing of $H^{\rm S}$ with the Higgs doublet fields $H^{\rm SM}$ and $H^{\rm NSM}$,  we have $\mathcal{R}_{11}=-\mathcal{R}_{22}=\sin(\beta-\alpha)$,
$\mathcal{R}_{12}=\mathcal{R}_{21}=\cos(\beta-\alpha)$, $\mathcal{R}_{33}=1$, and all other matrix elements of $\mathcal{R}$ vanish.  Inserting these values above yields the standard 2HDM Type-II Yukawa couplings of the neutral CP-even Higgs bosons.

Current experimental data on measurements and searches in the $WW$ and $ZZ$ channels already place strong constraints on the entries of the squared-mass matrix given in Eq.~(\ref{eq:higgsbasism2}). In addition, under the assumption of Type-II Yukawa couplings, the Higgs data in the fermionic channels will also yield additional constraints. 

It is convenient to rewrite the rotation matrix $\mathcal{R}$ [defined in \eq{rmatrix}] as
\be 
\left(\begin{array}{c}
h \\
H \\
h_S
\end{array}\right) = 
\left(\begin{array}{ccc}
\kappa^h_{\rm SM} & \kappa^h_{\rm NSM} & \kappa^h_{\rm S} \\
\kappa^H_{\rm SM} & \kappa^H_{\rm NSM} & \kappa^H_{\rm S} \\
\kappa^{h_S}_{\rm SM} & \kappa^{h_S}_{\rm NSM} & \kappa^{h_S}_{\rm S}
\end{array}\right)  
\left(\begin{array}{c}
H^{\rm SM} \\
H^{\rm NSM}\\
H^{\rm S}
\end{array}\right) \ .
\label{eq:mixingmatrixMain}
\ee 
Explicit expressions for the entries of the mixing matrix of \eq{eq:mixingmatrixMain} are given in Appendix~\ref{app:eigenstates}, following the procedure used to derive \eqs{x}{y}.

On the one hand, the non-SM components of the 125 GeV Higgs will be constrained by the LHC measurements of the properties of the 125 GeV Higgs boson. On the other hand, a small, non-zero component in $H^{\rm SM}$ of the non-SM-like Higgs bosons induces a small coupling to $W$ and $Z$ bosons, which can be constrained by searches for exotic resonances in the $WW$ and $ZZ$ channels.  In the notation of \eq{eq:mixingmatrixMain},
the couplings of the three CP-even Higgs states $\phi=\{h, H, h_S\}$ to  the gauge bosons $VV$ [cf.~\eqst{hV}{SV}] and the up and down-type fermions [cf.~\eq{YUK}], normalized to those of the SM Higgs boson are given by,
\bea 
g_{\phi VV} & = & \kappa^\phi_{\rm SM}\,,\label{eq:phicouplings1}
\\
g_{\phi t \bar{t}} & = & \kappa^\phi_{\rm SM} + \kappa^\phi_{\rm NSM}\cot\beta\,,\label{eq:phicouplings2}
\\
g_{\phi b \bar{b}} & = & \kappa^\phi_{\rm SM} - \kappa^\phi_{\rm NSM} \tan\beta\,. 
\label{eq:phicouplings3}
\eea
These couplings may be used to obtain the production cross section, such as in the gluon fusion channel, of these states, which is mostly governed by $g_{\phi t \bar{t}}$, as well as the branching ratios, under the assumption that the decay into non-standard particles is suppressed.  Although a more detailed study of the Higgs phenomenology will be presented below, it is useful to obtain first an understanding of the bounds on these  components based on current Higgs measurements as well as  searches for exotic Higgs resonances. 

\begin{figure}[t!]
\subfloat[]{\includegraphics[width=3.in, angle=0]{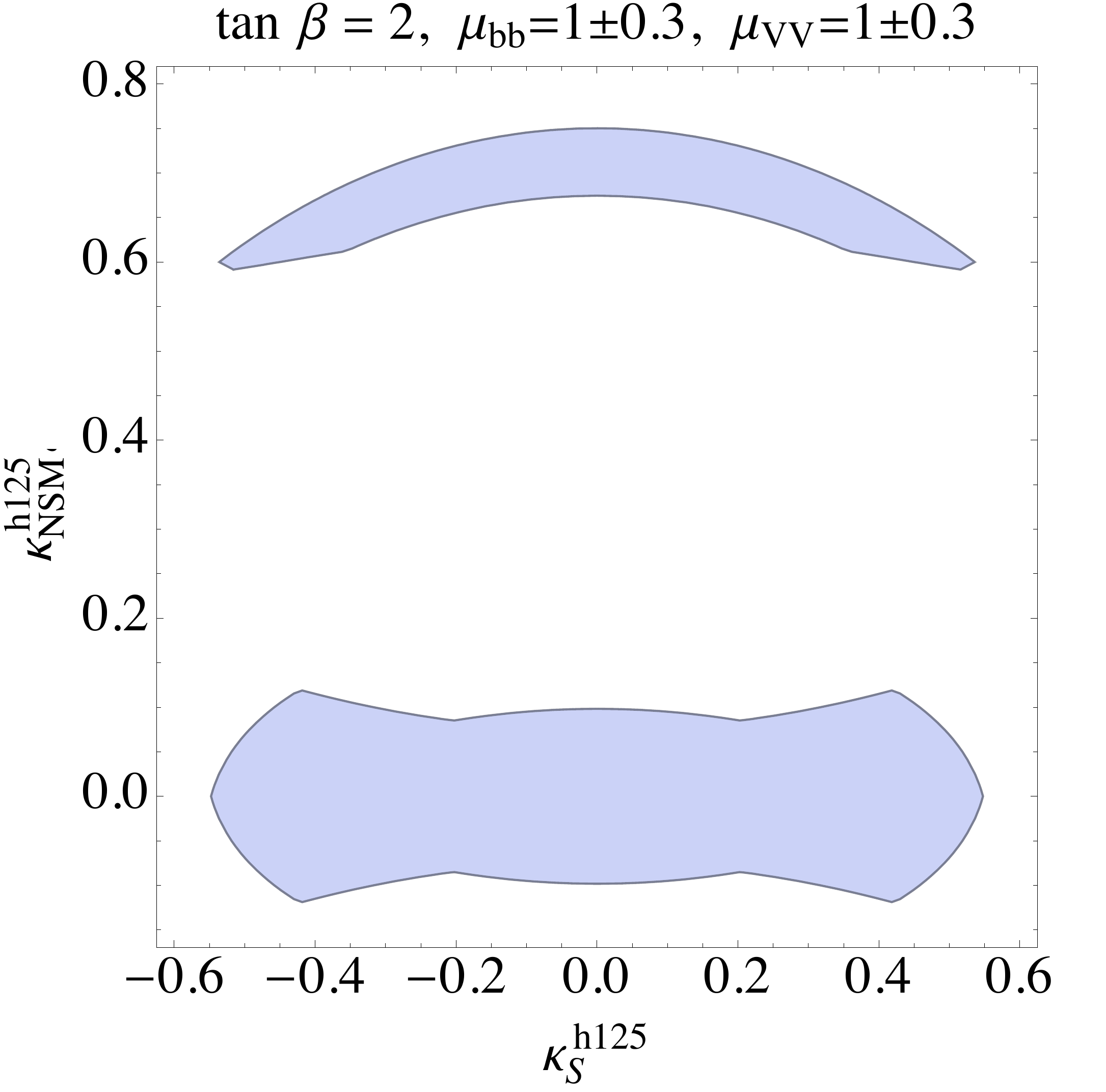}}~~
\subfloat[]{\includegraphics[width=3.in, angle=0]{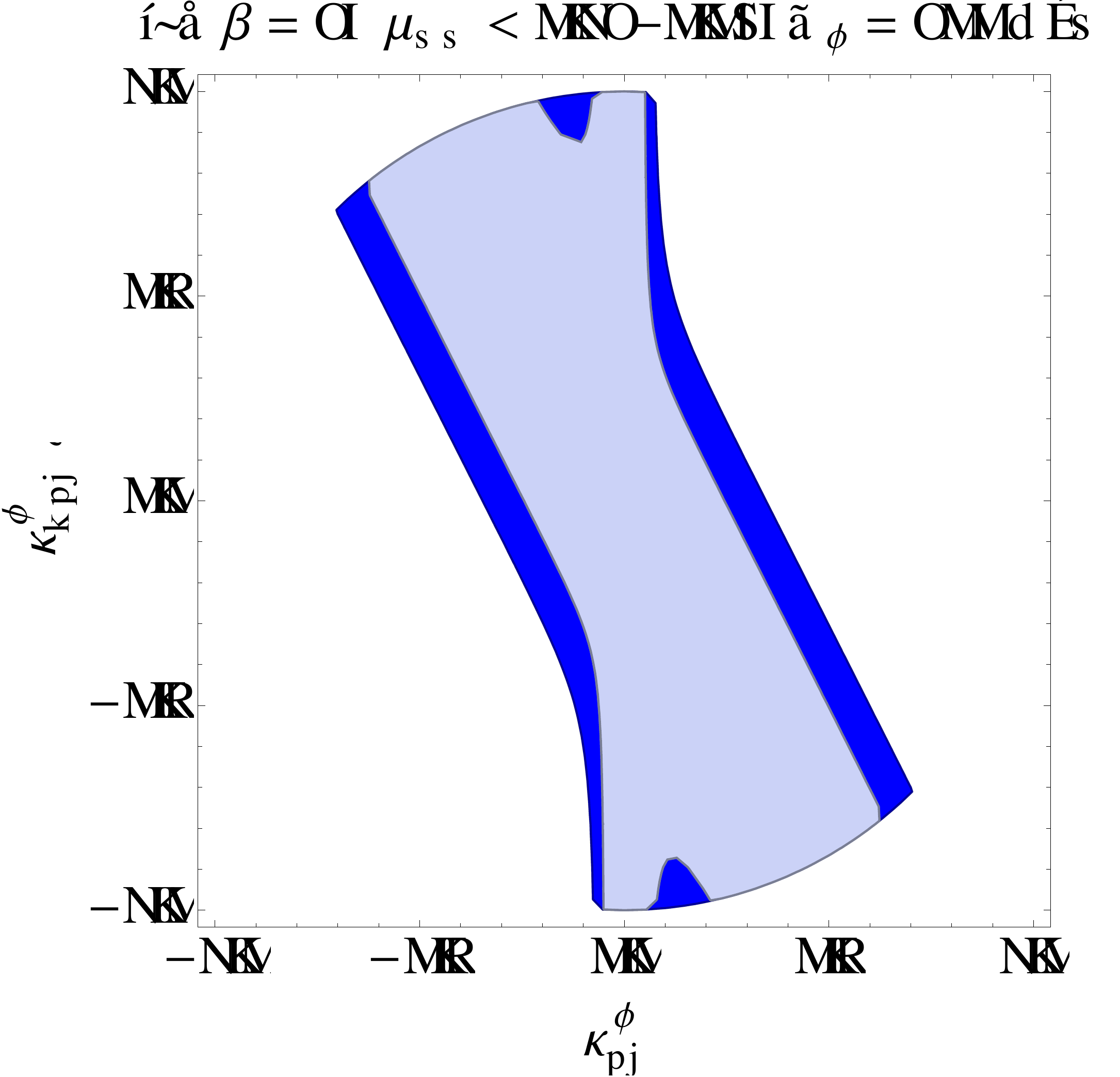}}
\caption{\label{fig:Component4}{\em In the left panel we show the constraints on the possible singlet and non-SM doublet  component of the 125 {\rm GeV} state derived from precision measurements on its production cross section and branching rations. In the right panel we show the constraints on the SM and non-SM doublet component of a Higgs state coming from the searches for Higgs bosons decaying into $W$ pairs, away from the SM Higgs mass values.}}
\end{figure}

In the left panel of Fig.~\ref{fig:Component4} we show the constraints on $\kappa^h_{\rm NSM}$ and $\kappa^h_{\rm S}$ for the 125 GeV Higgs boson $h$, derived from the LHC Run 1 measurements on its production cross section and branching ratios. Here we have assumed that the decay branching fractions into bottom quarks and massive vector boson cannot deviate by more than 30\% from their SM values. In anticipation of our focus on the NMSSM, we concentrate on small values of $\tan\beta$. In the right panel of Fig.~\ref{fig:Component4}, we consider the constraints on $\kappa^{H/h_S}_{\rm SM}$ and $\kappa^{H/h_S}_{\rm NSM}$ for the non-SM-like scalars from resonance searches in the $WW$ and $ZZ$ channels~\cite{Khachatryan:2015cwa},  assuming production from gluon fusion processes.  

First note that the singlet component of the observed 125 GeV scalar, which is only constrained by its unitarity relationship with the SM and NSM components, is allowed to be quite large. However, the NSM component is restricted to be small, except in the narrow region of parameter space where the $g_{h b \bar{b}}$ coupling is approximately equal in magnitude but with opposite sign as the SM bottom Yukawa \cite{Ferreira:2014naa}.  This can only occur far away from alignment and we shall not explore this region. 

On the other hand, the search for exotic Higgs resonances puts  strong constraints on the value of $\kappa^{H/h_S}_{\rm SM}$.  These constraints are satisfied when $\kappa^{H/h_S}_{\rm SM}$ is very small, so that the decay into $WW/ZZ$ is suppressed, or when the linear combination of $\kappa^{H/h_S}_{\rm SM}$  and $\kappa^{H/h_S}_{\rm NSM}$ is such that the coupling to top quarks in \eq{eq:phicouplings2} is small, resulting in a small production rate in the gluon fusion channel.

\subsection{The $\boldsymbol{\mathbb{Z}_3}$ Invariant NMSSM}
\label{nmssm}

In this paper, we shall analyze the NMSSM under the assumption that there are no mass parameters in the superpotential, which can be ensured by imposing a $\mathbb{Z}_3$ symmetry under which all chiral superfields are transformed by a phase $e^{2\pi i/3}$.  The superpotential then must contain only cubic combinations of superfields.  The coefficients of the possible cubic terms include the usual matrix Yukawa couplings $h_d$, $h_u$ and $h_e$, the coupling $\lambda$ of the singlet to the doublet Higgs superfields, and the singlet Higgs superfield self-coupling parameter $\kappa$,
\begin{equation}
W = \lambda \widehat{S} \widehat{H}_u\cdot \widehat{H}_d + \frac{\kappa}{3} \widehat{S}^{\,3}  + h_u \widehat{Q}\cdot \widehat{H}_u\, \widehat{U}_R^c + h_d \widehat{H}_d\cdot\widehat{Q}\, \widehat{D}_R^c + h_\ell \widehat{H}_d \cdot\widehat{L}\,\widehat{E}_R^c\,,
\end{equation}
where we are following the notation for superfields given in Ref.~\cite{Ellwanger:2009dp}.  In particular, we employ the dot product notation for the singlet combination of two SU(2) doublets.  For example,
\be
 \widehat{H}_u\cdot \widehat{H}_d \equiv \epsilon_{ij}\widehat{H}_u^i \widehat{H}_d^j
 =\widehat{H}_u^+ \widehat{H}_d^--\widehat{H}_u^0 \widehat{H}_d^0\,.
 \ee
 
All Higgs mass parameters are associated with soft supersymmetry-breaking terms appearing in the scalar potential,
\bea
V_{\rm soft}  &=&  m_S^2 S^\dagger S + m_{H_u}^2 H_u^\dagger H_u + m_{H_d}^2 H_d^\dagger H_d  +  \left ( \lambda A_\lambda S H_u\cdot H_d + \tfrac{1}{3}\kappa A_\kappa S^3 + {\rm h.c.} \right)\nonumber\\
&&+ m_Q^2 Q^\dagger Q + m_U^2 U_R^{c\,\dagger} U_R^c 
+ m_D^2 D_R^{c\,\dagger} D_R^c +M_L^2 L^\dagger L+m_E^2 E_R^{c\,\dagger}E_R^c 
 \nonumber \\
&& +\left(h_u A_u Q\cdot H_u\, U_R^c + h_d A_d H_d\cdot Q\, D_R^c + h_\ell A_\ell H_d \cdot L \,E_R^c + {\rm h.c.} \right), \label{soft}
\eea
where the scalar component of the corresponding superfield is indicated by the same symbol but without the hat.  For completeness, we also include the soft supersymmetry-breaking terms that are associated with the squark fields (where generation labels are suppressed).

The Higgs scalar potential receives contributions from (i) soft supersymmetry-breaking terms in the scalar potential given in \eq{soft}, (ii) from the supersymmetry-conserving $D$-terms, which depend quadratically on the weak gauge couplings, and (iii) from the supersymmetry-conserving $F$-terms associated with the scalar components of the derivatives of the superpotential with respect to the Higgs, quark and lepton superfields.  Explicitly, the supersymmetry-conserving contributions to the Higgs scalar potential are given by 
\bea
V_{\rm SUSY} &=& 
\tfrac{1}{8}(g^2+g^{\prime\,2})(H_u^\dagger H_u-H_d^\dagger H_d)^2+\half g^2 |H_d^\dagger H_u|^2
+|\lambda|^2|H_u\cdot H_d|^2 \nonumber \\
&&+|\lambda|^2 S^\dagger S (H_u^\dagger H_u + H_d^\dagger H_d) + |\kappa|^2 (S^\dagger S)^2 + 
(\kappa^* \lambda S^{*\,2} H_u\cdot H_d + {\rm h.c.}) \,.
\eea

In the MSSM, the quartic terms of the Higgs scalar potential are proportional to gauge couplings. As a result, the tree-level mass of the observed SM-like Higgs boson can be no larger than $m_Z$.  To obtain the observed Higgs mass of 125 GeV, significant radiative loop corrections (dominated by loops of top quarks and top squarks) must be present. A novel feature of the NMSSM is the appearance of tree-level contributions to the Higgs doublet quartic couplings that do not depend on the gauge couplings.  The new quartic couplings in $V_{\rm SUSY}$ play a very important role in the Higgs phenomenology.  Moreover, they provide a new tree-level source for the mass of the SM-like Higgs boson such that the observed 125 GeV mass can be achieved without the need of large radiative corrections. The structure of the scalar potential of the NMSSM allows for the alignment of one of the mass eigenstates of the CP-even Higgs bosons with the Higgs basis field $H_1$ (which possesses the full Standard Model VEV), while at the same time yielding a sizable tree-level contribution to the observed Higgs mass naturally, without resorting to large radiative corrections.

To simplify the analysis, we henceforth assume that the Higgs scalar potential and vacuum are CP-conserving.  That is, given the Higgs potential,
\be \label{vhiggs}
\mathcal{V}=V_{\rm SUSY}+V^{(1)}_{\rm soft}\,,
\ee
where $V^{(1)}_{\rm soft}$ is the first line of \eq{soft}, we assume that all the parameters
of $\mathcal{V}$ can be chosen to be real. The CP conservation of the vacuum can be achieved by assuming that the product $\lambda\kappa$ is real and positive, as shown in \cite{Ellis:1988er}. Minimizing the Higgs potential, the neutral Higgs fields acquire VEVs denoted by \eqs{svev}{hvev}. The non-zero singlet VEV, $v_s$, yields effective $\mu$ and $B$ parameters,
\be \label{mub}
\mu\equiv \lambda v_s\,,\qquad B\equiv A_\lambda+\kappa v_s\,.
\ee

Conditions for the minimization of the Higgs potential allows one to express the quadratic mass parameters $m_S^2$, $m_{H_u}^2$ and $m_{H_d}^2$ in terms of  the VEVs $v_u$, $v_d$, $v_s$, the $A$-parameters $A_\lambda$ and $A_\kappa$, and the dimensionless couplings that appear in the Higgs potential.  Using \eq{mub}, 
\bea
m^2_{H_d}&=&\mu B\frac{v_u }{v_d}-\mu^2-\half\lambda^2 v_u^2+\tfrac{1}{8}(g^2+g^{\prime\,2})(v_u^2-v_d^2)\,,\label{hm1}\\
m^2_{H_u}&=&\mu B\frac{v_d}{v_u}-\mu^2-\half\lambda^2 v_d^2+\tfrac{1}{8}(g^2+g^{\prime\,2})(v_d^2-v_u^2)\,,\label{hm2}\\
m^2_S&=&\half\mu B\frac{v_d v_u}{v^2_s}+\half\lambda\kappa v_d v_u-\half\lambda^2(v_d^2+v_u^2)-\kappa A_\kappa v_s-2\kappa^2 v_s^2\,.\label{hm3}
\eea

The Higgs mass spectrum can now be determined from \eq{vhiggs} by expanding the Higgs fields about their VEVs.  Eliminating the Higgs squared-mass parameters using \eqst{hm1}{hm3}, we obtain squared-mass matrices for the CP-even and the CP-odd scalars, respectively.  

To analyze the alignment conditions of the NMSSM Higgs sector, we compute the squared-mass matrices of the CP-even and the CP-odd neutral Higgs bosons in the Higgs basis.  It is convenient to introduce the squared-mass parameter $M_A^2$, which corresponds to the squared-mass of the CP-odd scalar in the MSSM,
\be \label{masq}
M_A^2\equiv\frac{\mu B}{\sinb\cosb}\,,
\ee
where $\sinb\equiv\sin\beta=v_u/v$ and $\cosb\equiv\cos\beta=v_d/v$. In the $\{H^{\rm SM}, H^{\rm NSM}, H^{\rm S}\}$ basis defined in Eq.~(\ref{eq:HSMdef}), the tree-level CP-even symmetric squared-mass matrix is given by
\begin{eqnarray}
\mathcal{M}^2_S & = & 
\!\!\! \left(
\begin{array}{ccc}
 \overline{M}_Z\lsup{\,2} c^2_{2\beta} +\frac{1}{2}\lambda ^2 v^2  & \quad -\overline{M}_Z\lsup{\,2}  s_{2\beta}
c_{2\beta} 
& \sqrt{2}\lambda v\mu
\left(1-\frac{M_A^2 }{4 \mu^2 }s^2_{2\beta}-\frac{\kappa }{2 \lambda }
s_{2\beta}\right) \\[15pt]
& \quad M_A^2+ \overline{M}_Z\lsup{\,2}  s^2_{2\beta} 
& -\frac{1}{\sqrt{2}}\lambda v \mu
c_{2\beta} \left(\frac{M_A^2}{2 \mu^2 } s_{2\beta}+\frac{\kappa
}{\lambda }\right) \\[15pt]
& & \ \ \tfrac{1}{4}\lambda ^2 v^2 s_{2\beta }\left(\frac{M_A^2} 
 {2\mu ^2} s_{2\beta }-\frac{\kappa }{\lambda }\right)+\frac{\kappa \mu}{\lambda}\left(
  A_{\kappa} +\frac{4 \kappa \mu}{\lambda}\right)
\end{array}
 \right),\nonumber\\
\label{eq:CPMM}
\end{eqnarray}
where we have introduced the squared-mass parameter,
\be
\overline{M}_Z\lsup{\,2}\equiv m_Z^2-\half\lambda^2 v^2\,,
\ee
and we have employed the shorthand notation, $c_{2\beta}=\cos 2\beta$ and $s_{2\beta}\equiv\sin2\beta$. The matrix elements below the diagonal have been omitted since their values are fixed by the symmetric property of $\mathcal{M}^2_S$.

Including the leading one-loop stop contributions, the elements of the CP-even Higgs squared-mass matrix  $\mathcal{M}_S^2$ involving the Higgs doublet components are~\footnote{For notational convenience, the subscript $S$ will be dropped when referring to the individual elements of the CP-even Higgs squared-mass matrix $\mathcal{M}^2_S$.}
\bea
\!\!\!\!\!\!\!\!
\mathcal{M}^2_{11}&=& \overline{M}_Z\lsup{\,2}  c^2_{2\beta}+\half\lambda^2 v^2
+\frac{3v^2 s_\beta^4 h_t^4}{8\pi^2}
\left[\ln \left(\frac{M_S^2}{m_t^2}\right)+\frac{X_t^2
   }{M_S^2}\left(1-\frac{X_t^2}{12 M_S^2}\right)\right] \label{mtwo11} \\[12pt]
   \!\!\!\!\!\!\!\!
   \mathcal{M}^2_{22}  & = & M_A^2+s_{2\beta}^2\left\{\overline{M}_Z\lsup{\,2} +\frac{3v^2 h_t^4}{32\pi^2}\left[\ln\left(\frac{M_S^2}{m_t^2}\right)+\frac{X_t Y_t}{M_S^2}\left(1-\frac{X_t Y_t}{12M_S^2}\right)\right]\right\}\,,
\label{mHcorr}\\[12pt]
\!\!\!\!\!\!\!\!
\mathcal{M}^2_{12}&=& -s_{2\beta}\left\{\overline{M}_Z\lsup{\,2}  c_{2\beta}-\frac{3v^2 s_\beta^2  h_t^4}{16\pi^2}\biggl[\ln\left(\frac{M_S^2}{m_t^2}\right)+\frac{X_t(X_t+Y_t)}{2M_S^2}-\frac{X_t^3 Y_t}{12 M_S^4}\biggr]\right\}\,, \label{mtwo12}
\eea   
where $M_S$ is the geometric mean of the two stop mass-eigenstates, $X_t = A_t - \mu\cot\beta$ and $Y_t = A_t + \mu \tanb$. 

In the CP-odd scalar sector, since we identify $A^{\rm SM}$ as the massless neutral Goldstone boson, the physical CP-odd Higgs bosons are identified by diagonalizing a $2\times 2$ symmetric matrix. In the $\{A^{\rm NSM},A^{\rm S}\}$ basis defined in Eq.~(\ref{eq:ASMdef}), the tree-level CP-odd symmetric squared-mass matrix is given by
\begin{eqnarray}
\mathcal{M}^2_{P} & = & \left(
\begin{array}{cc}
M_A^2 &\qquad\frac{1}{\sqrt{2}}\lambda v \left(\displaystyle\frac{M_A^2 }{2\mu }s_{2\beta}-\displaystyle\frac{3
  \kappa \mu }{\lambda }\right) \\[12pt]
  \frac{1}{\sqrt{2}}\lambda v \left(\displaystyle\frac{M_A^2 }{2\mu }s_{2\beta}-\displaystyle\frac{3
  \kappa \mu }{\lambda }\right)
& \qquad\frac{1}{2} \lambda ^2 v^2 s_{2\beta} \left(\displaystyle\frac{M_A^2}{4\mu ^2} s_{2\beta}+\displaystyle\frac{3 \kappa }{2\lambda }\right)-\displaystyle\frac{3 \kappa A_{\kappa }
  \mu }{\lambda }\\
\end{array}\right) \,.
\end{eqnarray}
We denote the CP-odd Higgs mass-eigenstate fields by $A$ and $A_S$, where $A$ is the dominantly doublet CP-odd scalar field and $A_S$ is the dominantly singlet CP-odd scalar field.

For completeness, we record the mass of the charged Higgs boson,
\be
m^2_{H^\pm}=M_A^2+m_W^2-\half\lambda^2 v^2\,,
\ee
in terms of the squared-mass parameter $M_A^2$ [cf.~\eq{masq}].
  
Exact alignment can be achieved if the following two conditions are fulfilled:  
\bea   
\mathcal{M}^2_{12}&=&\frac{1}{\tanb}\left[\mathcal{M}^2_{11}-c_{2 \beta } m_Z^2-\lambda ^2 v^2 s_{\beta }^2\right] +\frac{3 v^2 s_\beta^2 h_t^4 \mu X_t}{16 \pi ^2 M_S^2} \left(1-\frac{X_t^2}{6
   M_S^2}\right)= 0\ ,\label{m12cond}\\
\mathcal{M}^2_{13}&=&\sqrt{2} \lambda v \mu\left(1-\frac{M_A^2 s_{2 \beta }^2}{4 \mu ^2}-\frac{\kappa  s_{2 \beta }}{2 \lambda } \right) = 0\; , \,
\label{eq:alignment}
\eea
after noting that $Y_t-X_t=\mu/(s_\beta c_\beta)$. In what follows, we will study under what conditions alignment can occur in regions of parameter space where no large cancellation is necessary to achieve the spontaneous breaking of electroweak symmetry.  

 \begin{figure}[t!]
\subfloat[]{\includegraphics[width=3.2in, angle=0]{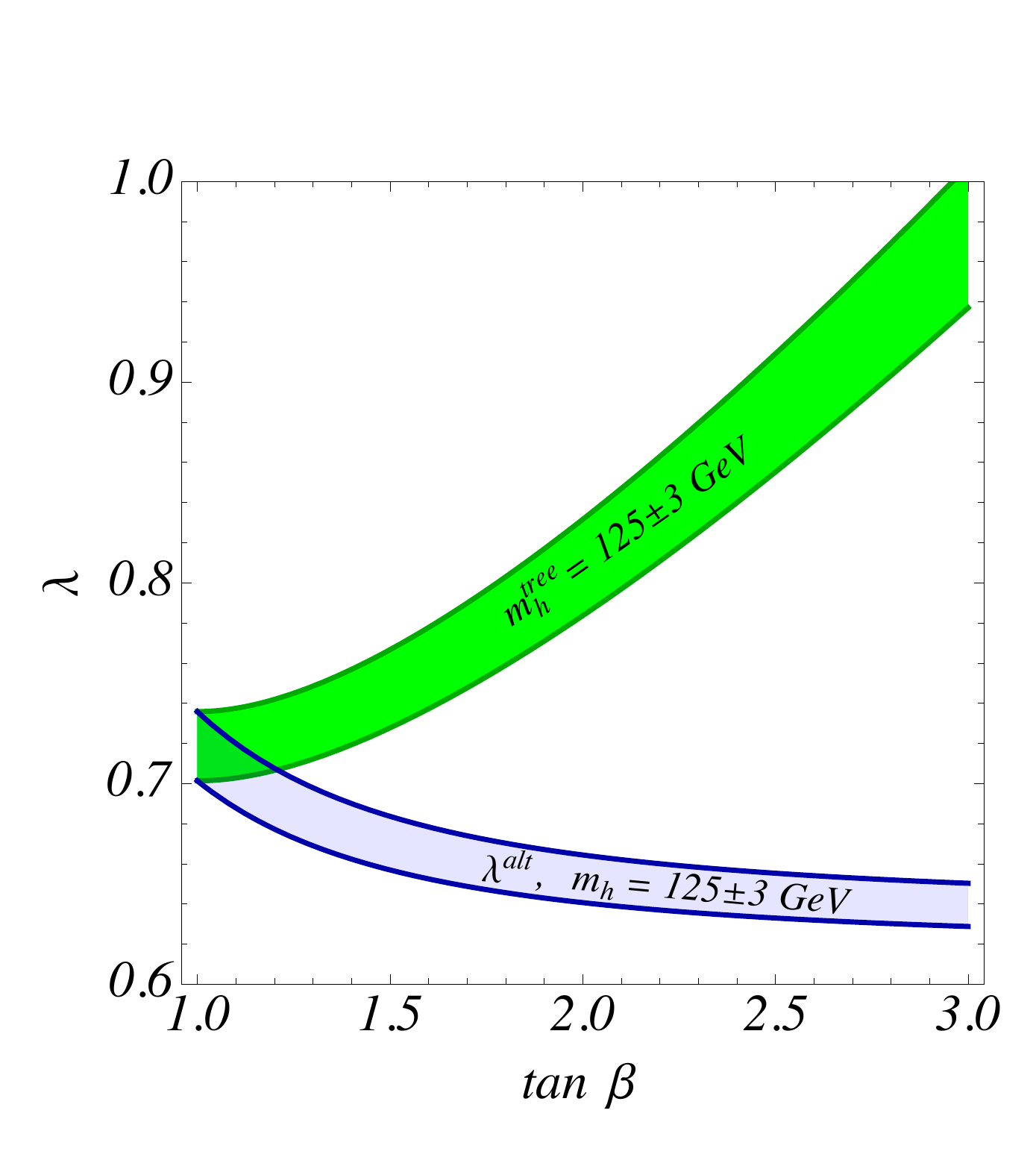}}~~
\subfloat[]{\includegraphics[width=3.2in, angle=0]{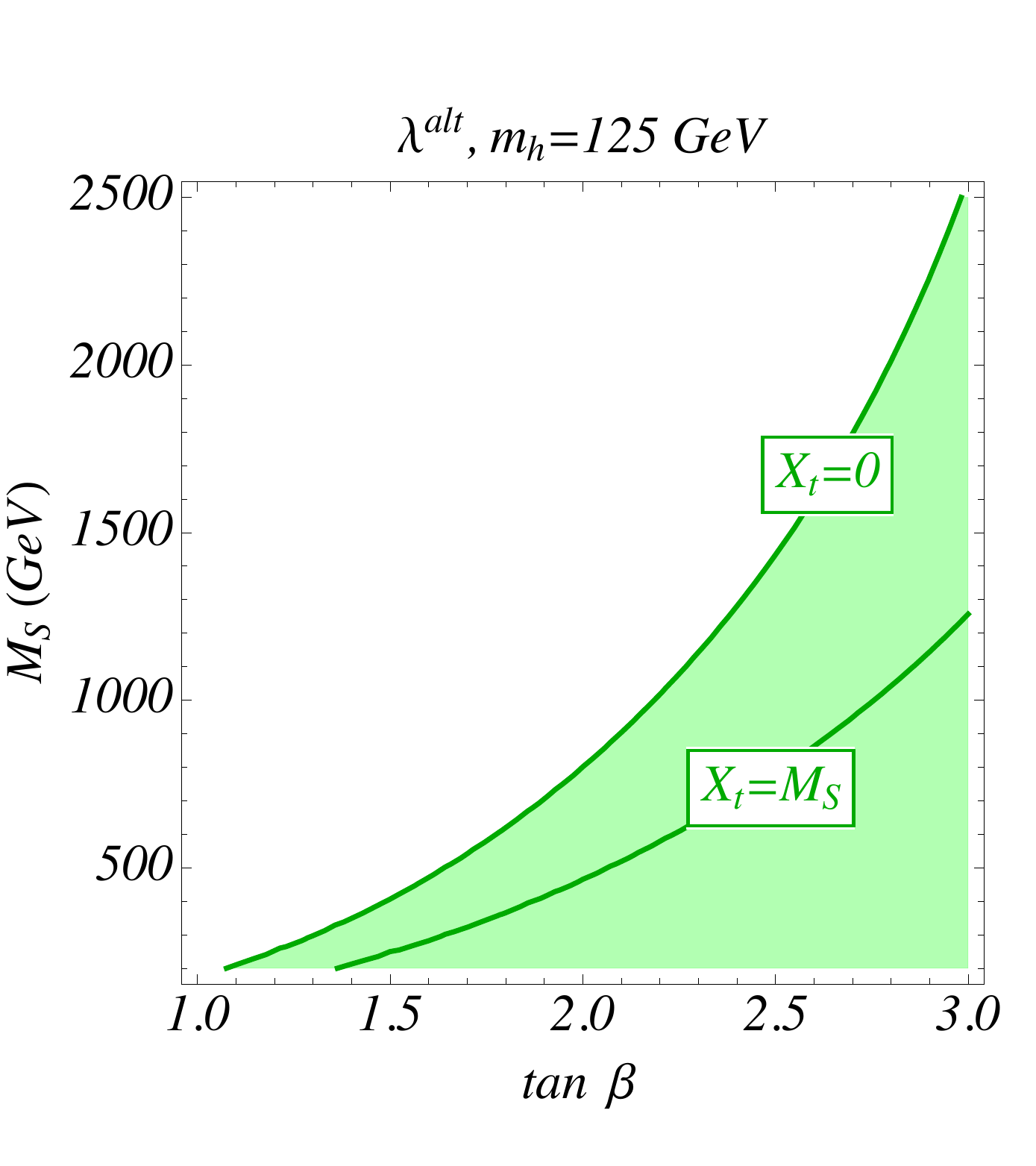}}
\caption{\label{fig:lambdafromh}{\em Left panel : The blue shaded band displays the values of $\lambda$ as a function of $\tan\beta$, necessary for alignment for $m_h = 125 \pm 3$~{\rm GeV}. Also shown in the figure as a green band are values of $\lambda$ that lead to a tree-level Higgs mass of $125 \pm 3$~{\rm GeV}.  Right panel : Values of $M_S$ necessary to  obtain a 125 {\rm GeV} mass for values of $\lambda$ fixed by the alignment condition and
stop mixing parameter $X_t = 0$ and $X_t = M_S$. The dominant two-loop corrections are included. \label{fig:lambda}
 }}
\end{figure}

Since $|\mu|^2$ is the diagonal Higgs squared-mass parameter at tree-level in the absence of supersymmetry breaking, it is necessary to demand that $|\mu| \ll M_S$. Furthermore, the SM-like Higgs mass in the limit of small mixing is approximately given by $\mathcal{M}^2_{11}$ [cf.~\Eq{mtwo11}]. The one-loop radiative stop corrections to  $\mathcal{M}^2_{12}$ exhibited in \Eq{mtwo12} that are not absorbed in the definition of $\mathcal{M}^2_{11}$ are suppressed by $\mu/M_S$ (in addition to the usual loop suppression factor), as shown in~\Eq{m12cond}, and thus can be neglected (assuming $\tanb$ is not too large) in obtaining the condition of alignment. Hence, satisfying \Eq{m12cond} fixes $\lambda$, denoted by $\lambda^{\rm alt}$, as a function of $m_h$, $m_Z$ and $\tanb$,
\be \label{lambdaalt}
(\lambda^{\rm alt})^2 = \frac{m_h^2 - m_Z^2 c_{2\beta}}{v^2 s_\beta^2}\ .
\ee
The above condition may only be fulfilled in a very narrow band of values of $\lambda = 0.6$ -- $0.7$ over the $\tan\beta$ range of interest. This is clearly shown in  Fig.~\ref{fig:lambda}, where the blue band exhibits the values of $\lambda$ that lead to alignment as a function of $\tan\beta$.  It is noteworthy that such values of $\lambda$ are compatible with the perturbative consistency of the theory up to the Planck scale, and lead to large tree-level corrections to the Higgs mass for values of $\tanb$ of order one. This is shown by the green band, which depicts the values of $\lambda$ necessary to obtain a tree-level Higgs mass  $m_h =125 \pm 3$~GeV as a function of $\tanb$.

The separation of the green and blue bands in Fig.~\ref{fig:lambda} for a given value of $\tanb$ is an indication of the required radiative corrections necessary to achieve a Higgs mass consistent with observations. In particular, for a given Higgs mass, the value of
the stop loop corrections $\Delta_{\tilde{t}}$ necessary to lift $\mathcal{M}^2_{11}$ to $m_h^2$, obtained from \eqs{mtwo11}{lambdaalt}, is given~by
\be
\Delta_{\tilde{t}} = -c_{2\beta} \left(m_h^2 - m_Z^2\right)\ .
\ee
In the right panel of Fig.~\ref{fig:lambda} we show the necessary values of $M_S$ as a function of $\tan\beta$ to obtain the required radiative corrections for $m_h = 125$~GeV, for two different values of the stop mass mixing parameter, $X_t = 0$ and $X_t = M_S$. We see that for moderate values of $X_t$ the values of $M_S$ relevant for the radiative corrections to the Higgs mass parameter remain below 1 TeV for values of $\tanb$ below about 3. In the following we shall  concentrate on this interesting region, which is complementary to the one preferred in the MSSM.

It should be noted that there are previous studies on the relation between fine-tunings and a SM-like Higgs boson in the NMSSM \cite{Farina:2013fsa, Gherghetta:2014xea}. These works focus on the regime where $\lambda$ is large and $m_h^{tree}=125$ GeV, i.e.~the green band region in Fig.~\ref{fig:lambdafromh}(a),  and conclude that a SM-like 125 GeV Higgs requires decoupling of supersymmetric particles, which in turn leads to more fine-tuning in the Higgs mass. In contrast, in the present work we allow for moderate contributions from the stop loops to raise the Higgs mass to keep $\lambda \sim 0.7$, which yields a SM-like Higgs boson via alignment without decoupling. The stop mass parameters do not need to be large, as can be seen in Fig.~\ref{fig:lambdafromh}(b), giving rise to natural electroweak symmetry breaking.

\begin{figure}[bth]
\subfloat[]{\includegraphics[width=3.2in, angle=0]{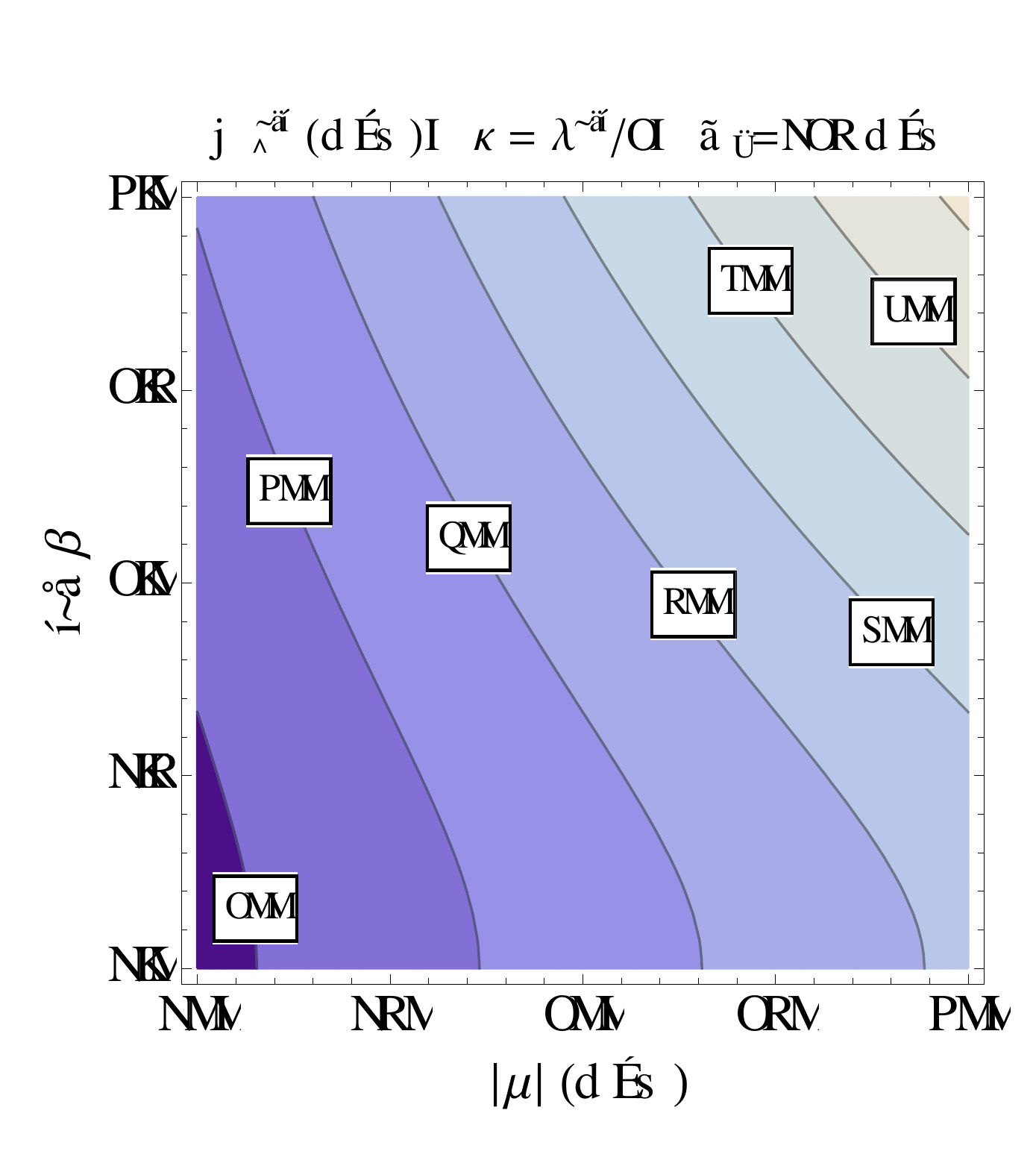}}
\subfloat[]{\includegraphics[width=3.45in, angle=0]{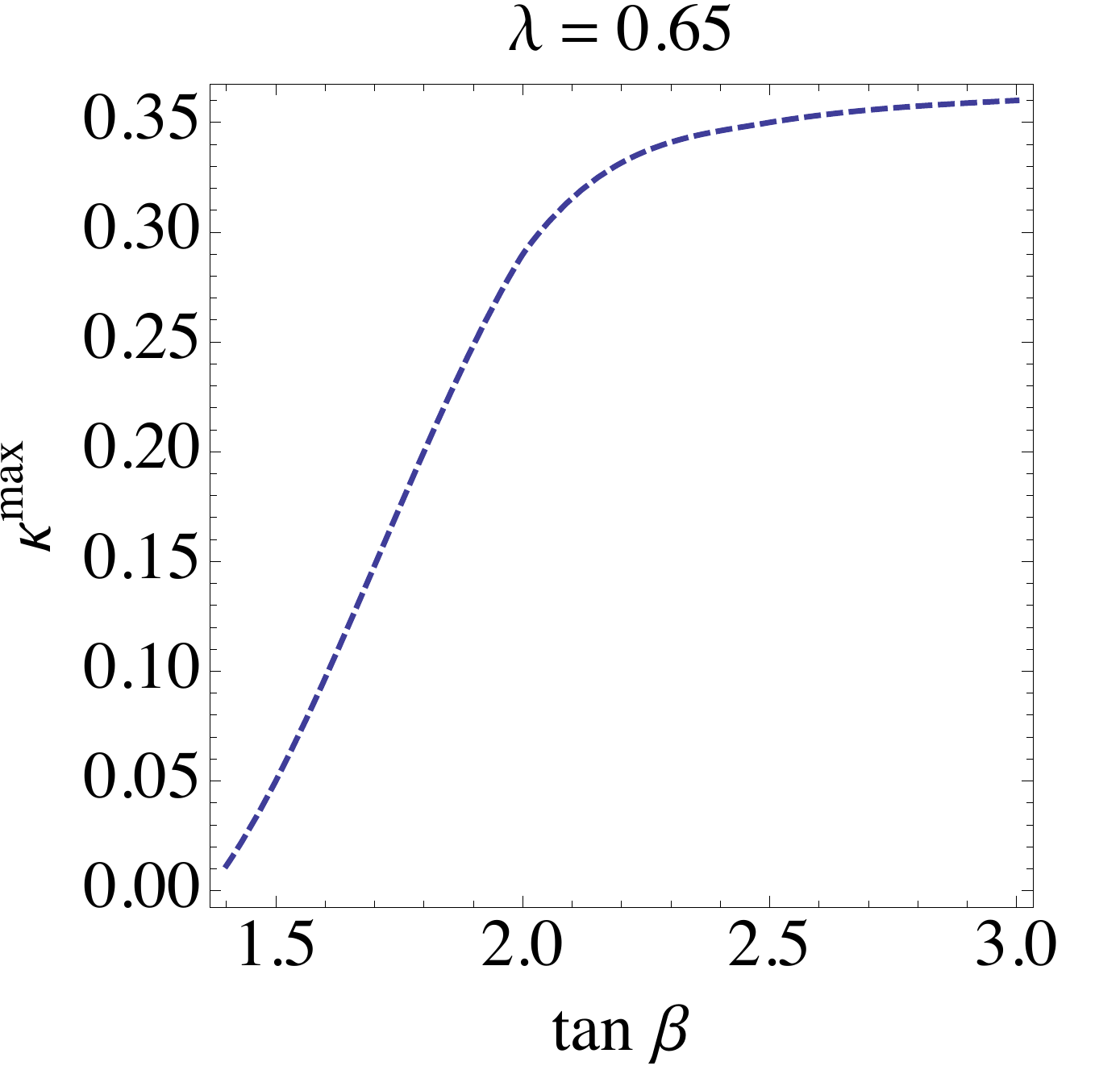}}
\caption{\label{fig:lambdaalt}{\em Left panel : Values of $M_A$ leading to a cancellation of the mixing of the singlet with the SM-like Higgs boson in the Higgs basis, shown in the $|\mu|$--$\tan\beta$ plane. The values of $\lambda$ were fixed so that the alignment condition among the doublet components is fulfilled. Values of $\kappa = \half\lambda$ close to the edge of the perturbativity consistency region were selected.
Right Panel: Maximum values of $\kappa$ consistent with perturbativity as a function of $\tan\beta$ for $\lambda=0.65$.
 }}
\end{figure}

The previous discussion assumed implicitly that the singlets are either decoupled or not significantly mixed with the  CP-even doublet scalars, which is why we only concentrated on the behavior  of the mass matrix element $\mathcal{M}^2_{12}$. If we now consider the case of a light singlet state, then the second condition of alignment, namely small mixing between the singlet and the SM-like CP-even Higgs boson, requires $\mathcal{M}^2_{13} \simeq 0$, as indicated in \eq{eq:alignment}. This yields the following condition:
\be
\frac{M_A^2 s_{2\beta}^2}{4 \mu^2} + \frac{\kappa s_{2\beta}}{2\lambda} = 1 \ .
\label{eq:mamu}
\ee
We shall take $\lambda \simeq 0.65$, as required by the alignment condition given in \eq{lambdaalt}, and  $\kappa\leq\half\lambda$, where the latter is a consequence of the
perturbative consistency of the theory up to the Planck scale, as shown in the right panel of Fig.~\ref{fig:lambdaalt}.  It follows that in order to satisfy Eq.~(\ref{eq:mamu}) the mass parameter $M_A$ must be approximately correlated with the parameter $\mu$, 
\be \label{masim}
M_A \sim \frac{2 |\mu|}{s_{2\beta}}\ .
\ee
In the parameter regime where $100\lsim |\mu|\lsim 300$~GeV (so that no tree-level fine tuning is necessary to achieve electroweak symmetry breaking) and $1\lsim\tanb\lsim 3$, we see that $M_A$ is somewhat larger than $|\mu|$. This is shown in the left panel of Fig.~\ref{fig:lambdaalt}, in which the values of $M_A$ leading to the cancellation of the mixing with the singlet CP-even Higgs state is shown in the $|\mu|$--$\tan\beta$ plane. Here, we have chosen a value of $\kappa \simeq \half\lambda$, which as mentioned above is about the maximal value of $\kappa$ that could be obtained for $\tan\beta \simgt 2$ if the theory is to remain perturbative up to the Planck scale.

The condition $\mathcal{M}^2_{13} = 0$ has implications for the value of $\mathcal{M}^2_{23}$, which governs the mixing between the singlet CP-even state and the non-standard CP-even component of the doublet states.  More precisely,  if $\mathcal{M}^2_{13}$ vanishes, as implied by the condition of alignment given in \eq{eq:alignment}, then
\be
\mathcal{M}^2_{23} = - \sqrt{2} \lambda v \mu \frac{c_{2\beta}}{s_{2\beta}}\ ,
\label{ms212}
\ee
leading to a non-vanishing mixing effect between the light singlet and the heaviest CP-even Higgs boson when $\tanb\neq 1$. For the range of values of the parameters employed in Fig.~\ref{fig:lambdaalt}, the ratio $\mathcal{M}^2_{23}/M_A^2 \ll1$.

In practice, for $\lambda \simeq 0.65$, the inequality $\mathcal{M}^2_{12} \ll \mathcal{M}^2_{13}$ holds unless the mass parameters $M_A$ and $\mu$ are tuned to obtain
almost exact alignment.  Hence, based on the discussion above, we shall henceforth assume that the following hierarchy among the elements of the CP-even Higgs squared-mass matrix is fulfilled close to the alignment limit,
\begin{equation} \label{hierarchy}
\mathcal{M}^2_{12} \ \ll \ \ \mathcal{M}^2_{13}\ \ \ll \ \ \mathcal{M}^2_{23}, \mathcal{M}^2_{11}, \mathcal{M}^2_{33}\ \ \ll \ \  \mathcal{M}^2_{22}.
\end{equation}

Given the above observations, it is not difficult to see that all mixing angles in the CP-even Higgs mixing matrix are small. Therefore, the mass eigenstate $h$, whose predominant component is $H^{\rm SM}$, is SM-like, whereas the predominant components of the other two eigenstates $H$ and $h_S$ are  $H^{\rm NSM}$ and $H^{\rm S}$, respectively. In particular,
\begin{equation}
m_h^2 \simeq \mathcal{M}^2_{11}\;,  \;\;\;\;\;\;\;\;  m_H^2 \simeq \mathcal{M}^2_{22}\;, \;\;\;\;\;\;\;\;m_{h_S}^2 \simeq \mathcal{M}^2_{33} \; ,
\end{equation}
and the hierarchy of masses 
\begin{equation} \label{ll}
m_H^2 \ \gg \ m_h^2, \  m_{h_S}^2
\end{equation}
is fulfilled in the region of parameter space under consideration. Using \eqst{x}{z} and ignoring terms of order $\epsilon_1\equiv \mathcal{M}^2_{12}/\mathcal{M}^2_{22}$
and $\epsilon_2\equiv\mathcal{M}_{13}^2\mathcal{M}_{23}^2/\mathcal{M}_{22}^4$, we derive the following approximate relationship between the interaction and mass eigenstates,
\be
\left(\begin{array}{c}
h \\
H \\
h_S
\end{array}\right)  \simeq 
\left(\begin{array}{ccc}
1 & \ -\eta \eta'  \ & \,\,\,\,\phm\eta' \\
{\cal{O}}(\epsilon) & -1 &  \,\,\,\,-\eta \\
-\eta' & \ -\eta & \,\,\,\, \phm 1
\end{array}\right)  
\left(\begin{array}{c}
H^{\rm SM} \\
H^{\rm NSM}\\
H^{\rm S}
\end{array}\right) \ ,
\label{eq:kappamatrix}
\ee 
where  the elements of the CP-even Higgs mixing matrix are expressed in terms of
\bea
-\kappa^H_{S} = \eta &=& \frac{\mathcal{M}^2_{23}}{m_H^2},  \label{kapprox1} \\[6pt]
 \kappa^h_S = \eta' &=& \frac{\mathcal{M}^2_{13}}{m_h^2 - \mathcal{M}^2_{33}} \,, \label{kapprox2}
\eea
and $\mathcal{O}(\epsilon)$ denotes a linear combination of terms of order $\epsilon_1$ and $\epsilon_2$, respectively. In \eq{eq:kappamatrix}, we have kept all terms in the mixing matrix up to quadratic in the small quantities $\eta$ and~$\eta'$.  Given the assumed hierarchy of \eq{hierarchy}, the $\mathcal{O}(\epsilon)$ terms are at best of the same order as quantities that are cubic in $\eta$ and $\eta'$ and hence are truly negligible. This then tells us that the following approximate relationship exists between the mixing angles defined in \eq{rmatrix},
\be
s_{12}\simeq -s_{13} s_{23}\,,
\ee
and the alignment limit in the hierarchy of \eq{hierarchy} is primarily governed by two small mixing angles.

Eq.~(\ref{eq:kappamatrix}) provides a useful guide for understanding the Higgs phenomenology in our numerical study. In particular, there are correlations among the different matrix elements. For example,
\bea
\kappa^h_{\rm S} & \simeq & - \kappa^{h_S}_{\rm SM}  \ ,\label{eq:kappacorr1}
\\
\kappa^h_{\rm NSM} & \simeq & \kappa^{h}_{\rm S} \kappa^H_{\rm S} \ ,\label{eq:kappacorr2}
\\
\kappa^H_{\rm S}  & \simeq & \kappa^{h_S}_{\rm NSM} \ .
\label{eq:kappacorr3}
\eea
In light of \eqs{masim}{ms212}, it follows that [cf.~\eqs{kapprox1}{kapprox2}]:
\be
\kappa^{H}_{\rm S}  \simeq \kappa^{h_S}_{\rm NSM} \simeq \frac{\lambda  v}{2\sqrt{2} \,\mu}  c_{2\beta} s_{2\beta}\; . \label{kappahShNSM}
\ee
We have previously argued that values of $\lambda \simeq 0.6-0.7$ are preferred from both the perspective of Higgs phenomenology as well as perturbative consistency of the NMSSM up to the Planck scale.  In addition we note that the range of $\mu$ is rather restricted:
$|\mu| > 100$~GeV, in order to fulfill the LEP chargino bounds, however $\mu$ cannot be too large in order to preserve a natural explanation for electroweak symmetry breaking. Hence, we can see from \Eq{kappahShNSM} that if for example $|\mu| \simlt 200$~GeV, then
\be
0.15 \simlt |\kappa_{\rm NSM}^{h_S}| \simlt 0.3\ ,
\label{eq:kappahsNSM}
\ee
which implies that all mixing angles are small if the conditions of alignment are imposed.
For the small values of $\kappa$ consistent with a perturbative extension of the theory up to
the Planck scale (see Fig.~\ref{fig:lambdaalt}), the above estimate continues to hold
even after the $\kappa$-induced effects  as well as the corrections associated with the $h_S$ mass are included.

\subsection{Spectrum of the Higgs Sector Near the Alignment Limit}
\label{spectrum}

Close to the alignment limit, the mass parameter $M_A \simeq 2 |\mu|/s_{2\beta}$. Since $|\mu|$ must be larger than about 100 GeV in order to fulfill the current LEP constraints on the chargino masses, it follows that for $\tanb \geq 2$, the CP-odd Higgs mass must be larger than about 250 GeV. We conclude that $M_A^2 \gg m_h^2$.  In light off this observation, the spectrum of neutral Higgs bosons near the alignment limit may be  approximated by~\footnote{Note that $m_A$ is the mass of the mostly doublet CP-odd neutral Higgs boson, whereas $M_A$ is the mass \textit{parameter} defined in Eq.~(\ref{masq}).  In this paper we always employ a lower case $m$ when referring to the physical mass of a particle.  In contrast, an upper case $M$ refers to some quantity with mass dimensions that is defined in terms of the fundamental model parameters.} 
\begin{itemize}
\item  A SM-like CP-even Higgs boson state of mass $m_h^2=(125~{\rm GeV})^2 \simeq \mathcal{M}^2_{11}  \ll M_A^2$.
\item A heavy CP-even Higgs boson state $H$ of mass $m_H \simeq M_A$.
\item A heavy CP-odd Higgs boson state $A$ of mass $m_A \simeq M_A$.
\item  Light, mostly singlet-like CP-even and CP-odd Higgs boson states with masses\footnote{\Eqs{mhsapp}{masapp} are obtained in an approximation that includes the first non-trivial corrections to $m^2_{h_S}\simeq \mathcal{M}^2_{33}$ and $m^2_{A_S}\simeq (\mathcal{M}^2_P)_{33}$ due to the off-diagonal elements of the corresponding squared-mass matrix.}
\bea
m_{h_S}^2 &  \simeq & \frac{\kappa \mu}{\lambda} \left( A_{\kappa} + \frac{4\kappa \mu}{\lambda} \right) + \frac{\lambda^2 v^2 M_A^2}{8\mu^2} s_{2 \beta}^4
- \tfrac{1}{4}{v^2} \kappa \lambda  ( 3 - 2 s_{2 \beta}^2) s_{2\beta}- \half v^2 \kappa^2 \frac{\mu^2}{M_A^2} c_{2 \beta}^2\,,\label{mhsapp}\nonumber \\
&& \\
m_{A_S}^2 & \simeq & 3 \kappa \left[ \tfrac{3}{4} \lambda v^2 s_{2\beta} - \mu \left( \frac{A_{\kappa}}{\lambda} +\frac{3 v^2 \kappa \mu}{2M_A^2} \right) \right] \,.\label{masapp}
\eea
\end{itemize}
It is interesting to note that the singlet-like Higgs masses depend on the parameter $A_\kappa$ which is not restricted by the conditions of alignment. As such, these masses are not correlated with the other Higgs boson masses.  For positive values of $\mu$ and $\kappa$, larger values of $A_\kappa$ lead to an increase in $m_{h_S}^2$ and a decrease in $m_{A_S}^2$.  Therefore, for fixed values of the other parameters, the value of $A_\kappa$
is restricted by the requirement of non-negative $m_{h_S}^2$ and $m_{A_S}^2$. In particular, due to the anti-correlation in the behavior of $m_{h_S}^2$ and $m_{A_S}^2$ with $A_\kappa$, the maximal possible value, $(m_{h_S}^2)_{\rm max}$, is achieved when $m_{A_S} ^2=0$.  Likewise, the maximal value, $(m_{A_S}^2)_{\rm max}$, is achieved when $m_{h_S}^2=0$. Using \eqs{mhsapp}{masapp} to eliminate $A_\kappa$, and making use
of \eq{eq:mamu} in the alignment limit to eliminate $\mu^2$,
\be
m^2_{A_S}+3m^2_{h_S}\simeq
\frac{3M_A^2 s_{2\beta}^2}{1-\half\kappa s_{2\beta}/\lambda}\left[\frac{\kappa^2}{\lambda^2}+\frac{\lambda^2 v^2}{2 M_A^2}
\left( 1 - \frac{\kappa^2}{\lambda^2} \right) \right] \,.
\label{eq:msbound}
\ee

\begin{figure}[t!]
\subfloat[]{\includegraphics[width=3.2in, angle=0]{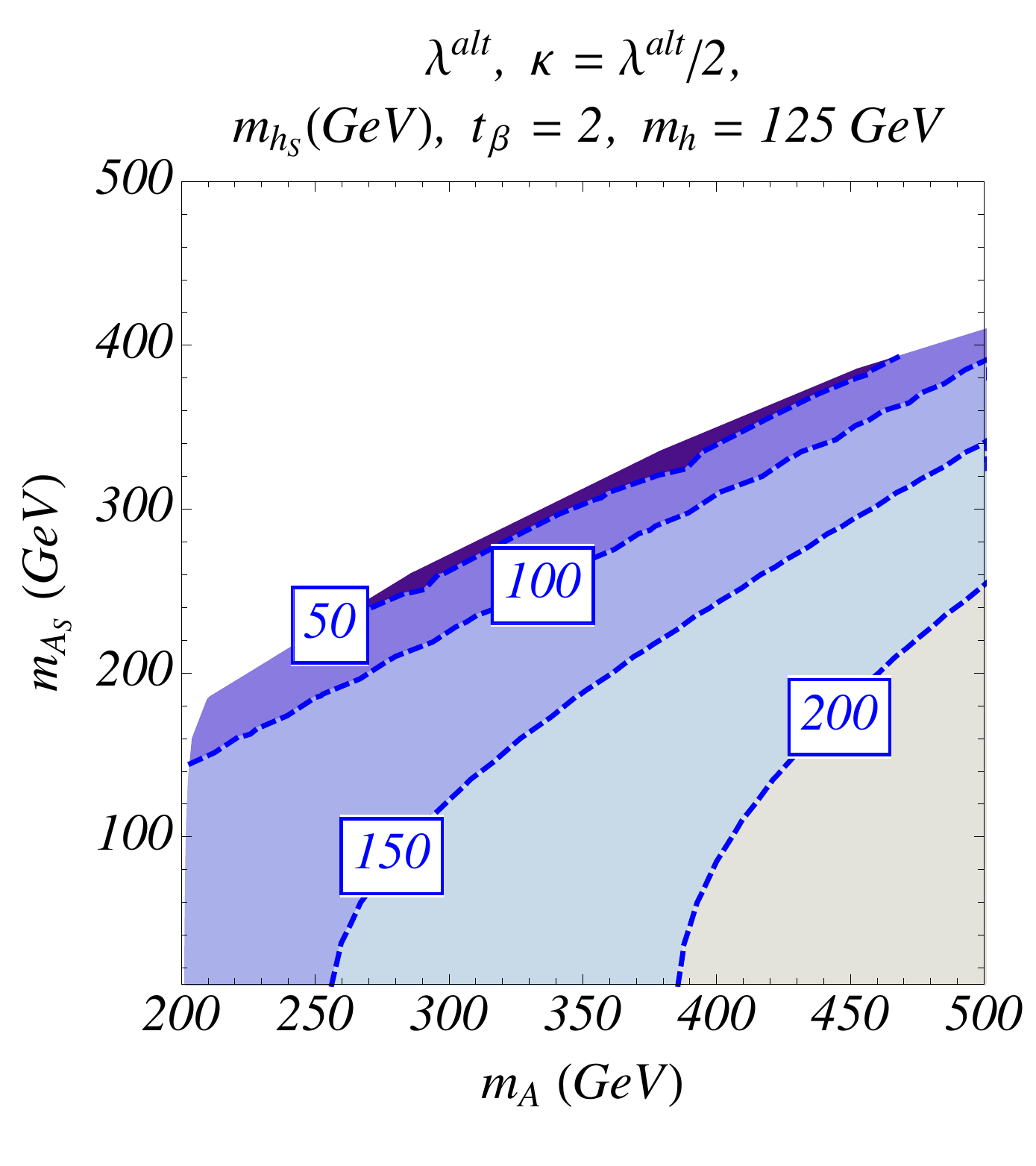}}~~
\subfloat[]{\includegraphics[width=3.2in, angle=0]{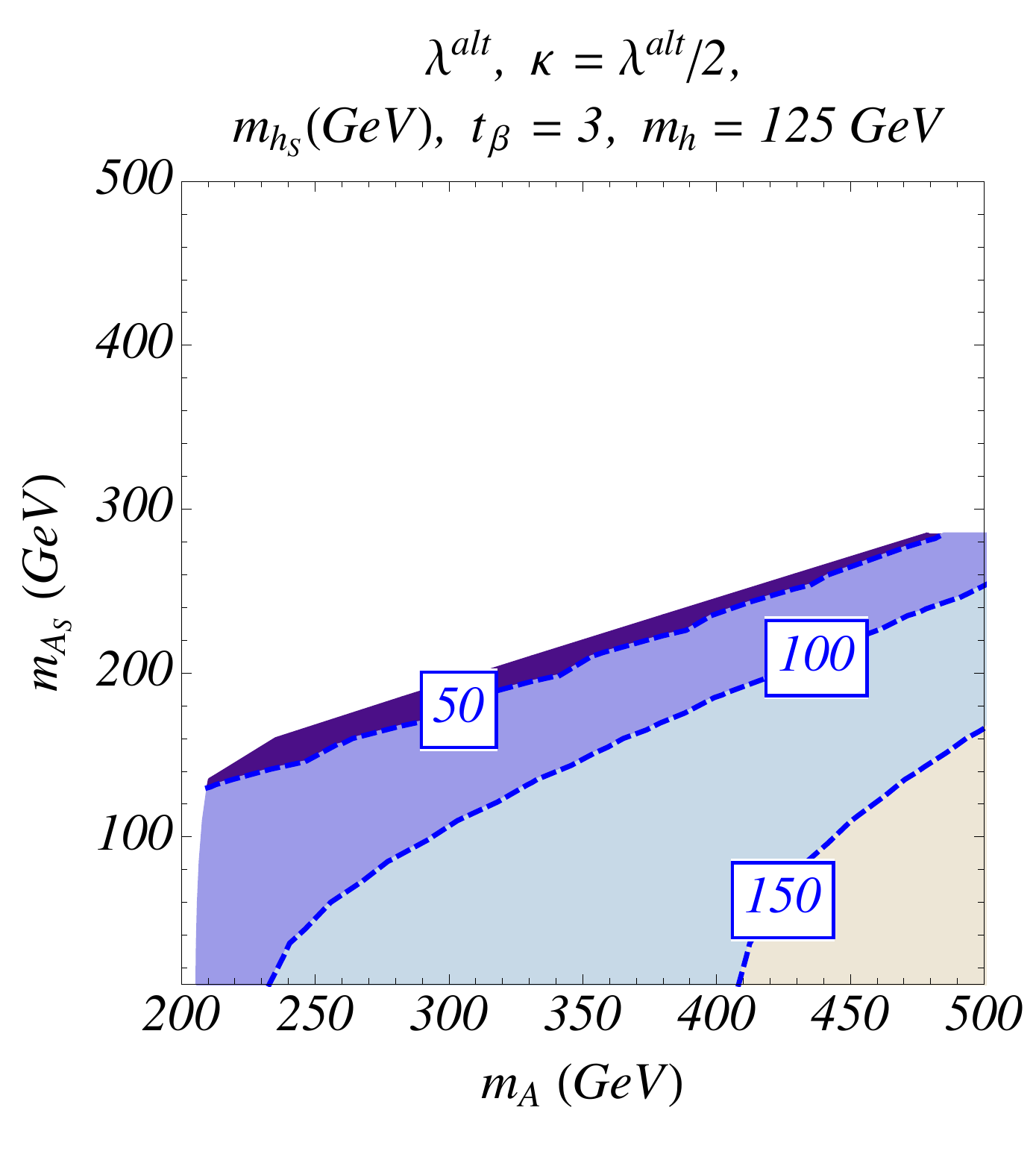}}
\caption{\label{fig:singlet}{\em Values of the singlet CP-even Higgs mass $m_{h_S}$
for $\tanb = 2$ (left panel) and $\tanb=3$ (right panel) in the plane of $m_A$  vs.~$m_{A_S}$,  imposing a SM-like Higgs boson with a  mass of 125~{\rm GeV}~(with  $\lambda$ and $\mu$ satisfying the alignment conditions and $\kappa=\half\lambda$).
 }}
\end{figure}

In the parameter region of interest, $\kappa \leq\half\lambda$ and $s_{2\beta}$ is near 1.
Close to the alignment limit (where $\lambda\simeq 0.65$), we have noted above that $m_A^2\simeq M_A^2\gg\half\lambda^2 v^2$, in which case $(m_{h_S}^2)_{\rm max} \simlt \tfrac{1}{3}m_A^2$ and $(m_{A_S}^2)_{\rm max} \lsim m_A^2$.  In the left and right panels of Fig.~\ref{fig:singlet}, we display the contours of the singlet-like CP-even Higgs mass in the $m_A$--$m_{A_S}$ plane for $\kappa \simeq \half\lambda^{\rm alt}$ and for $\tanb = 2$  and $\tanb = 3$, respectively. Whereas $m_{A_S}$  may become of order $m_A$ for low values of $\tanb$ (i.e.~for $s_{2\beta} \simeq 1$),  the singlet CP-even Higgs mass remains below $\half m_A$  over most of the parameter space, in agreement with \eq{eq:msbound}.

\section{Numerical Results of Masses and Mixing Angles}
\label{Higgspheno}

In the previous section, we have performed an analytical study of the implications of the alignment limit on the masses and mixing angles of the Higgs mass eigenstates.  To obtain a more accurate picture, we complement the above analysis with the results obtained from a numerical study, including all relevant corrections to the CP-even and CP-odd Higgs squared-mass matrix elements. In our numerical evaluation we have used the code {\tt NMHDecay}~\cite{Ellwanger:2005dv} and the code {\tt Higgsbounds} \cite{Bechtle:2013wla} included in  {\tt NMSSMTools} \cite{NMSSMTools}. We keep parameter points that are consistent with the present constraints coming from  measurements on properties of the 125 GeV Higgs boson $h$, as well as those coming from searches for the new Higgs bosons $H$ and $h_S$, which impose constraints on  $\kappa^\phi_i$ similar to those shown in Fig.~\ref{fig:Component4}. We do not impose flavor constraints since they depend on the flavor structure of the supersymmetry breaking parameters, which has only a very small impact on Higgs physics. Moreover, in most of our analysis we have assumed the gaugino mass parameters to be large by fixing the electroweak gaugino masses to 500~GeV and the gluino mass to 1.5~TeV.  Since the Higgsino mass parameter is small in our region of interest, 
the resulting dark matter relic density due to the lightest supersymmetric particle (LSP) tends to be smaller than the observed value, which implies that other particles outside of
the NMSSM (e.g.~axions) must contribute significantly to the dark matter relic density.
Alternatively, it is possible to saturate the observed relic density with the LSP by lowering the value of the electroweak gaugino masses chosen above. We have also fixed  $\mu >0$, but we have checked that the generic behavior discussed in this work does not depend on
the sign of $\mu$, as can be understood analytically from the expressions given in Section~\ref{alignment}. The implications of lowering the gaugino masses to obtain the proper relic density will be briefly discussed below.

\begin{figure}[htb]
\subfloat[]{\includegraphics[width=3.2in, angle=0]{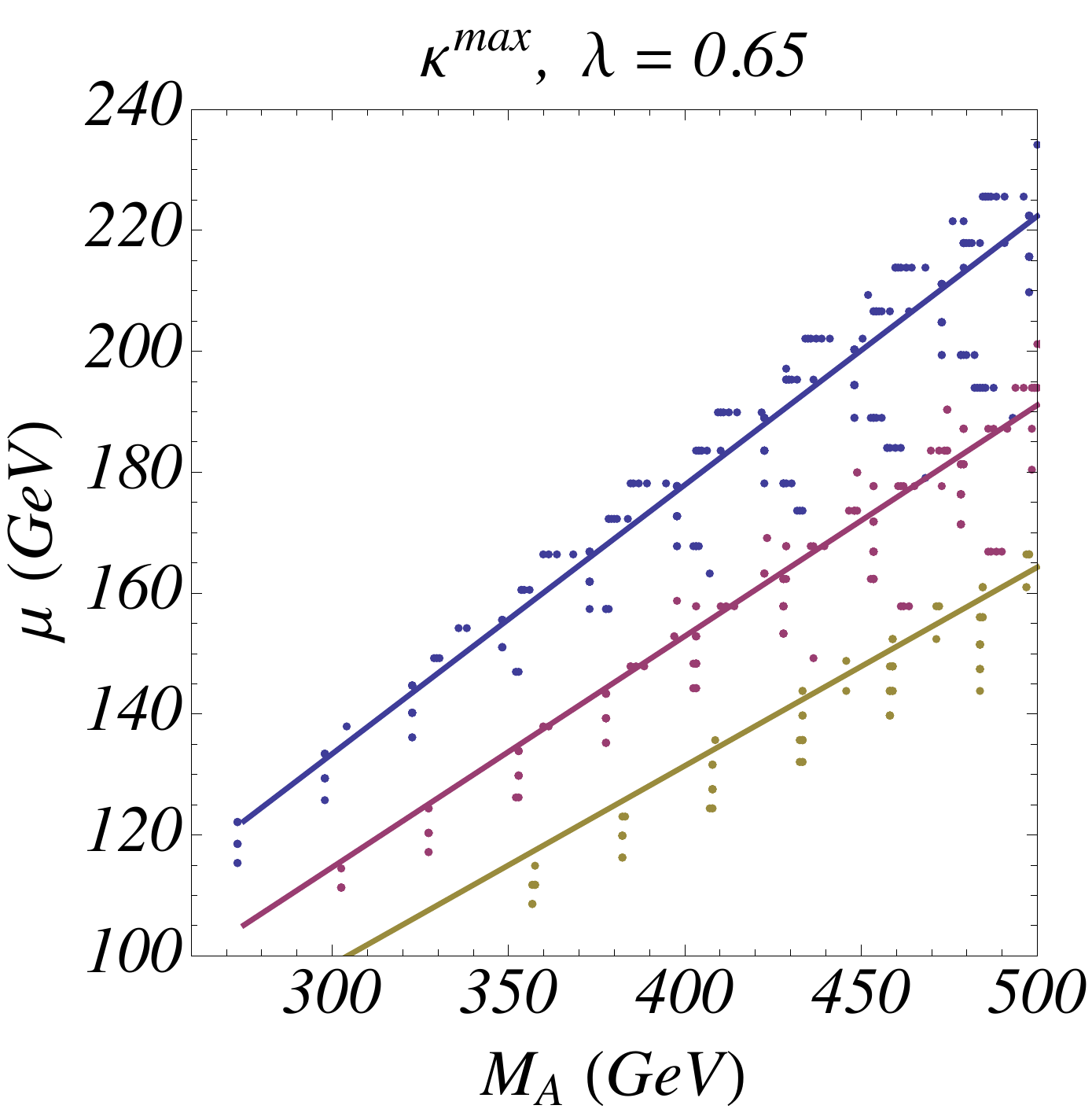}}~~
\subfloat[]{\includegraphics[width=3.2in, angle=0]{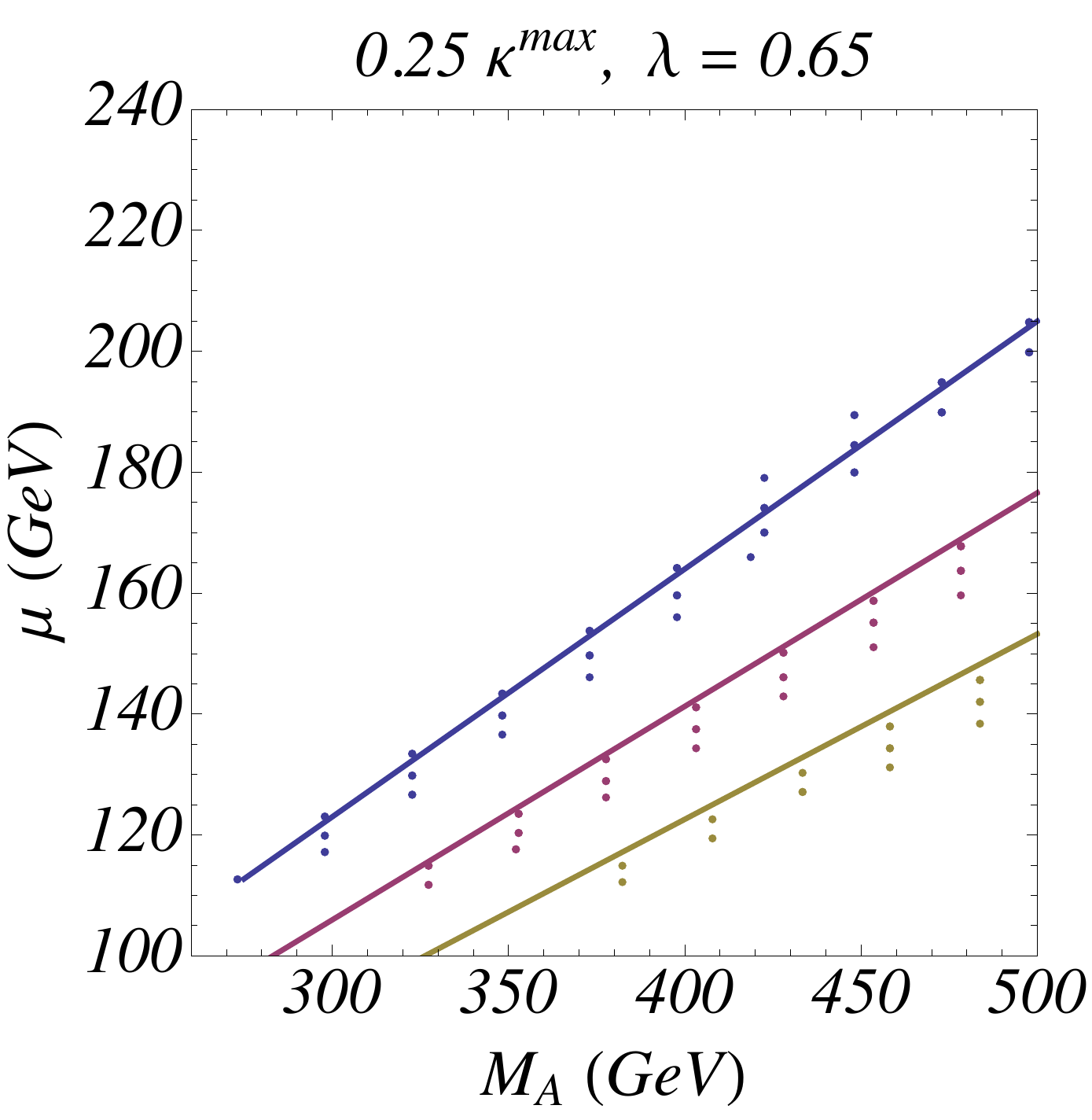}}
\caption{\label{fig:NumAlign}{\em Values of the mass parameters $M_A$  and $\mu$ consistent with the current LHC constraints on the SM-like Higgs properties, for values of $\kappa = \kappa^{\rm max}$, the maximal value of $\kappa$ leading to perturbative consistency of the theory up to the Planck scale (left panel) and for $\kappa = \tfrac{1}{4}\kappa^{\rm max}$. Solid lines represent the alignment condition, Eq.~(\ref{eq:mamu}), and the colors blue, red and yellow represent values of $\tanb = 2$, 2.5 and 3, respectively.}}
\end{figure}

In our numerical study, we have chosen $\lambda = 0.65$. The stop spectrum has been determined to reproduce the observed Higgs mass, assuming small stop mixing, and we have varied all other relevant parameters allowed by the above constraints. As shown in Fig.~\ref{fig:lambdafromh}, for $\tanb \geq 2$ and common values of the left- and right-stop supersymmetry breaking parameters, the assumption of small stop mixing leads to stops that
are heavier than about 600 GeV and essentially decouple from  Higgs phenomenology. For $\tanb \simeq 2$, larger values of the stop mixing may lead to lighter stops, resulting in a variation of the Higgs production cross section in the gluon fusion channel. We shall briefly comment on such possible effects below. 

As discussed in Section~\ref{nmssm}, a value of $\lambda \simeq 0.65$ leads to an approximate cancellation of the mixing between the SM and non-SM doublet components for  all moderate or small values of $\tanb$ and allows a perturbative extension of the theory up to energy scales of order the Planck scale.  Moreover, close to the alignment limit, the SM-like Higgs mass receives a  significant tree-level contribution, which reduces the need for large radiative corrections associated with heavy stops, as shown in Fig.~\ref{fig:lambdafromh}. Due to the strong perturbativity constraints shown in Fig.~\ref{fig:lambdafromh}, we focus on the NMSSM parameter regime with $\tanb = 2, 2.5$ and 3, which we henceforth display in blue, red and yellow colors, respectively.

The allowed  values of $M_A$  and $\mu$ are shown in Fig.~\ref{fig:NumAlign}, for the values of $\kappa= \kappa^{\rm max}$,  the maximal values consistent with the perturbative consistency of the theory up to the Planck scale (left panel) and for values of $\kappa = \tfrac{1}{4}\kappa^{\rm max}$  (right panel).  The solid lines represent the correlation 
between $M_A$ and $\mu$ necessary to get alignment at tree-level [cf.~Eq.~(\ref{eq:mamu})]. The dots represent the allowed values of these parameters as obtained from {\tt NMSSMTools}.  We find that the present constraints on the SM-like Higgs properties allow values of $M_A$  and $\mu$ that deviate not more than 10\% from the alignment condition specified in Eq.~(\ref{eq:mamu}).

\begin{figure}[t!]
\subfloat[]{\includegraphics[width=3.in, angle=0]{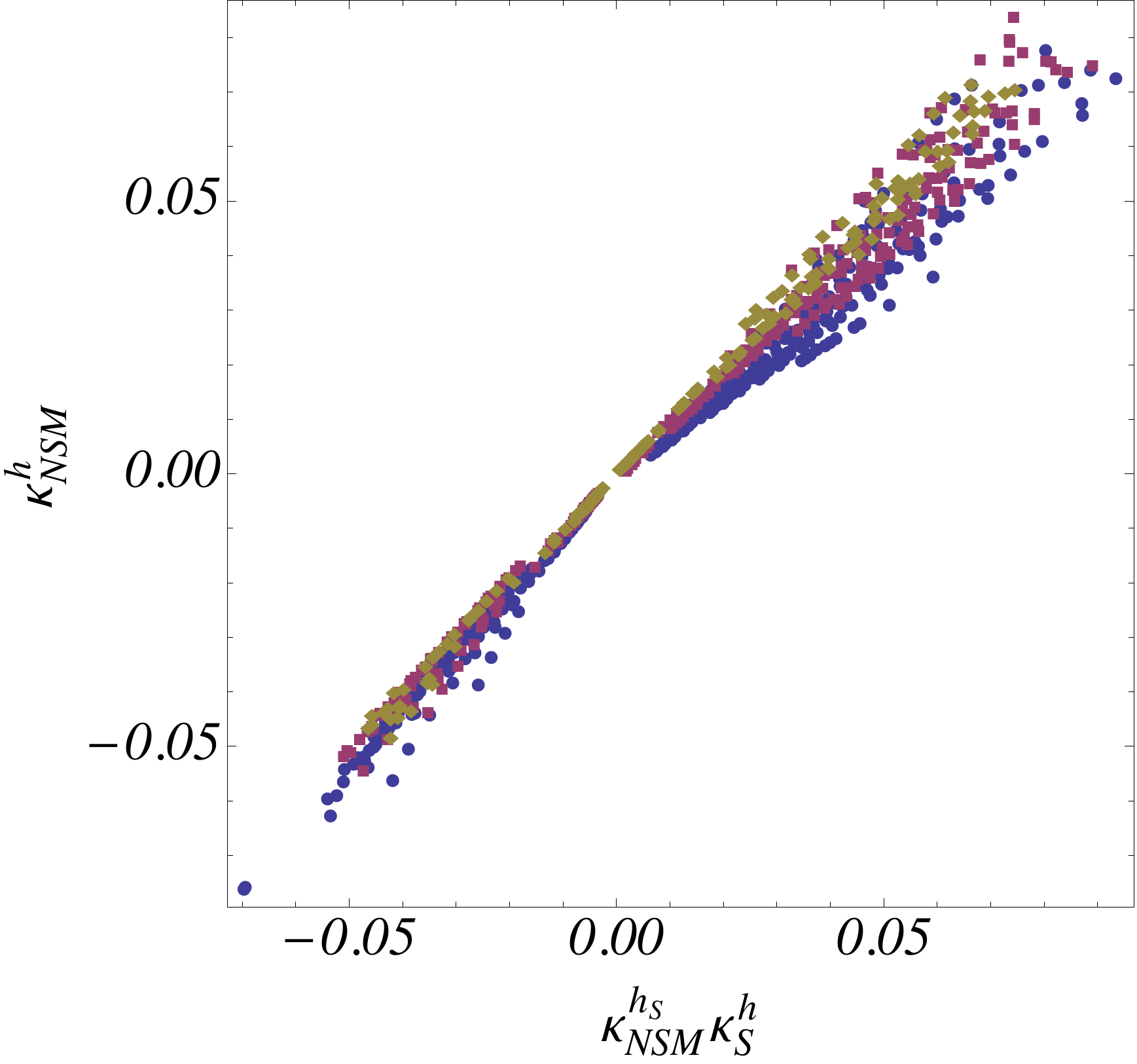}}~~
\subfloat[]{\includegraphics[width=3.in, angle=0]{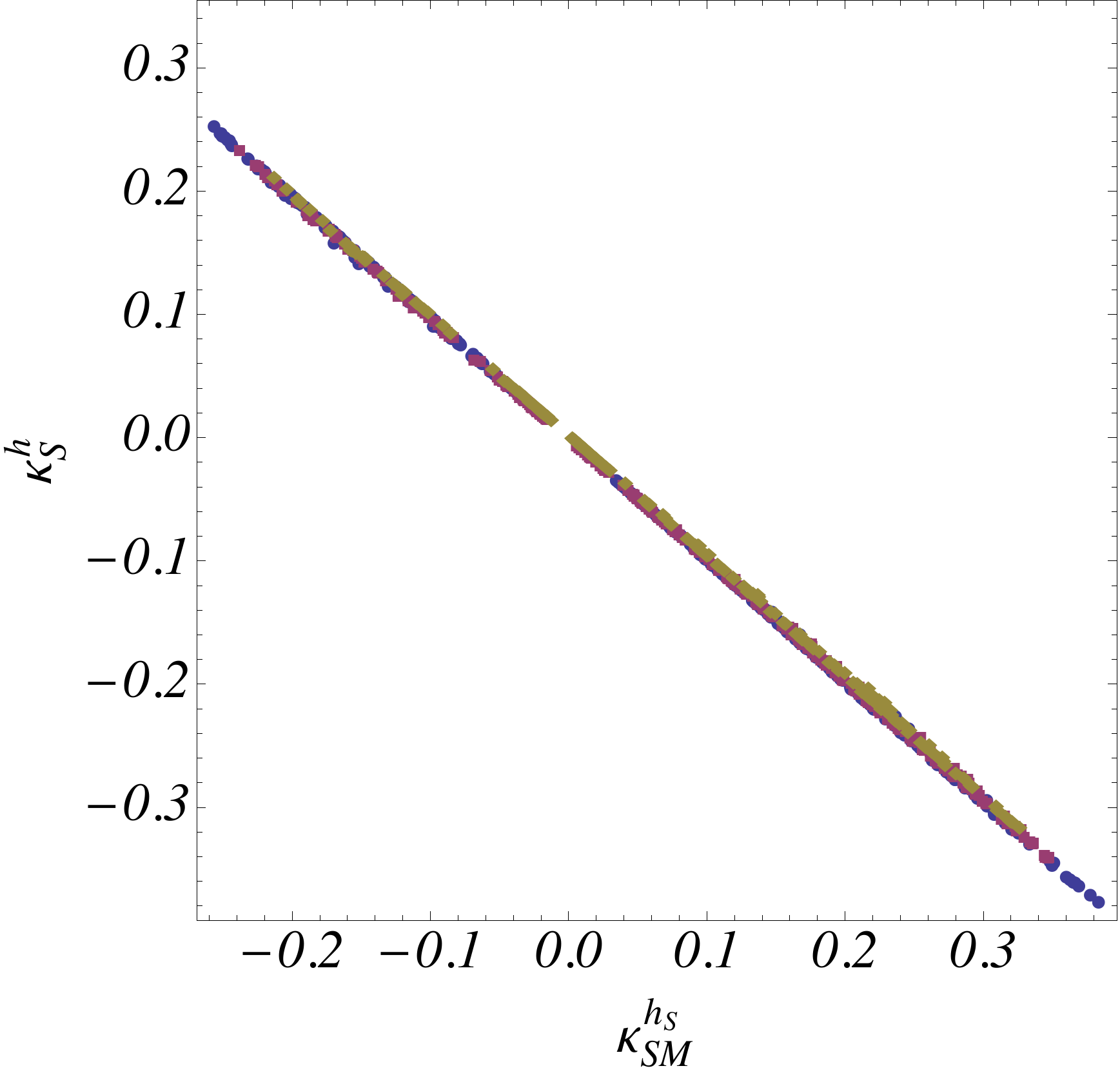}}
\caption{\label{fig:Component1}{\em For the points allowed by LHC constraints in the left panel we plot the correlation between the non-SM doublet component of the 125 {\rm GeV} Higgs state with the product of the non-SM doublet component of the mainly singlet state and the singlet component of the 125 {\rm GeV} Higgs state. In the right panel we plot the correlation between the SM doublet component of the singlet state with the singlet component of the 125 {\rm GeV} Higgs state. Blue, red and yellow represent values of $\tanb = 2$, 2.5 and 3, respectively.
 }}
\end{figure}

\begin{figure}[t!]
\subfloat[]{\includegraphics[width=3.in, angle=0]{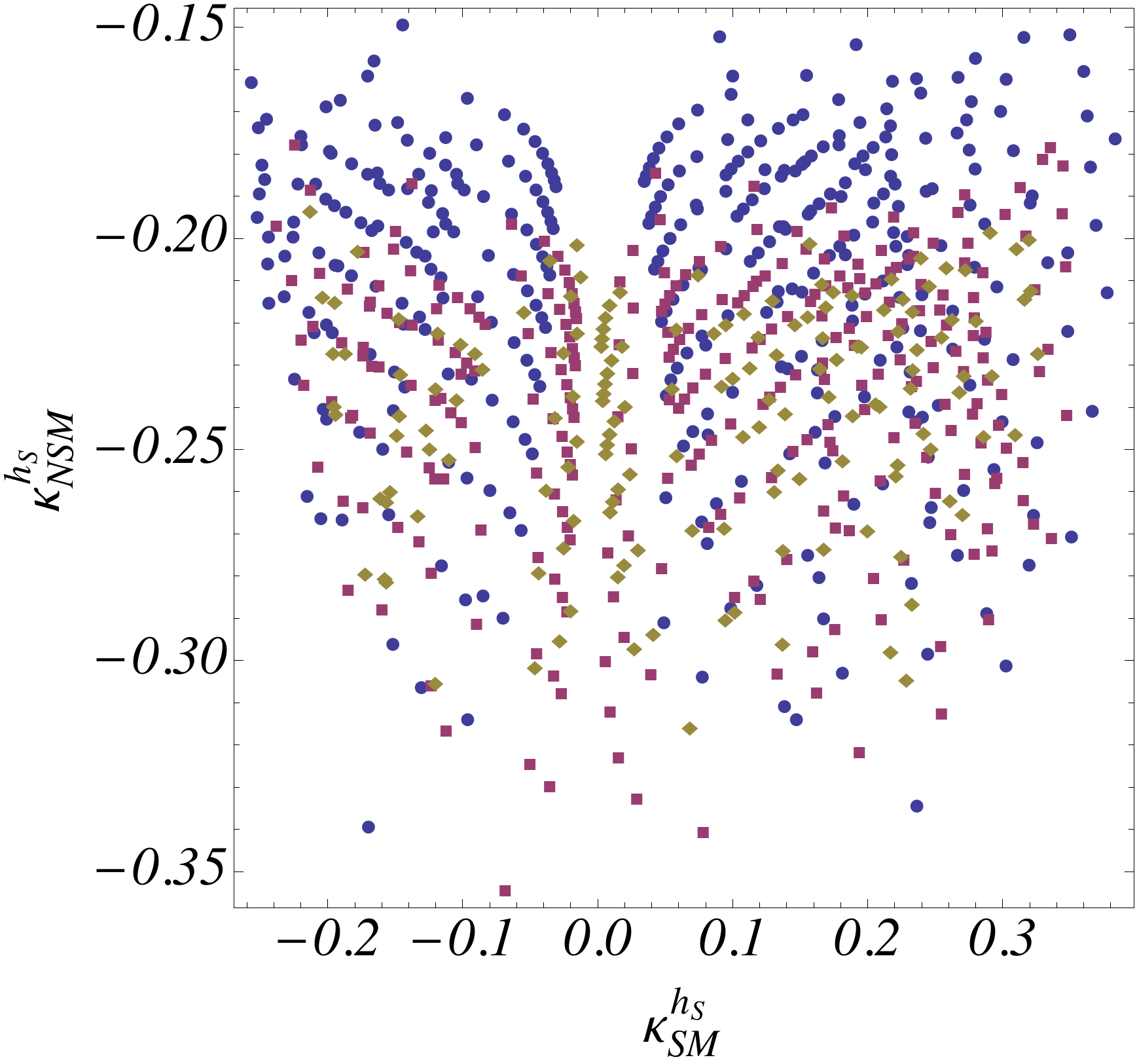}}~~
\subfloat[]{\includegraphics[width=3.in, angle=0]{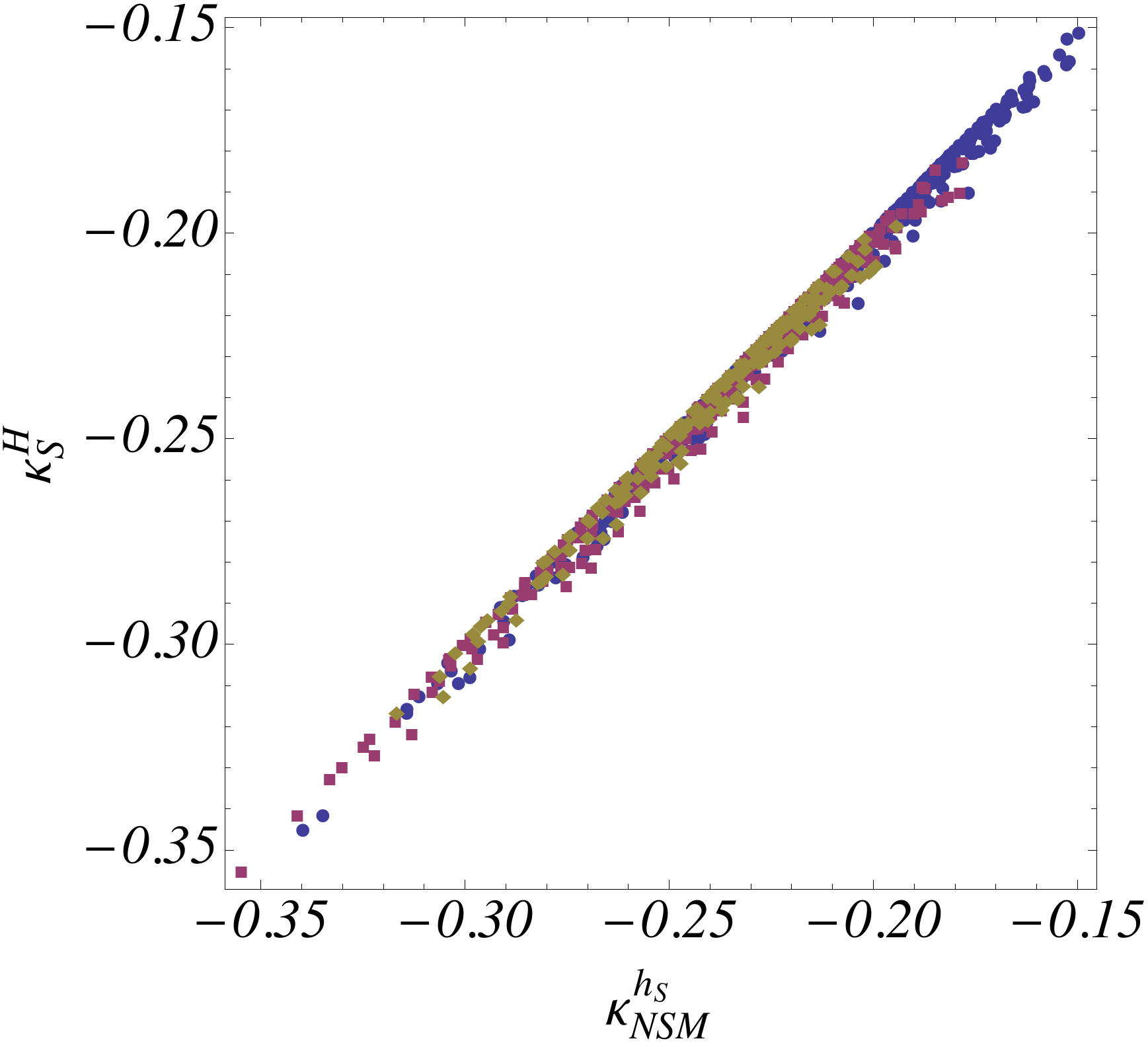}}
\caption{\label{fig:Component3}{\em For the points allowed by LHC constraints, we show the
 correlation between the non-SM doublet component and the SM doublet component of the mainly singlet Higgs state. Blue, red and yellow represent values of $\tanb = 2$, 2.5 and 3, respectively.\\[-20pt]
 }}
\end{figure}

\begin{figure}[h!]
\subfloat[]{\includegraphics[width=3.in, angle=0]{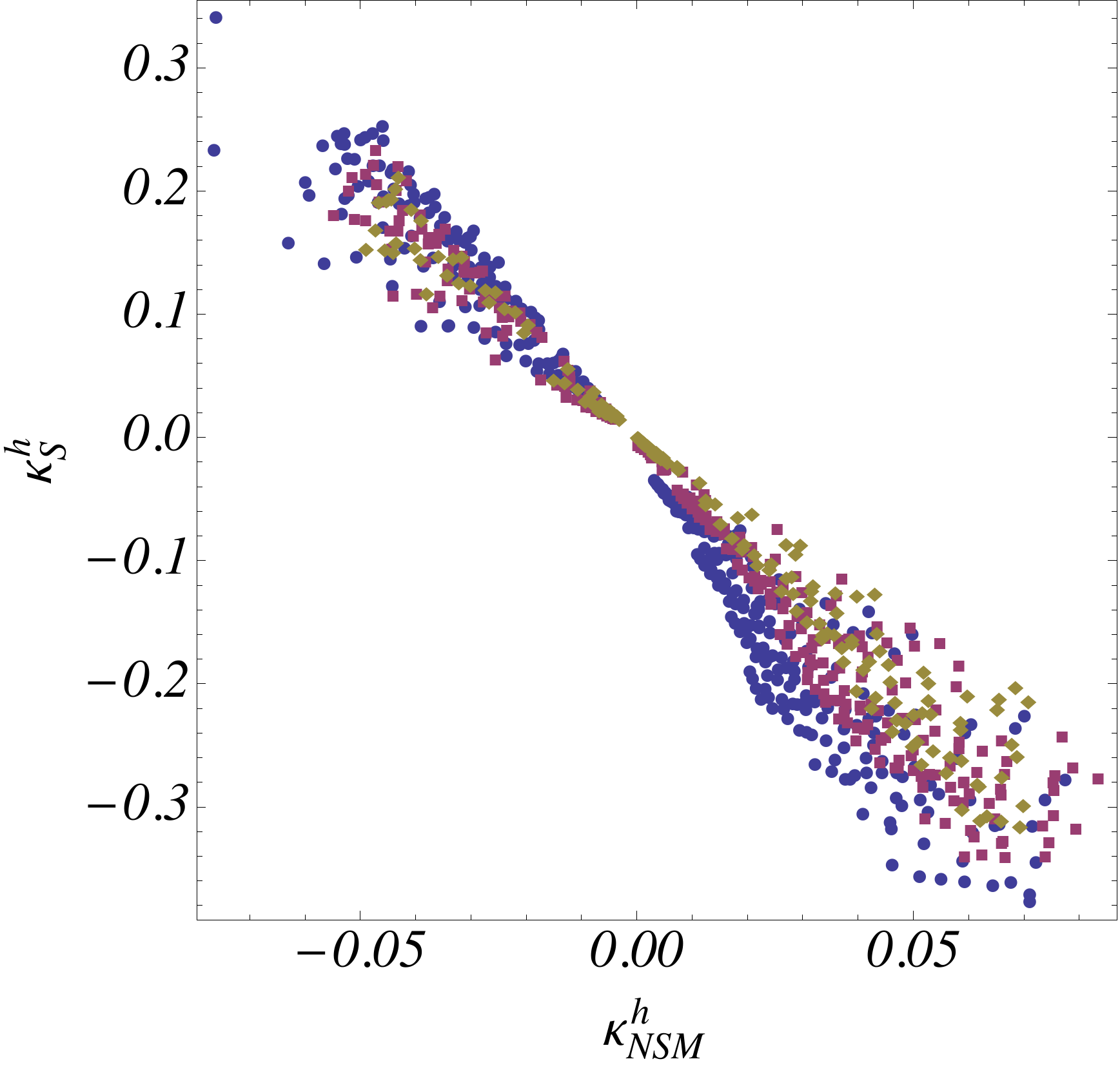}}~~
\subfloat[]{\includegraphics[width=3.in, angle=0]{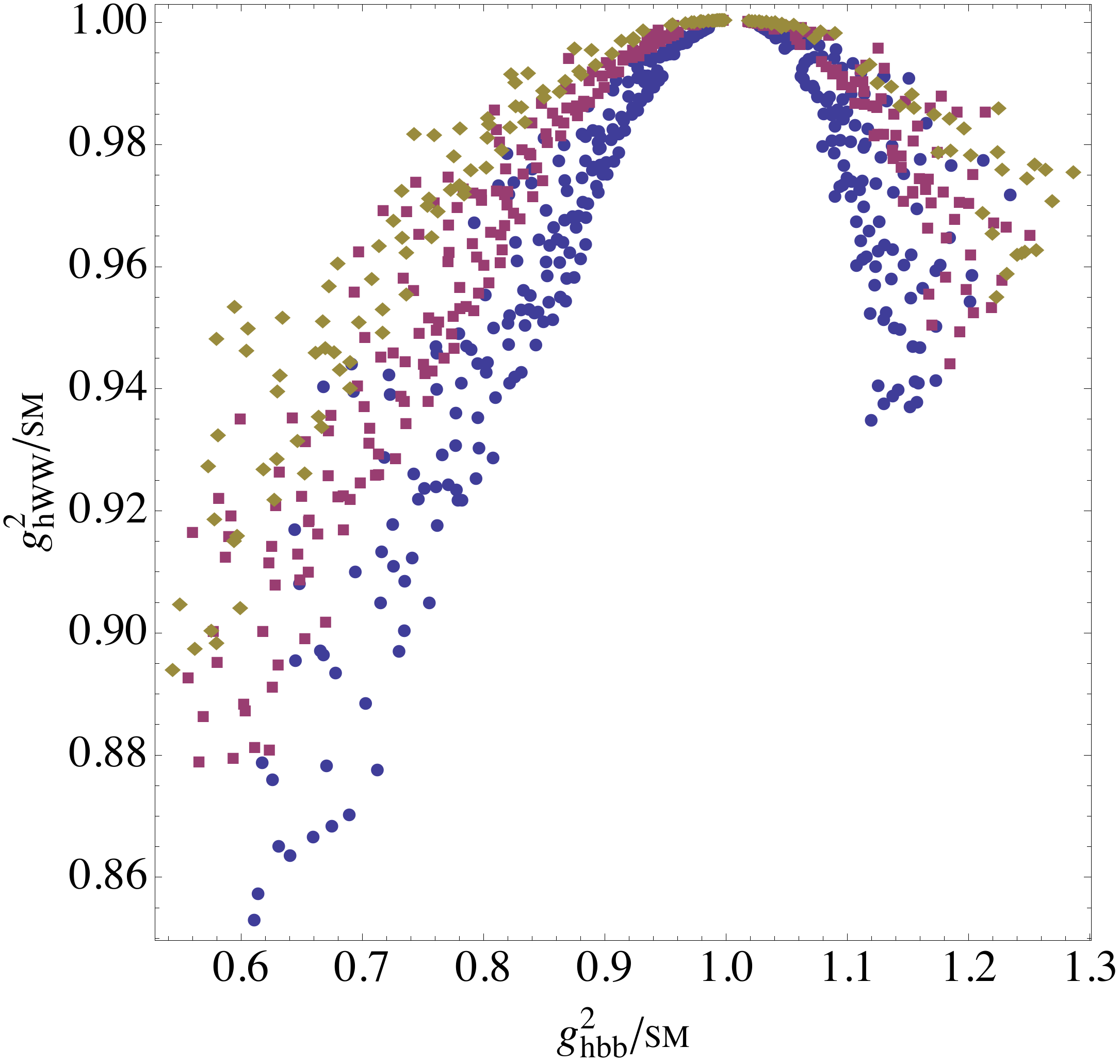}}
\caption{\label{fig:Component2}{\em For the points allowed by LHC constraints, in the left panel we plot the correlation between the non-SM doublet component and the singlet component of the 125 {\rm GeV} Higgs state. In the right panel we plot the square of the couplings of the 125 {\rm GeV} Higgs state, normalized to its SM value. Blue, red and
yellow represent values of $\tanb =2$, 2.5 and 3, respectively.
 }}
\end{figure}

The correlations obtained in \eqst{eq:kappacorr1}{eq:kappacorr3} among the interaction eigenstate components, $H^{\rm SM}$, $H^{\rm NSM}$ and $H^{\rm S}$, of the mass-eigenstate neutral Higgs bosons are clearly displayed in Figs.~\ref{fig:Component1}, \ref{fig:Component3} and \ref{fig:Component2}. The right panel of Fig.~\ref{fig:Component1} displays the correlation between the singlet component of $h$ with the SM component of the mostly singlet state $h_S$, whereas  the left panel exhibits the correlation between $\kappa^h_{\rm NSM}$ with $\kappa^h_{S} \kappa^{h_S}_{\rm NSM}$. We see the 
correlation given in \eq{eq:kappacorr1} is preserved over most of the parameter space, however there are small departures from the correlations exhibited in \eqs{eq:kappacorr1}{eq:kappacorr2}  due to neglected terms that are higher order in $\eta$ and $\eta^\prime$.

%
%

Similarly,  the right panel of Fig.~\ref{fig:Component3} shows the correlation between $\kappa^h_{\rm S}$ and $\kappa^{h_S}_{\rm NSM}$, with values of $0.15 \simlt |\kappa^h_S| \simlt 0.35$, as anticipated in Eq.~(\ref{eq:kappahsNSM}).  In the left panel we show the values of $\kappa^{h_S}_{\rm NSM}$ and $\kappa^{h_S}_{\rm NSM}$, which are proportional to $\eta$ and $\eta'$ respectively.  Whereas $\kappa^{h_S}_{\rm SM}$ can become very small in the region of alignment, there is no strong correlation between these two singlet components.  There is  only a weak correlation associated with the dependence of the singlet production cross section on the doublet components, which leads to negative values of $\kappa^{h_S}_{\rm SM}$ being somewhat more restricted than positive ones for $\mu >0$, as could be anticipated from the behavior exhibited in Fig.~\ref{fig:Component4}.

Due to the specific values of  $\kappa^{h_S}_{\rm NSM}$ under consideration, and the correlation between $\kappa^h_{\rm NSM}$ and the product of $\kappa^h_{\rm S} \kappa^{h_S}_{\rm NSM}$, a mild correlation appears between the non-SM components of  the SM-like Higgs, which is displayed in the left panel of Fig.~\ref{fig:Component2}.  The largest singlet components are associated with the smallest SM component and hence the smallest values of the couplings to vector bosons. The bottom-quark coupling can be visibly 
suppressed in this region, but the branching ratios and signal strengths remain in the allowed region due to the suppression of the vector bosons coupling and a small enhancement of the up-quark couplings, as follows from \eqst{eq:phicouplings1}{eq:phicouplings3}.  In
contrast, as shown in the right panel of Fig.~\ref{fig:Component2}, enhancements of the bottom couplings are more restricted due to a suppression of the $h$ branching ratios to photons and vector bosons and an additional suppression of the gluon fusion production cross section associated with smaller top-quark  couplings.

\section{Higgs Boson Production and Decay}
\label{HiggsDecays}

The study of the properties of the 125 GeV Higgs boson and their proximity to SM expectations  has been the subject of intensive theoretical and experimental analyses, and will remain  one of the most 
important research topics at  the LHC.
Close to the alignment limit, and in the absence of beyond-the-SM light charged or colored particles,
the properties of one of the neutral scalars (identified with the observed Higgs boson of mass 125 GeV) 
are nearly identical to those of the SM Higgs boson. 
However, as demonstrated  
 in the right panel of Fig.~\ref{fig:Component2}, the current Higgs data
allow for variations, of up to about 30\%,  of the 125 GeV Higgs boson production and decay rates 
with respect to the SM predictions.
Such deviations can be understood as a function of the mixing of the observed SM-like Higgs boson with additional non-SM-like Higgs scalars that could be searched for at  the LHC.

In this section, we shall mainly concentrate on the non-SM-like Higgs boson production and decay rates and their
possible search channels at the LHC.  It is noteworthy that, due to the smallness of $\kappa^H_{\rm SM}$, the couplings of 
the heavy Higgs bosons to the up and down-quarks are close to the MSSM decoupling values. 
In particular, using \eq{eq:kappamatrix},  the ratio of their
couplings  to the SM ones given by \eqst{eq:phicouplings1}{eq:phicouplings3} are
\begin{eqnarray}
g_{HVV} & \simeq & \mathcal{O}(\epsilon)\,,
\nonumber\\
g_{Htt} & \simeq &  -\frac{1}{\tan\beta} + \mathcal{O}(\epsilon)\,,
\nonumber\\
g_{Hbb} & \simeq & \tan\beta + \mathcal{O}(\epsilon) \,,
\end{eqnarray}
where the terms of $\mathcal{O}(\epsilon)\ll 1$ represent a linear combination of terms of order
$\mathcal{M}^2_{12}/\mathcal{M}^2_{22}$
and $\mathcal{M}_{13}^2\mathcal{M}_{23}^2/\mathcal{M}_{22}^4$ [cf.~the discussion below \eq{ll}].

Similarly, the CP-odd couplings are given approximately by their MSSM expressions,
\begin{eqnarray}
g_{Att}  & \simeq & \frac{1}{\tan\beta}\,,
\nonumber \\
g_{Abb}  & \simeq & \tan\beta\,.
\end{eqnarray}
Finally, the $h_S$ couplings are given by
\begin{eqnarray}
g_{h_SVV} & = & - \eta' \,,
\nonumber\\
g_{h_Stt} & = & - \eta' - \frac{\eta}{\tan\beta}\,,
\nonumber\\
g_{h_Sbb} & = & -\eta' + \eta \tan\beta\,.
\end{eqnarray}
Considering the typical values of the mixing angles $\eta$ and $\eta'$, we see that
the production cross section of $h_S$ via top-quark induced gluon fusion
is generally at least an order of magnitude lower than the one for a SM-Higgs boson of the
same mass. Due to the smallness of the bottom Yukawa couplings and the small
values of $\tanb$ considered in this work, the decay branching ratios are
mainly determined by the $h_S$ mass and will be of order of the SM ones. Therefore
the present constraint on the signal strength of the production of $h_S$ decaying to
vector bosons, $\mu_{VV}\simlt 0.1$ discussed in Section \ref{sect:general} is not expected to strongly constrain this model.

\begin{figure}[h]
\subfloat[]{\includegraphics[width=3.25in, angle=0]{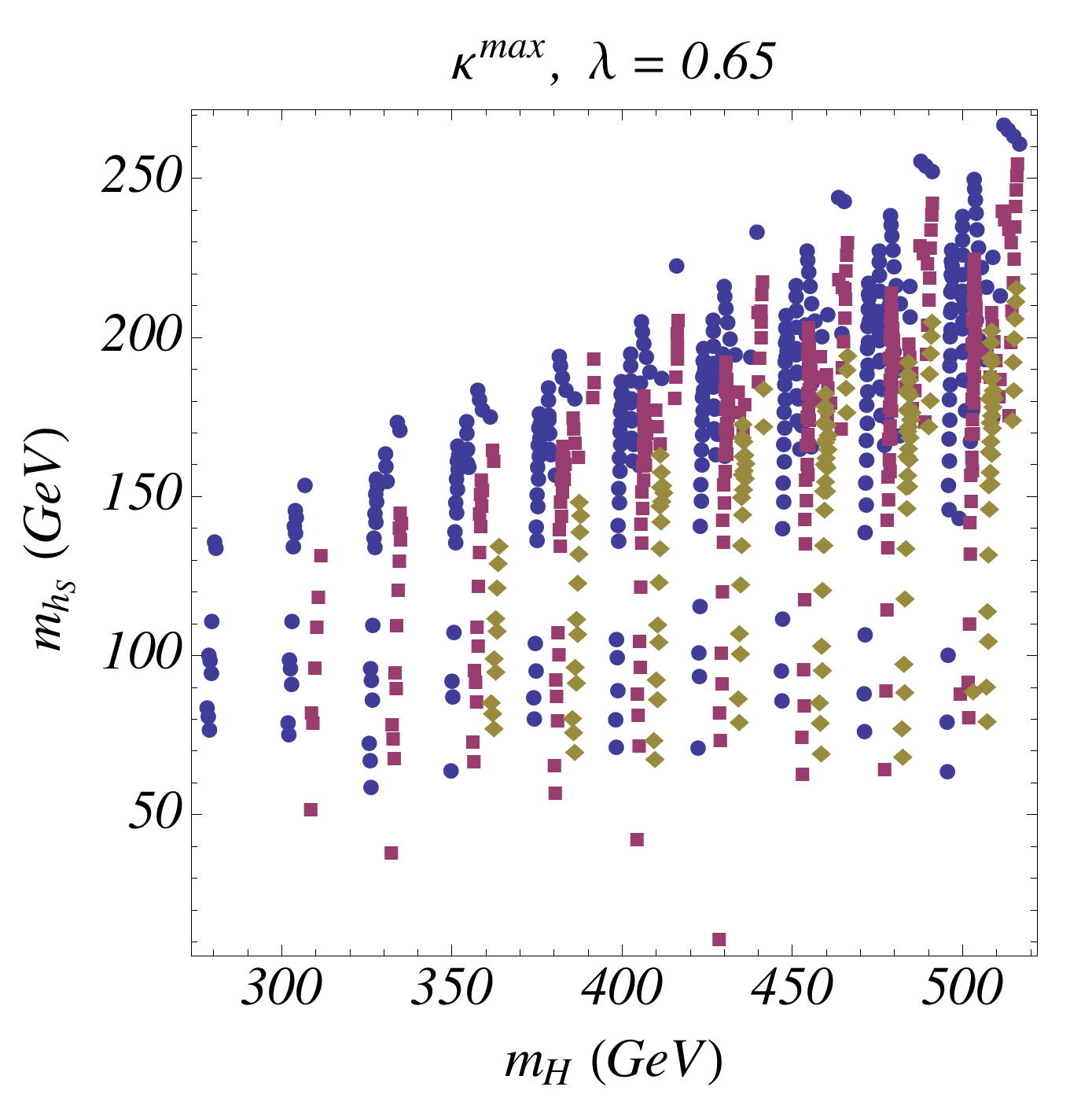}}~~
\subfloat[]{\includegraphics[width=3.25in, angle=0]{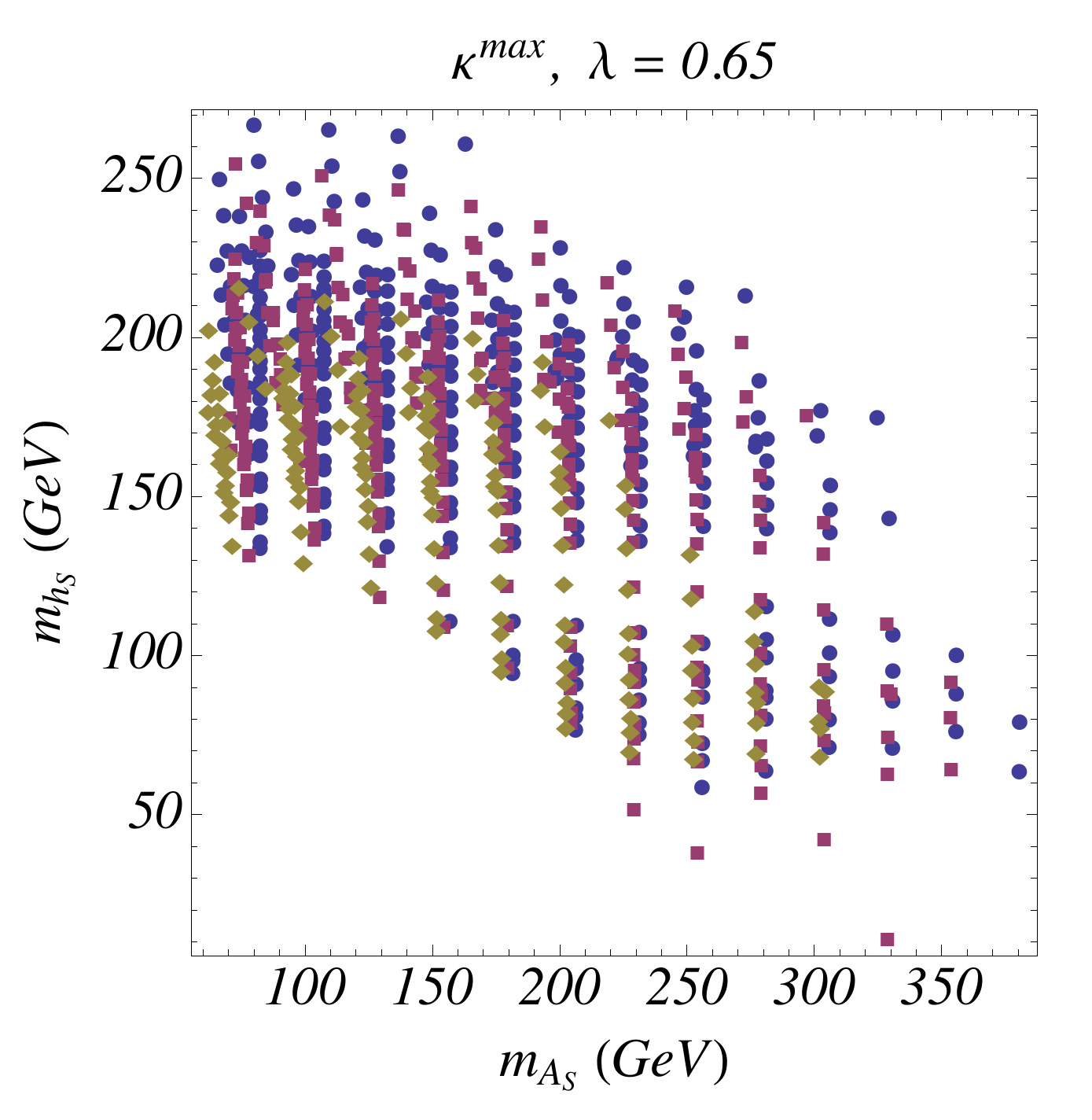}}
\caption{\label{fig:NumSpectr}{ \em Correlation between $m_H \simeq m_A$ and the lightest non-SM-like CP-even  Higgs boson mass (left panel) and anti-correlation between the masses of
the lightest non-SM-like CP-even Higgs boson and the lightest, mostly singlet CP-odd
Higgs boson (right panel), for values of $\kappa = \kappa^{\rm max}$. Blue, red and
yellow represent values of $\tanb = 2$, 2.5 and 3, respectively.}}
\end{figure}

In Section~\ref{spectrum} we discussed the analytical constraints on the Higgs spectrum that play a crucial role in the phenomenology of the non-SM-like Higgs bosons.
In
Fig.~\ref{fig:singlet} we showed contours of the singlet CP-even Higgs mass for different 
values of 
$M_A \simeq m_{A} \simeq m_H$ and the lightest CP-odd Higgs mass, which tends to be mostly singlet in this region of parameter space. 
In Fig.~\ref{fig:NumSpectr}, we exhibit the correlation between the mass  $m_{A} \simeq m_H$ of the heaviest CP-odd/even Higgs bosons (which possess a significant doublet component) and the lighter mostly singlet CP-even Higgs boson mass (left panel), and the anti-correlation between the mass of the  lightest CP-odd Higgs boson (which possesses a significant singlet component) and the
mostly singlet like CP-even Higgs boson (right panel).  These numerical result verify the expectations based on the analytical analysis of Section~\ref{spectrum}.
In particular, these singlet-like Higgs boson masses are always smaller than $m_A$ and the relation
\be
m_{A} \geq  2 \ m_{h_S}
\ee
is fulfilled. On the other hand, the anti-correlation between the CP-odd/even mainly singlet Higgs boson masses implies that 
values of  $ m_{A_S} \simlt 150$ GeV  constrain $m_{h_S}$ to be larger than about 120 GeV, while values of   $m_{h_S} \simlt 120$ GeV  imply $ m_{A_S} \simgt 150$ GeV.

In general, large values of $M_A \simeq m_{A} \simeq m_{H}$
are allowed, as in the usual decoupling regime, but in this work we are mostly interested in having a SM-like Higgs boson for values of $M_A \lsim 500$~GeV, where the non-SM-like Higgs bosons are not heavy.
Given that we are interested in values of $\tanb\sim 2$ and 
$M_A \simeq  |\mu|/s_\beta c_\beta $, this leads also to low values of $\mu$, improving the naturalness of the theory. 
Considering the LEP lower bound on $|\mu|$, the above relation also implies  $M_A \simgt 250$~GeV. Therefore,
the decays \be
H \to h \ h_S, \;\;\;\;\;\;  H \to h_S  h_S   \;\;\;\;\;\; {\rm and}  \;\;\;\;\;  H \to h h
\ee
are always allowed. However, since the coupling $g_{Hhh}$ approaches zero in the alignment limit [cf.~Appendix~\ref{couplings}], the first two
decay rates are in general more significant than the decay into pairs of SM-like Higgs bosons. Moreover,
from Table~\ref{masseigenstatescalars} of Appendix~\ref{couplings}, it follows that when $M_A\simeq 2|\mu|/s_{2\beta}$ and $\kappa$ is small,
\begin{equation}
g_{Hhh_S} \simeq  \sqrt{2}\ \lambda \mu\, \cot {2\beta}    \;\;\;\;\;\;   g_{H h_S h_S} \simeq 4 \sqrt{2}\ \eta \lambda \mu.
\end{equation}
Hence, for $0.15 \simlt \eta \simlt 0.35$, these couplings are of the same order as $|\mu|$ for $2 \simlt \tanb \simlt 3$,
 which implies that these decay channels may contribute significantly to the $H$ decay width.

 On the other hand, mixing between the doublet and singlet states in the CP-odd sector is also dictated by $\eta$ and hence
non-vanishing.  Therefore, the decay channels  
\be
H \to A_{S} Z    \;\;\;\;\;  {\rm and}  \;\;\;\;\;\;  A \to h_{S} Z
\ee
may become significant.
In particular, for values of  
the heavy Higgs states below the $t \bar{t}$ threshold, the decay branching ratio in these channels may become dominant at low values of $\tanb$, for which the  
couplings to down-quark fermions and charged leptons are small, of the order of the  
corresponding SM Yukawa couplings.  

\begin{figure}[t!]
\subfloat[]{\includegraphics[width=3.1in, angle=0]{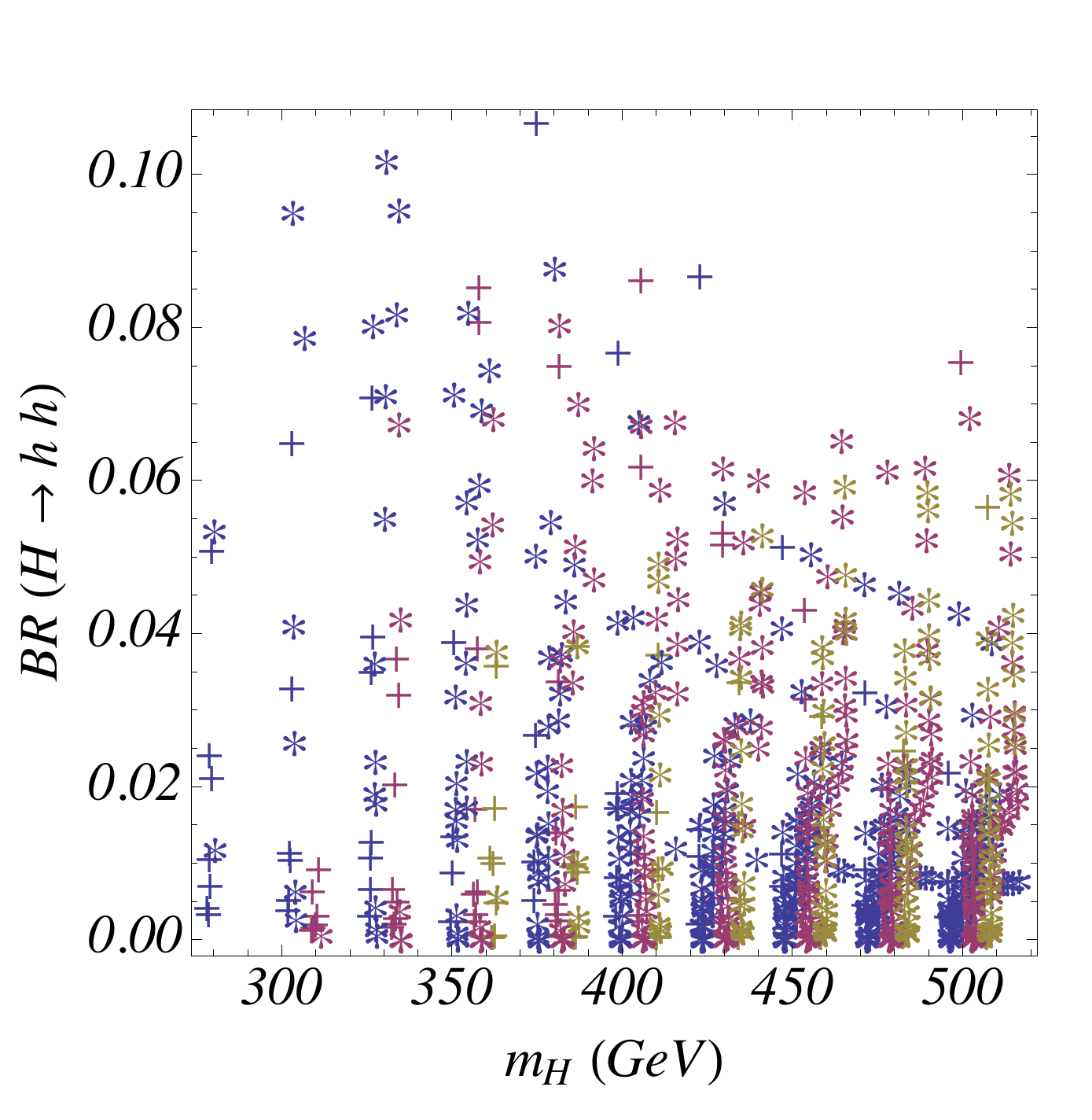}}~~
\subfloat[]{\includegraphics[width=3.1in, angle=0]{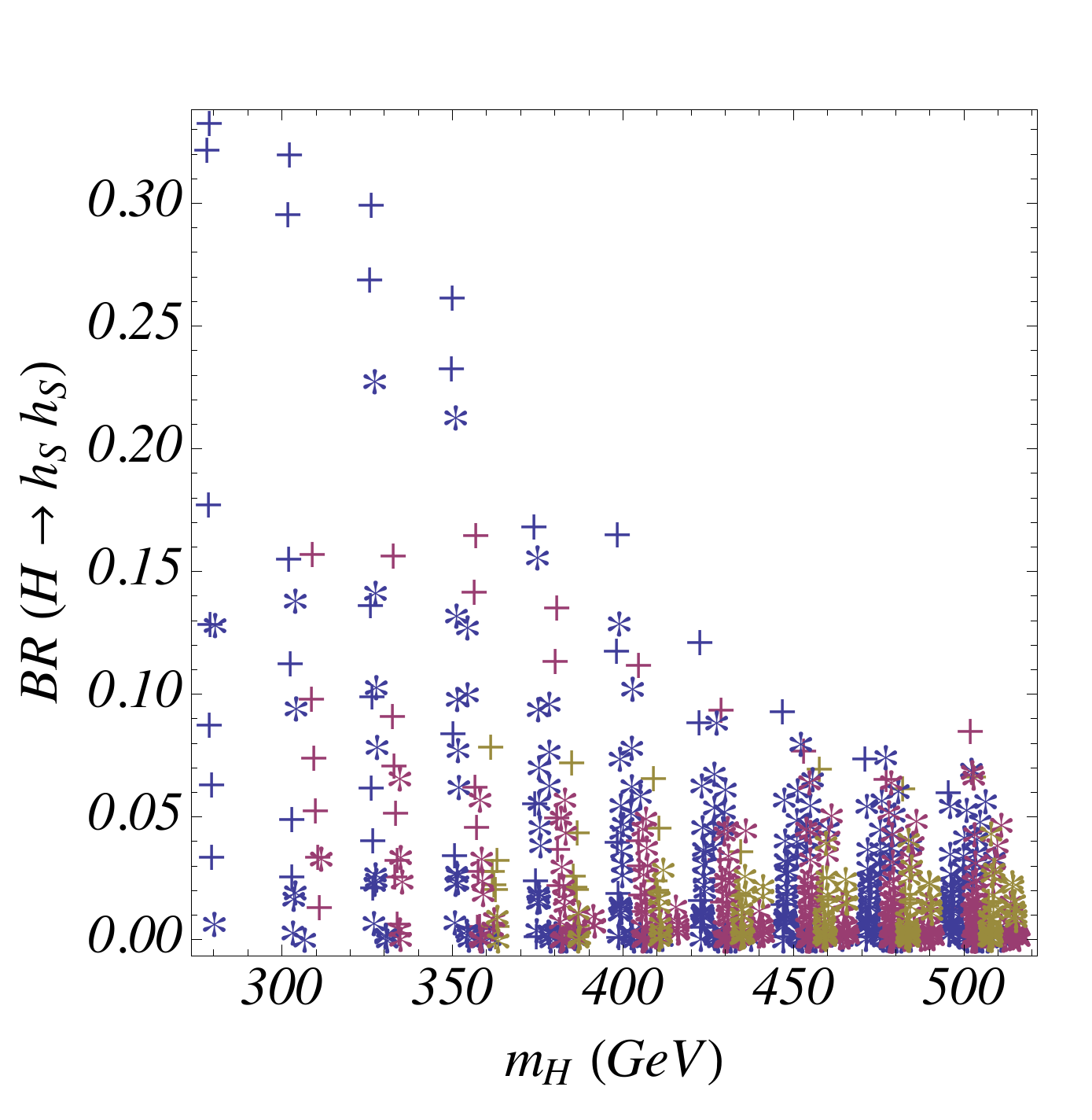}}
\caption{\label{fig:toh1h1h2h2}{\em Branching ratio of the decay of the heaviest CP-even Higgs boson into pairs of identical CP-even Higgs bosons. Blue, red and
yellow represent values of $\tanb = 2$, 2.5 and 3, respectively. \\[-40pt]}}
\end{figure}
%
\begin{figure}[h!]
\subfloat[]{\includegraphics[width=3.1in, angle=0]{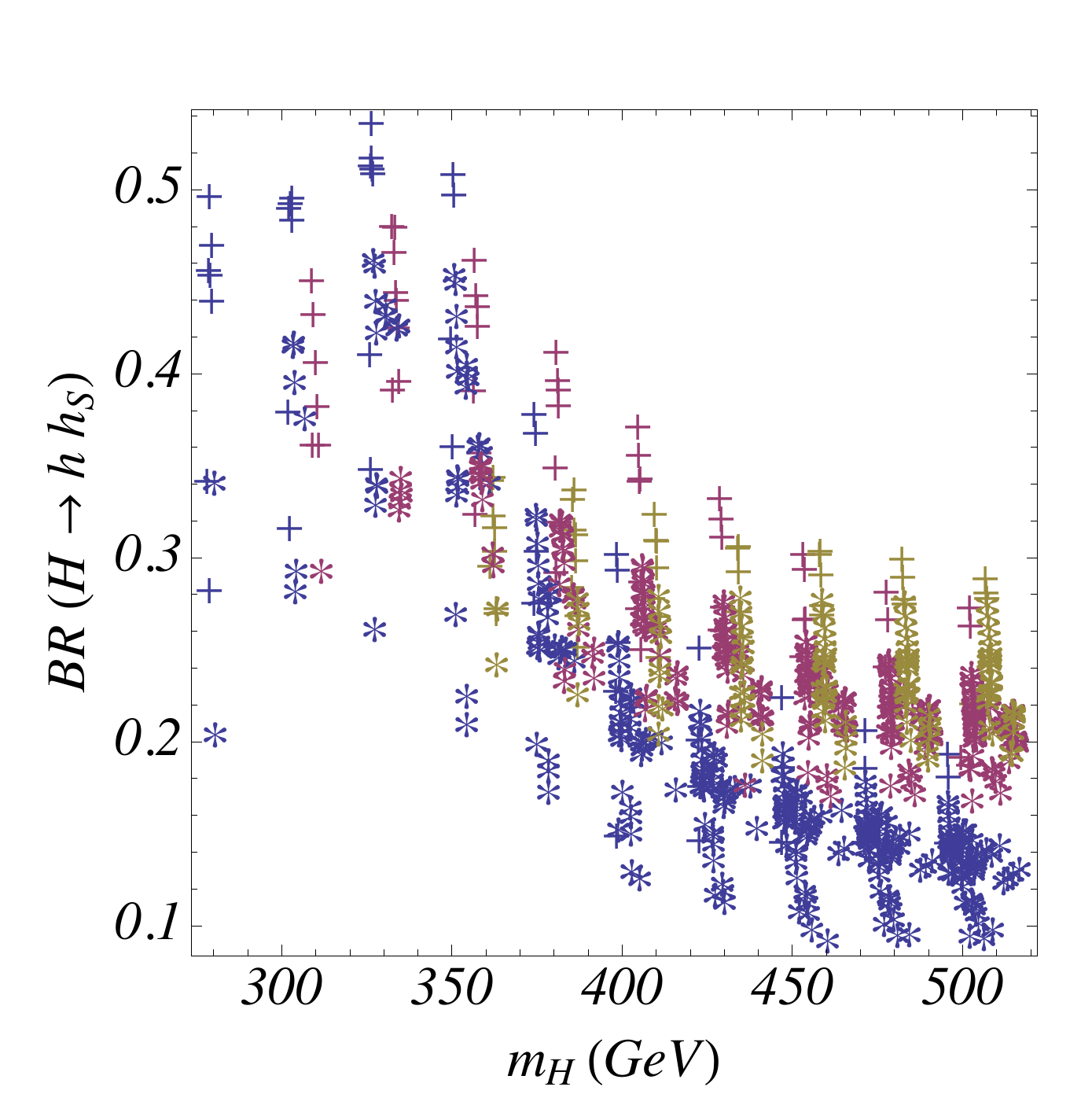}}~~
\subfloat[]{\includegraphics[width=3.1in, angle=0]{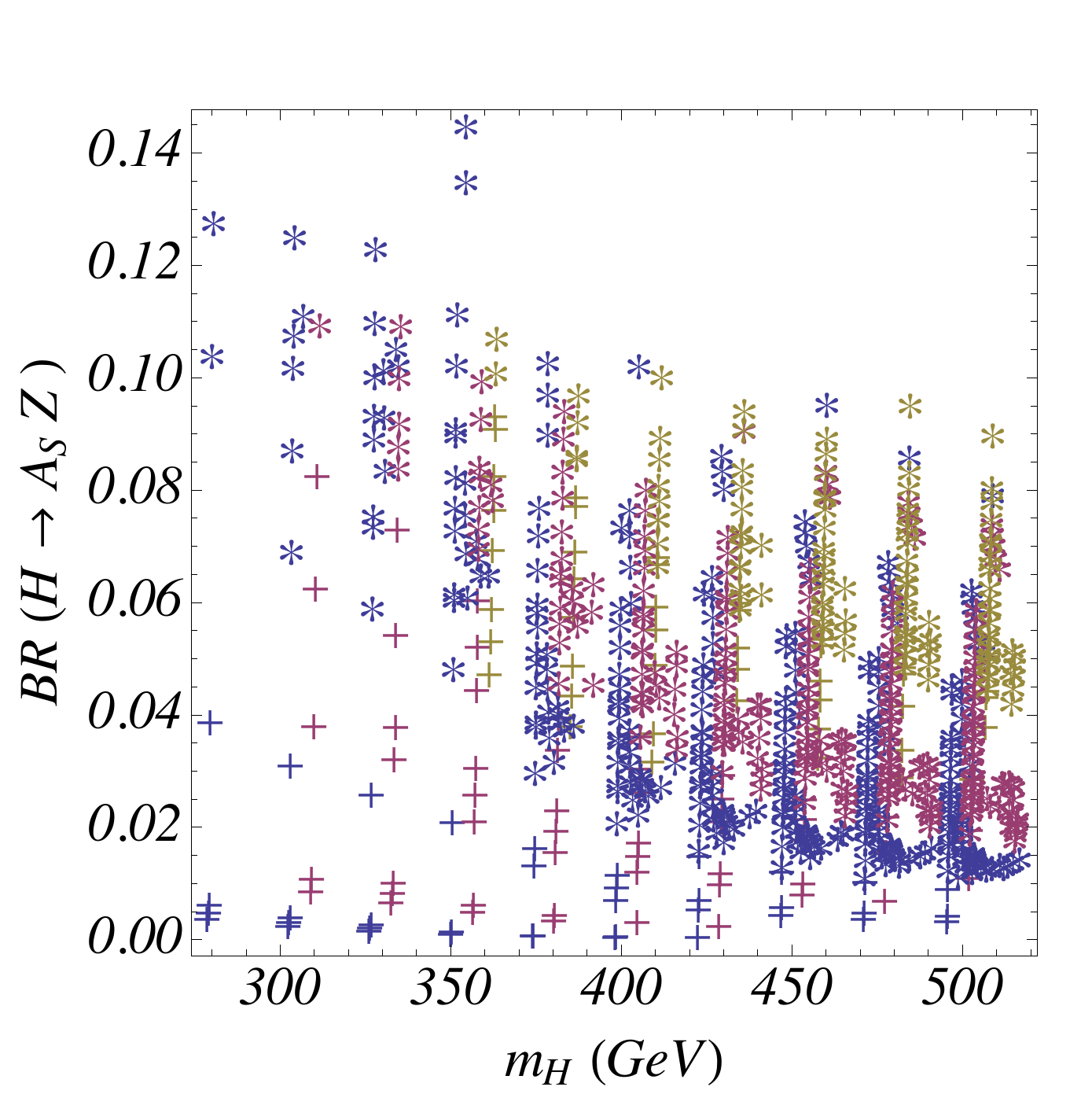}}
\caption{\label{fig:toh12h2a1z}{\em Branching ratios of the decay of the heavy CP-even Higgs boson into a pair of non-identical lighter CP-even
Higgs bosons, $H \to h h_S$ (left panel) and into the lightest CP-odd Higgs boson and a $Z$ boson (right panel). Blue, red and
yellow represent values of $\tanb = 2$, 2.5 and 3, respectively. }}
\end{figure}

%
\begin{figure}[t!]
\subfloat[]{\includegraphics[width=3.2in, angle=0]{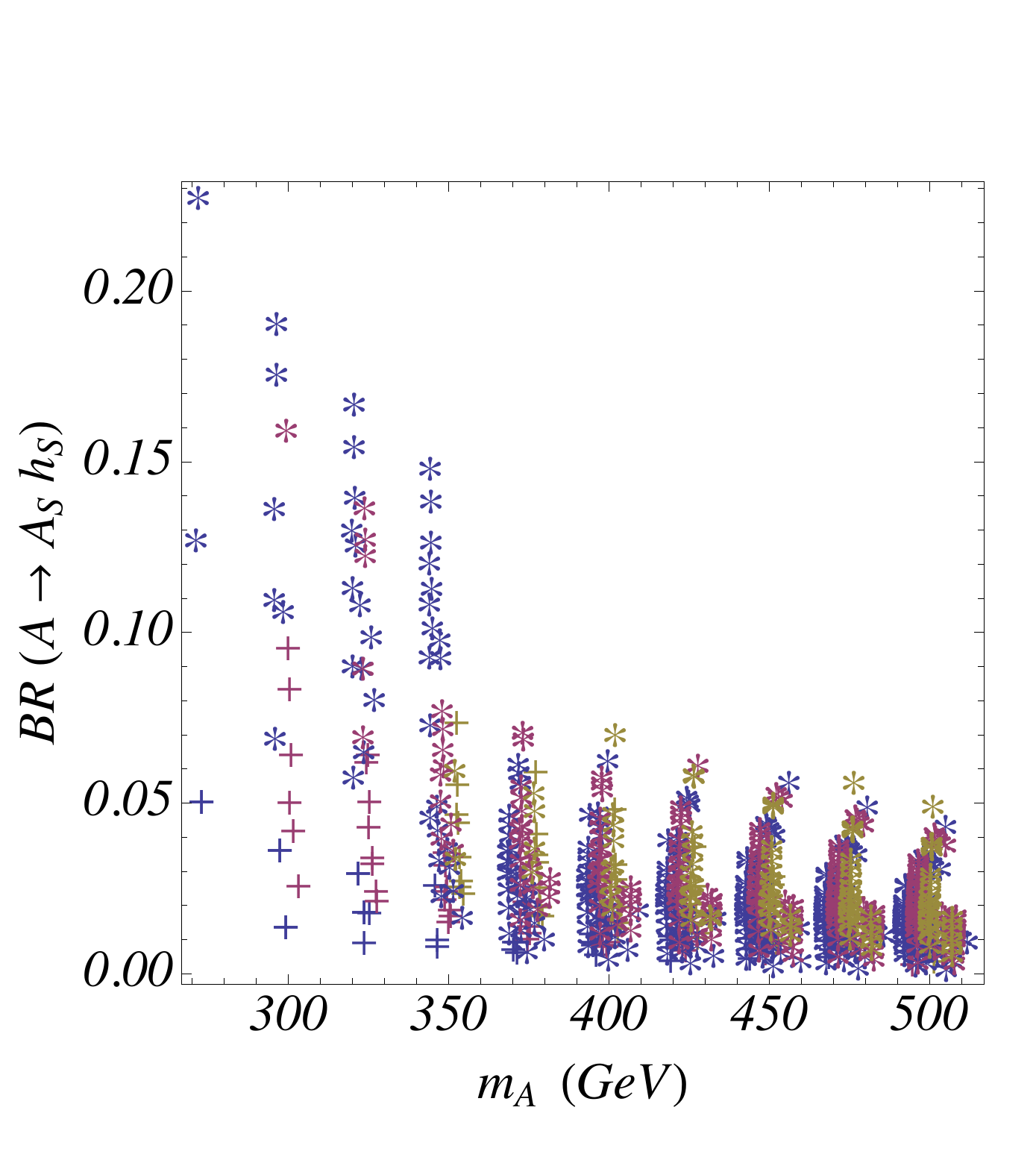}}
\subfloat[]{\includegraphics[width=3.2in, angle=0]{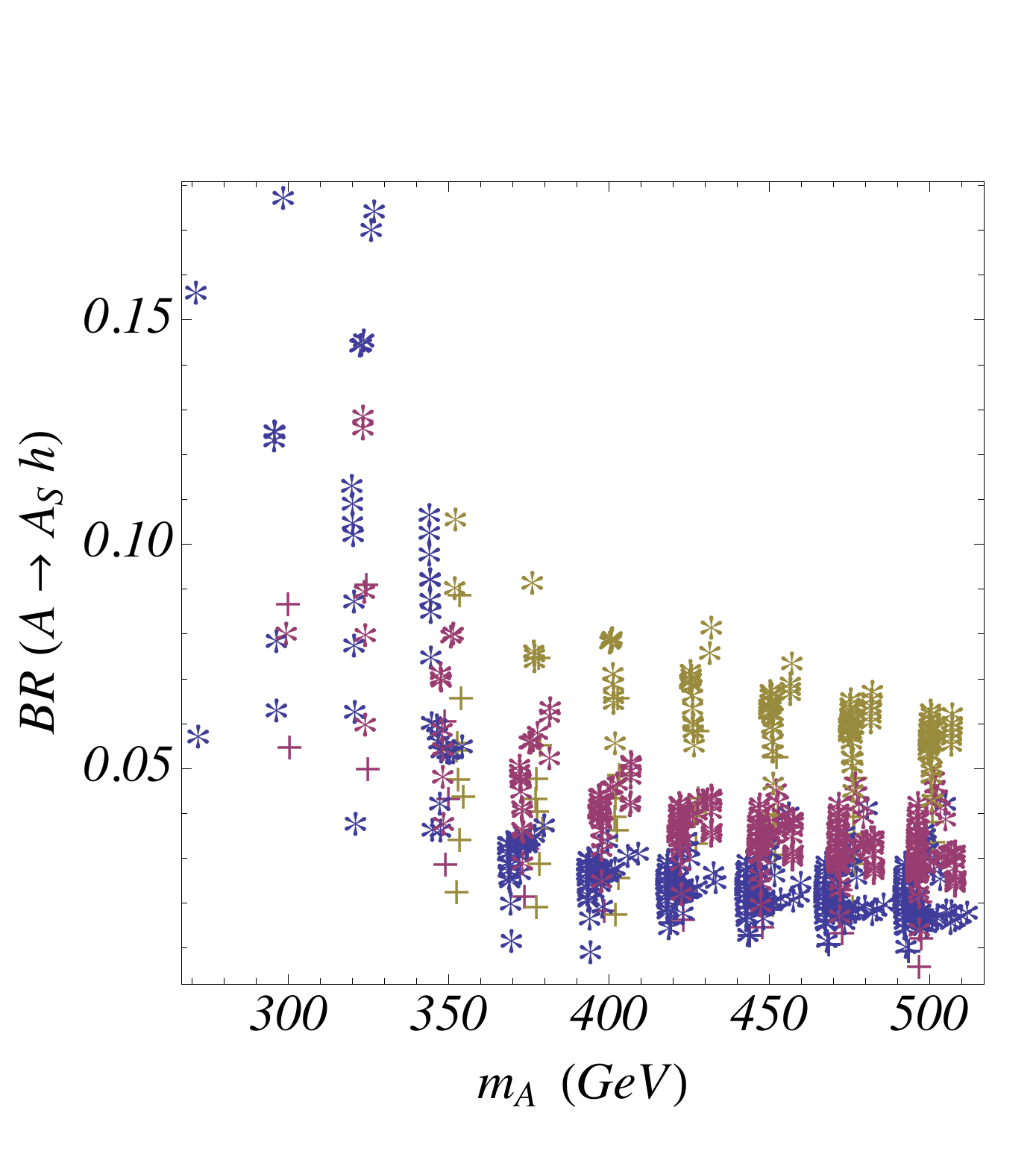}}
\caption{\label{fig:toa1h12}{\em Branching ratio of the decay of the heaviest CP-odd Higgs boson into
a pair of non-identical Higgs bosons consisting of
the lightest CP-odd Higgs boson and one of the two lighter
CP-even Higgs bosons. 
Blue, red and
yellow represent values of $\tanb = 2$, 2.5 and~3, respectively.}}
\end{figure}
%
\begin{figure}[h!]
\subfloat[]{\includegraphics[width=3.2in, angle=0]{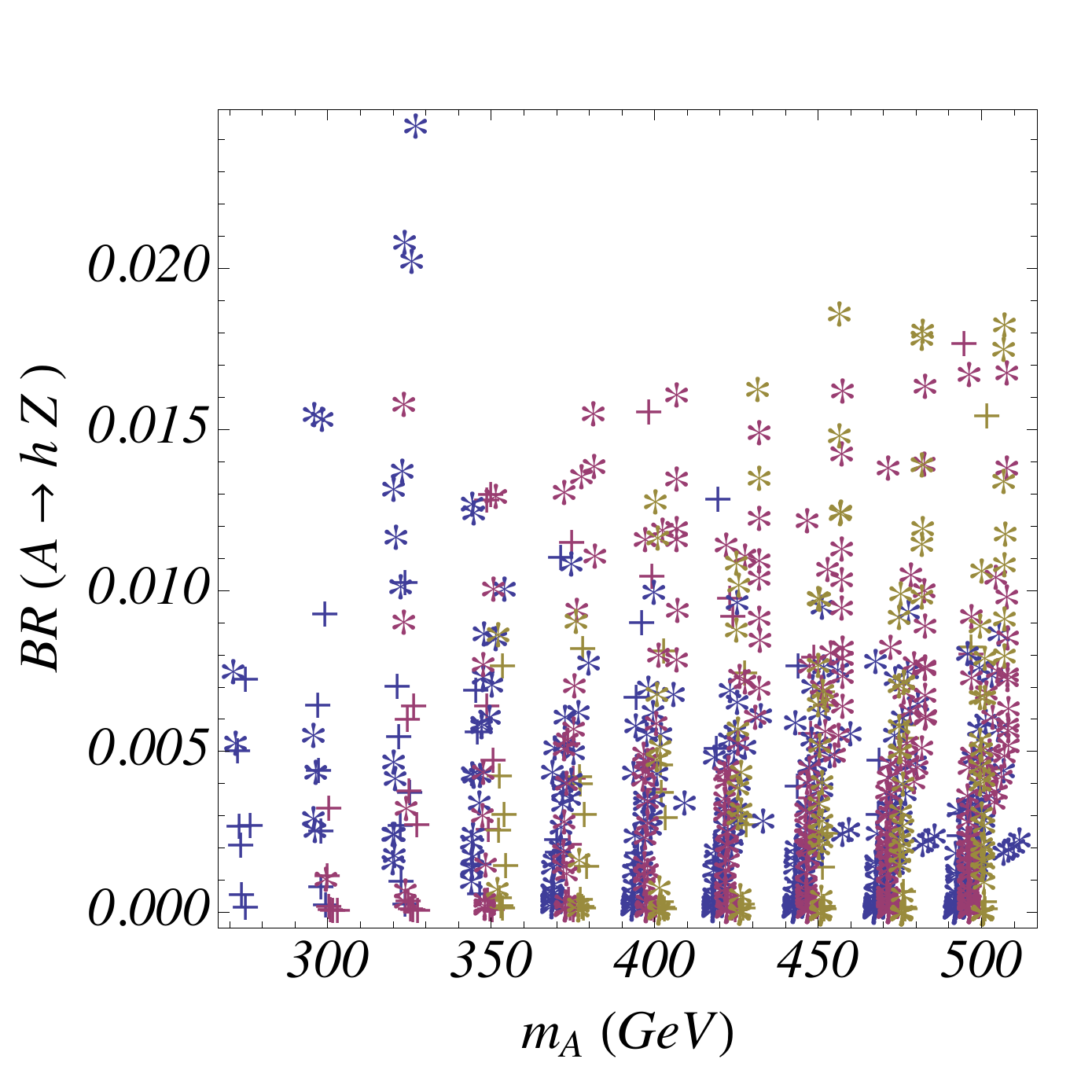}}
\subfloat[]{\includegraphics[width=3.2in, angle=0]{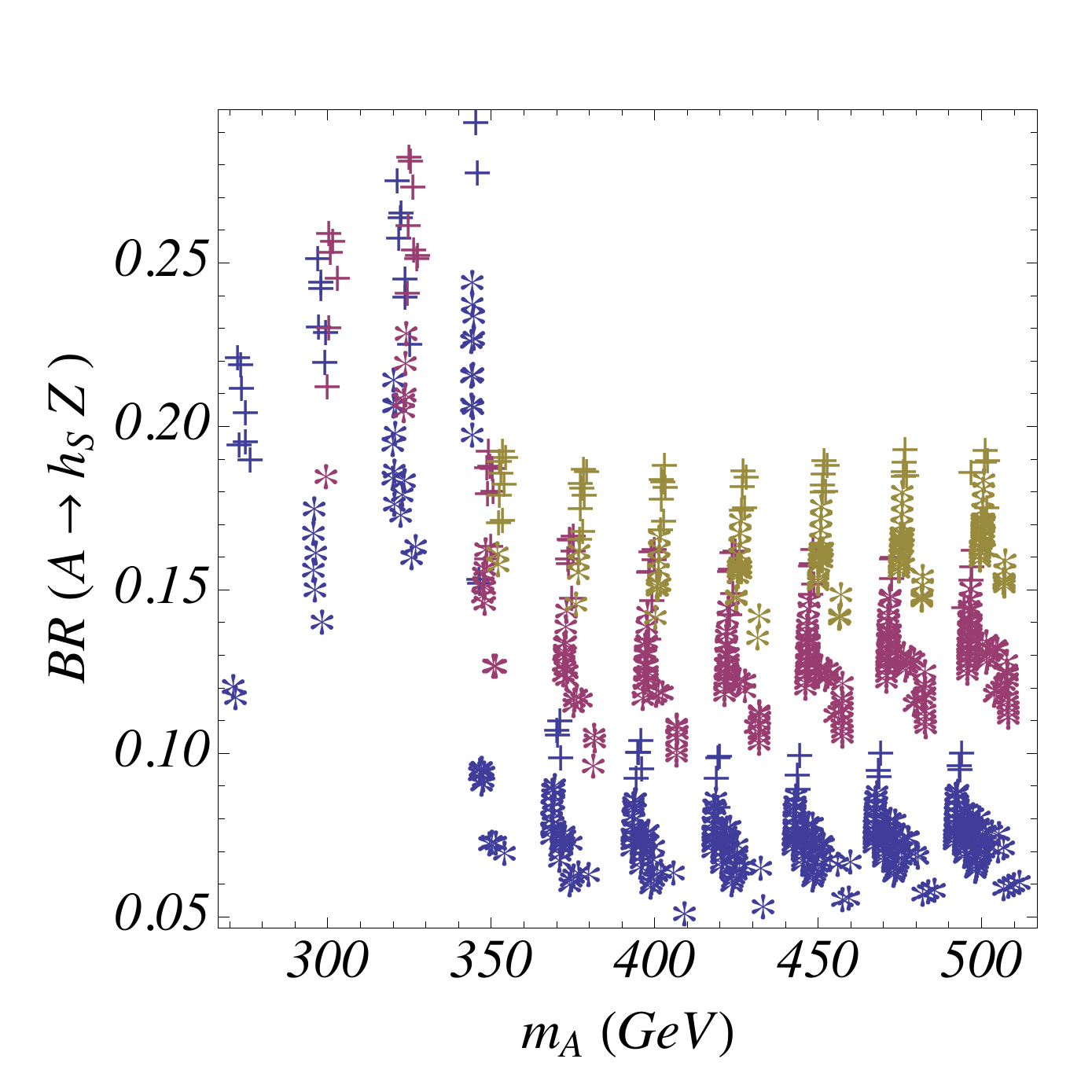}}
\caption{\label{fig:tozh}\em Branching ratio of the decay of the heaviest CP-odd Higgs boson into a $Z$ and the lightest CP-even 
Higgs bosons, $h$ (left panel) and $h_S$ (right panel). Blue, red and
yellow represent values of $\tanb = 2$, 2.5 and 3, respectively.}
\end{figure}

Figs.~\ref{fig:toh1h1h2h2}  and \ref{fig:toh12h2a1z} show the branching ratio for the decay of the heaviest CP-even Higgs boson into lighter bosons.   We observe that these branching ratios are appreciably large for values of the heaviest Higgs boson masses smaller than 350~GeV,
for which the decay into  top-quark pairs is forbidden, and remain significant for larger value of $m_A$, particularly for the largest values of
$\tanb$ considered.  In particular, the decay of $H$ into a pair of non-identical lighter CP-even Higgs bosons is the most important one.
In Figs.~\ref{fig:toh1h1h2h2}  and \ref{fig:toh12h2a1z} we differentiate between the cases in which the lightest CP-even Higgs boson is identified with the SM-like Higgs boson with mass 125 GeV, represented by snowflakes, from the
case in which the lightest CP-even Higgs boson is singlet-like (hence with mass below 125 GeV), represented by crosses. We clearly see from Fig.~\ref{fig:toh1h1h2h2}  
that the decay of $H \to h h$ is suppressed, being at most of order of 10\% as a result of the proximity to alignment.

Similarly, in Figs.~\ref{fig:toa1h12} and \ref{fig:tozh} we exhibit the branching ratios of the decay of the heaviest CP-odd Higgs boson
into the lighter CP-odd and CP-even Higgs bosons,  and the branching ratio of its decay into a CP-even Higgs boson and a $Z$ 
boson.
From Fig.~\ref{fig:tozh} 
we can see that the branching ratio into a $Z$ and the SM-like Higgs boson $h$ is always suppressed. 
However, the decay of the heavy CP-odd scalar into a $Z$ and $h_S$ is unsuppressed and hence may serve as  a good discovery
channel.  This possibility will be addressed later in this Section.

As a result of the approximate alignment condition, $M_A \simeq 2 |\mu|/s_{2\beta}$ [cf.~\eq{masim}], decays of the heavy CP-even and CP-odd Higgs bosons into pairs of
charginos and neutralinos are kinematically allowed. In contrast to the case of the MSSM, where heavy gauginos imply a suppression of the coupling
of the Higgs bosons to Higgsino pairs, in the NMSSM the coupling $\lambda$ induces a non-negligible coupling to charginos via the singlet component of~$H$. Moreover, the coupling $\lambda$ and the self-coupling parameter $\kappa$ induce new decays in the neutralino sector due to the mixing of the singlinos and Higgsinos.
Indeed, the  singlino  mass
\be
{m_{\tilde{S}}}   \simeq  \frac{2 \kappa}{\lambda} \mu
\ee
 is constrained to be below the Higgsino mass $\mu$ due to the condition of perturbative consistency up to the Planck scale (see Fig.~\ref{fig:lambdafromh}), implying that the decays 
\be
H, A  \to \chi^{0,\pm}_i \chi^{0,\mp}_j
\ee
are likely to have sizable rates in the region of parameters under consideration.

\begin{figure}[t!]
\subfloat[]{\includegraphics[width=3.2in, angle=0]{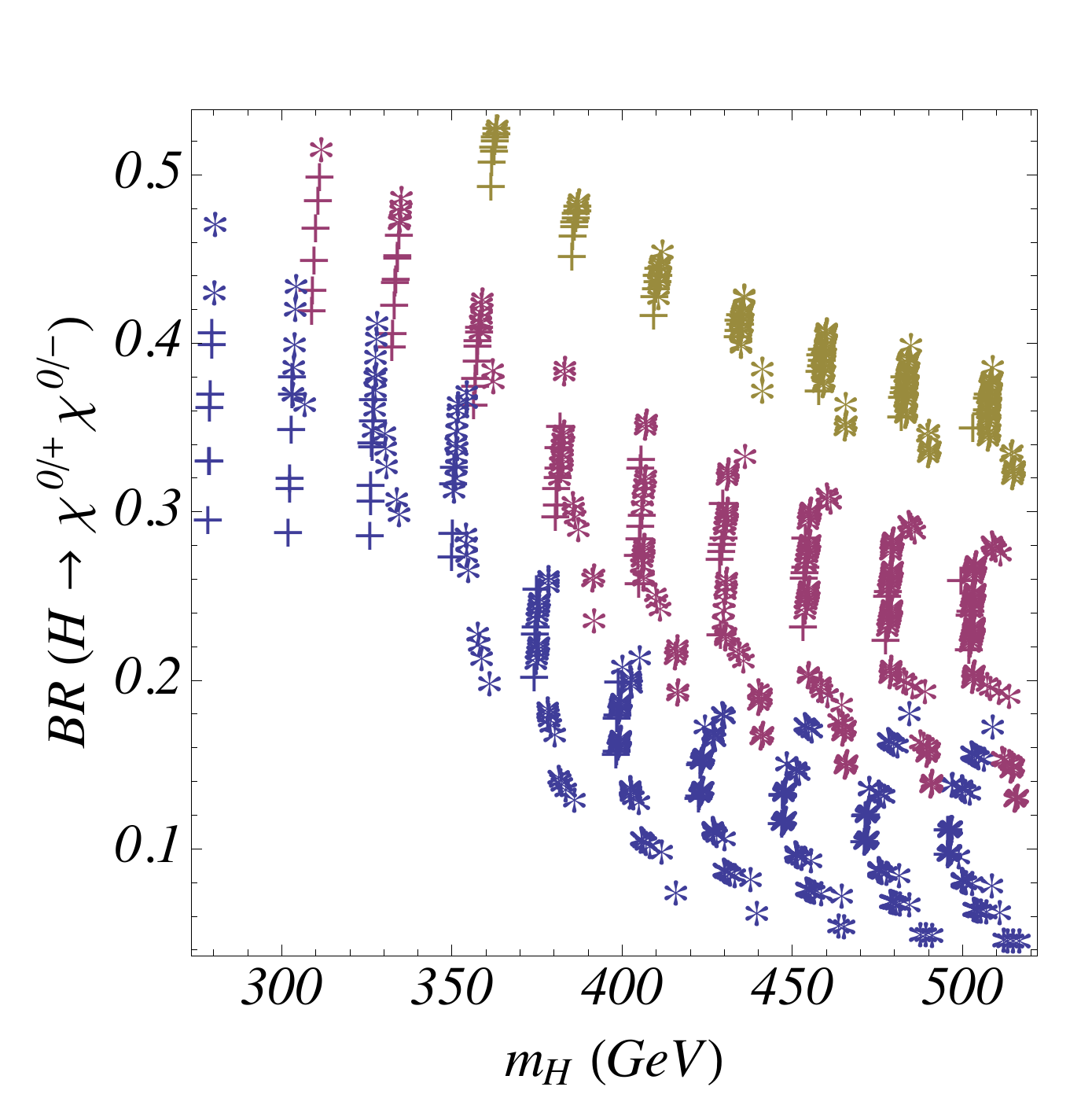}}~~
\subfloat[]{\includegraphics[width=3.2in, angle=0]{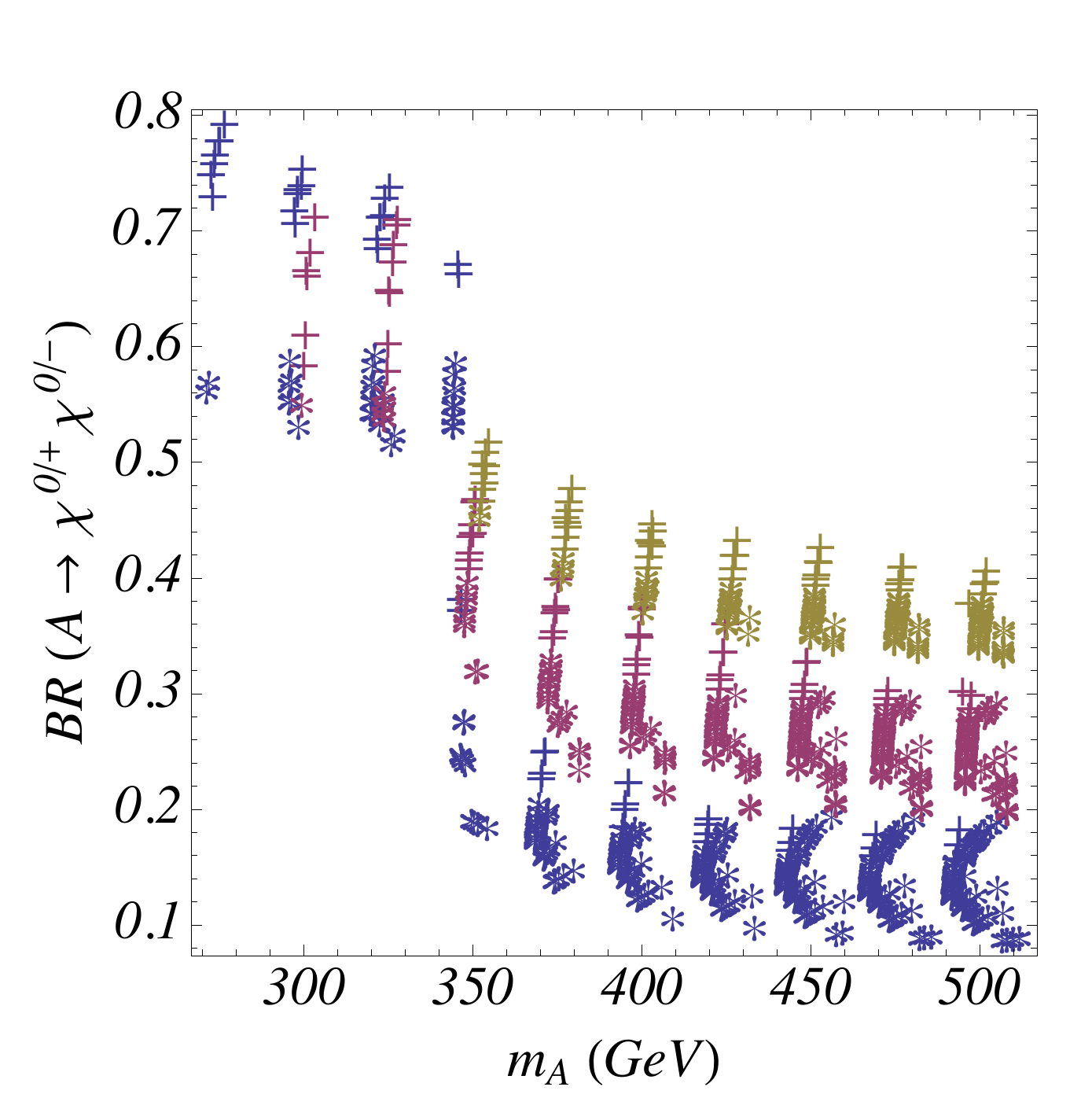}}
\caption{\label{fig:toinos}{ Branching ratio of the decay of the heaviest CP-even (left panel) and CP-odd (right panel) Higgs bosons into
charginos and neutralinos. Blue, red and
yellow represent values of $\tanb = 2$, 2.5 and 3, respectively.}}
\end{figure}
\begin{figure}[h!]
\subfloat[]{\includegraphics[width=3.2in, angle=0]{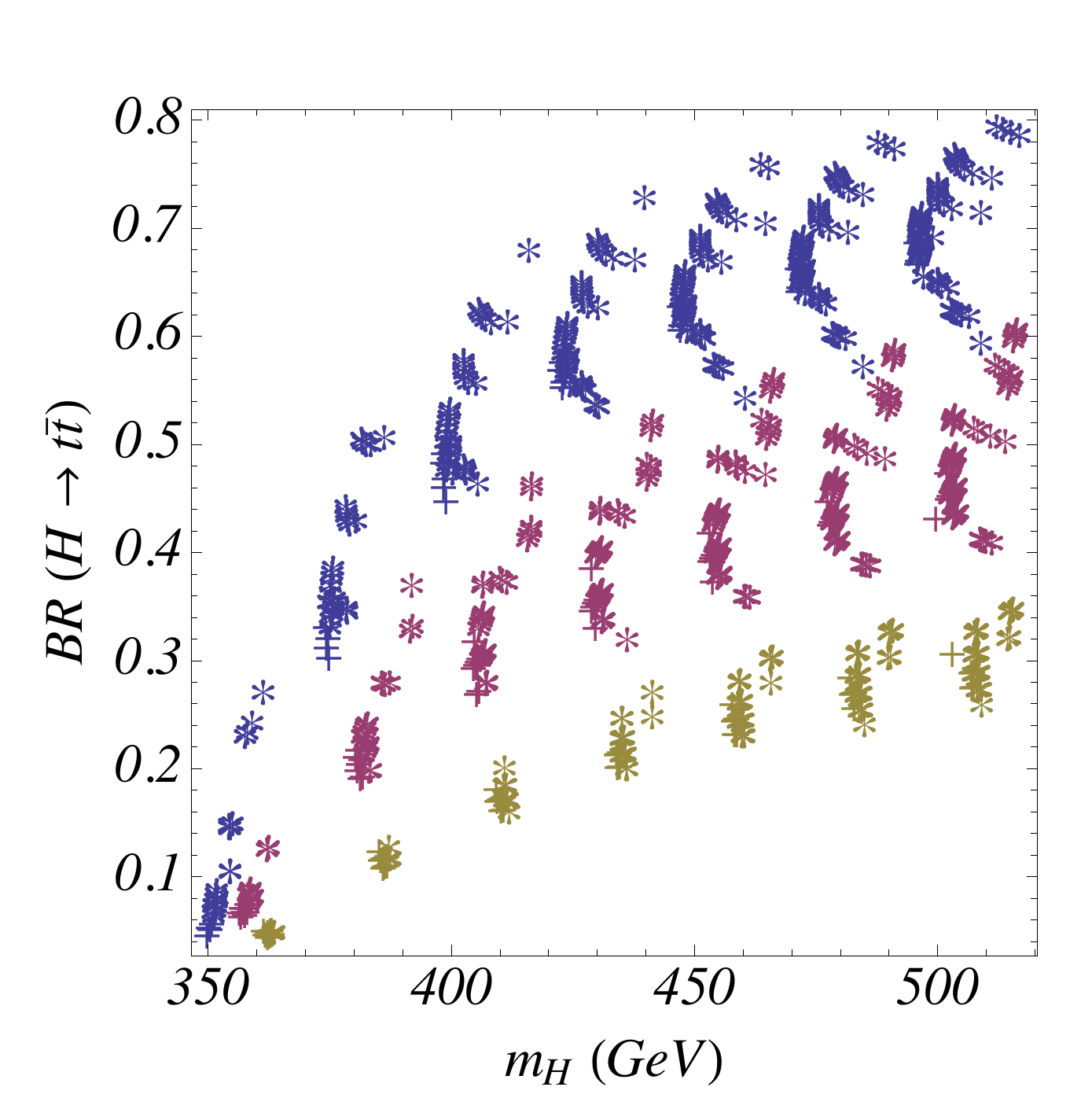}}
\subfloat[]{\includegraphics[width=3.2in, angle=0]{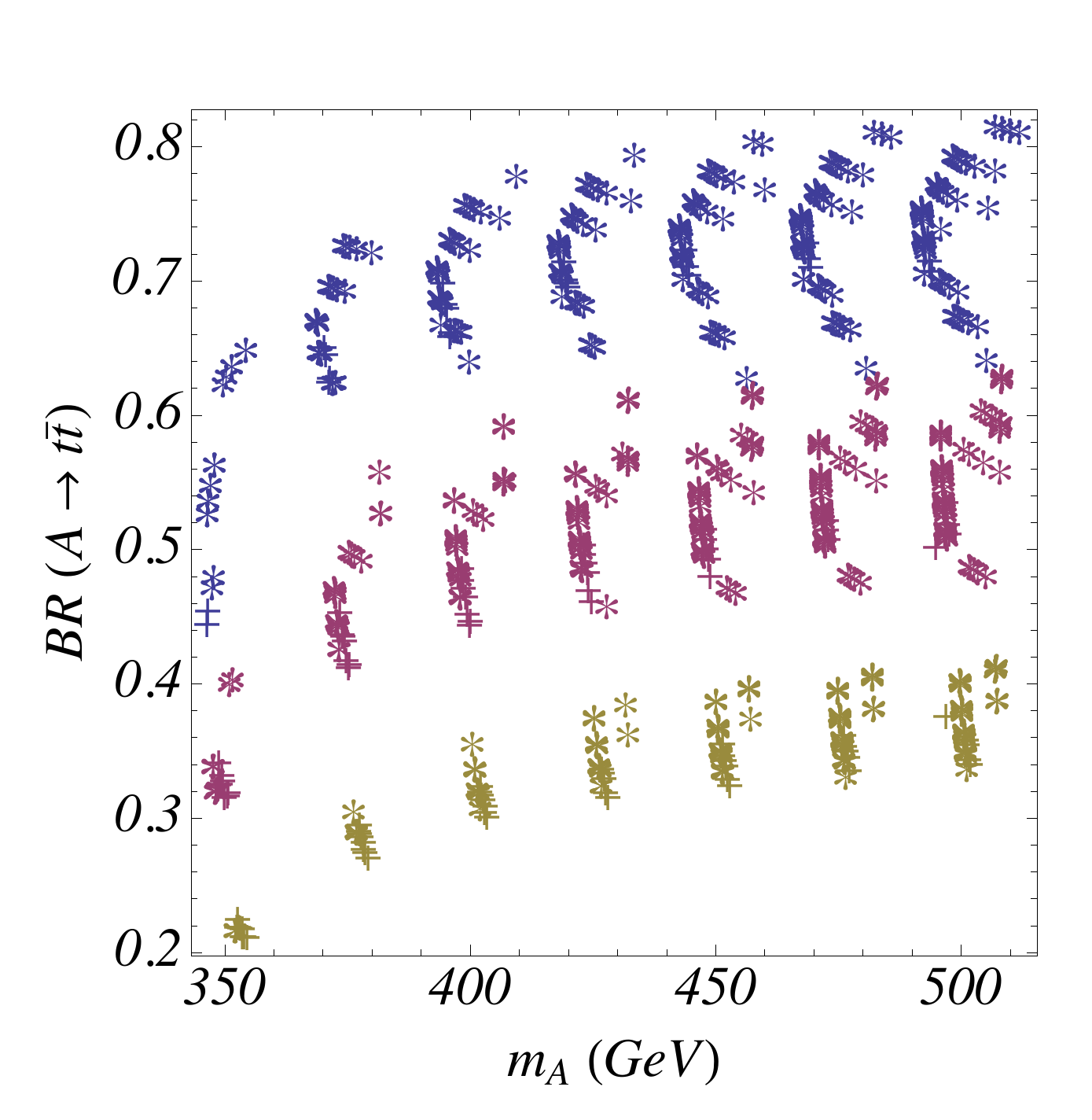}}
\caption{\label{fig:tott}{\em Branching ratio of the decay of the heaviest CP-even Higgs boson (left panel) and the heaviest
CP-odd Higgs  boson (right panel) into pairs of top quarks. Blue, red and
yellow represent values of $\tanb = 2$, 2.5 and 3, respectively. }}
\end{figure}

Fig.~\ref{fig:toinos} illustrates that  the heavy Higgs bosons $H$ and $A$ have sizable decay branching ratios into
charginos and neutralinos.  These branching ratios become more prominent for larger values of $\tanb$ and for masses below 350~GeV where
the decays into top quarks are suppressed.

For completeness, we present the branching ratio of the heaviest CP-even and CP-odd Higgs bosons into top quarks in Fig.~\ref{fig:tott}.
As expected, this branching ratio tends to be significant for masses larger than 350 GeV and becomes particularly
important at low values of $\tanb$, for which the couplings of the heaviest non-SM-like Higgs bosons to the top quark
are enhanced.  In spite of being close to the alignment limit, this branching ratio is always significantly lower than 1,
due to the decays of the Higgs bosons to final states consisting of the lighter Higgs bosons and chargino and/or neutralino pairs, as noted above. 

\begin{figure}[t!]
\subfloat[]{\includegraphics[width=3.2in, angle=0]{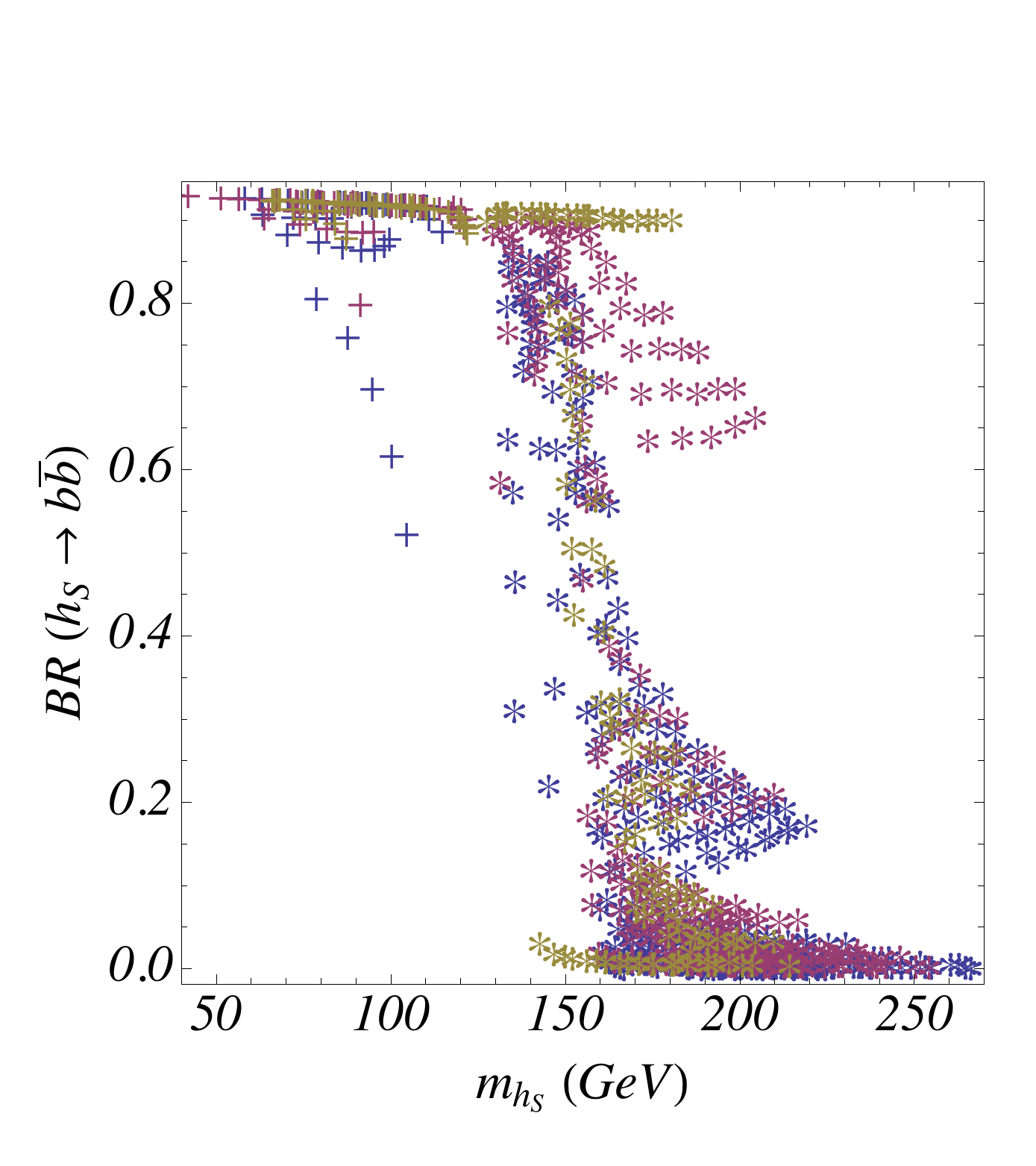}}
\subfloat[]{\includegraphics[width=3.2in, angle=0]{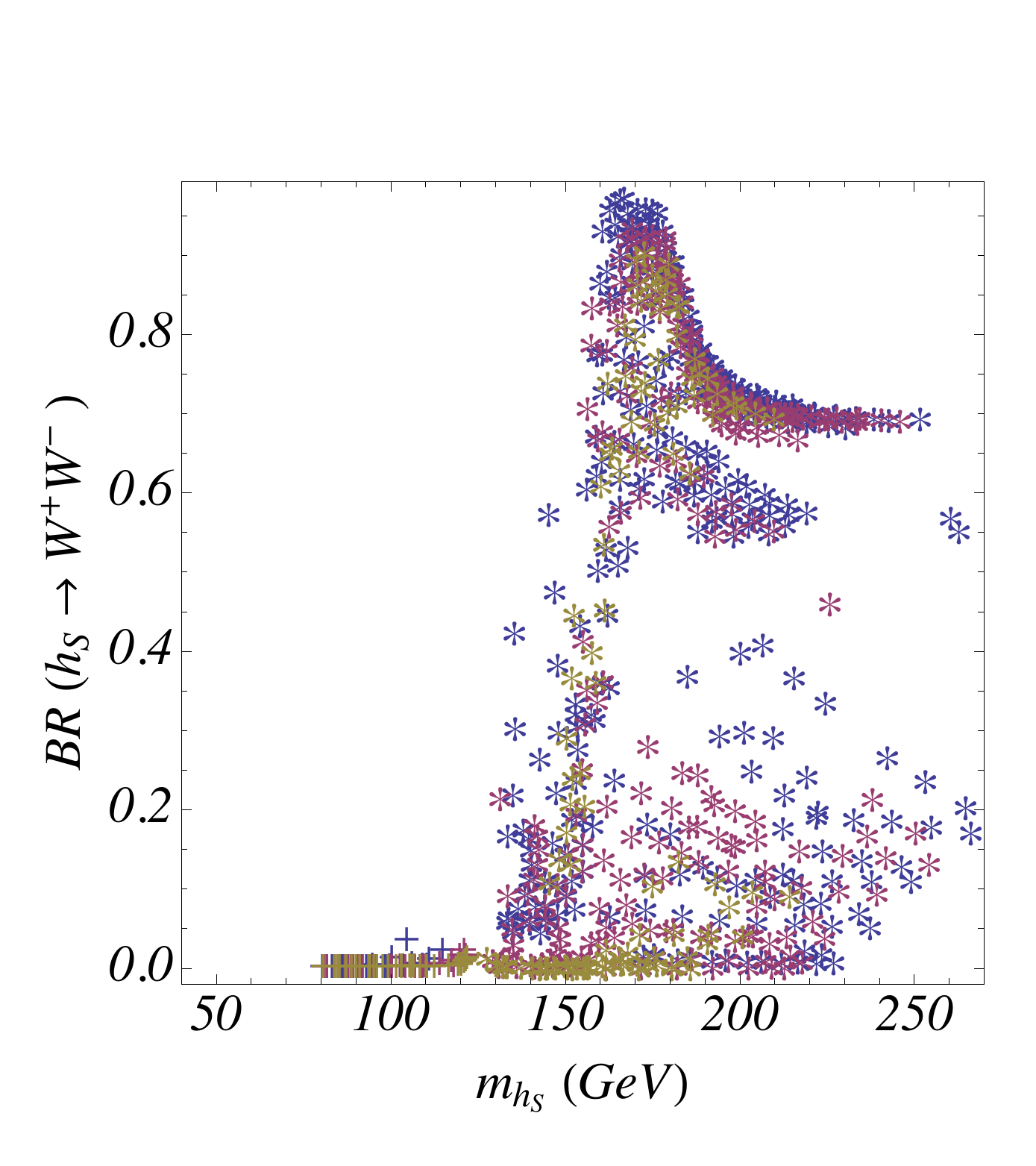}}
\caption{\label{fig:hstobbWW}{ \em Branching ratio of the lightest non-SM-like CP-even Higgs boson into bottom quarks (left panel)
and pair of $W$ gauge bosons (right panel). Blue, red and
yellow represent values of $\tanb = 2$, 2.5 and 3, respectively. }}
\end{figure}

Indeed, apart from the decays into top-quark pairs, whose observability demands a good top reconstruction method  and 
is quite challenging \cite{Craig:2015jba, Hajer:2015gka}, 
the heaviest Higgs bosons exhibit prominent branching ratios into lighter Higgs bosons (as in the case of
generic 2HDMs \cite{Coleppa:2014hxa}). 
Moreover, in light of the  large gluon fusion $A/H$ production cross sections,
the heavy Higgs decays into charginos and neutralinos are also relevant and yield production rates that are of the same order of magnitude as the chargino/neutralino Drell-Yan production cross sections. Unfortunately, the subsequent decays of the charginos into $W/Z$ and missing energy renders these search modes challenging.

In order to ascertain the constraints on the heavy non-SM-like Higgs bosons arising from their decays into the lightest Higgs bosons, one must analyze the decay branching
ratios of $h_S$ and $A_S$. Since these particles are  singlet-like, their couplings are controlled via the mixing with the doublet states.
As shown in Fig.~\ref{fig:Component3},
the CP-even singlet state has small mixing with the SM-like Higgs boson, $\kappa^{h_S}_{\rm SM}  = -\eta'$, which  is small and can be no larger than 0.3.
On the other hand, 
the mixing with the non-SM doublet component $\kappa^{h_S}_{\rm NSM}$
is small but non-vanishing.  Therefore, as shown in Fig.~\ref{fig:hstobbWW},  the bottom quark decays are clearly dominant for masses below 130~GeV, while the $WW$ and eventually
$ZZ$ decay branching ratios may become dominant for masses above 130~GeV, depending on the proximity to alignment. For mass values above about 150 GeV, decays into two CP-odd singlet-like Higgs bosons open up for certain regions of parameter space.\footnote{For sufficiently heavy $h_S$ and light neutralinos, the decays into neutralinos could also open, although such a channel does not show up in the benchmarks to be discussed later.}
The singlet-like CP-odd Higgs boson has dominant decay into bottom quark pairs  for masses up to about 200 GeV, whereas decays into $Z h_S$ and into neutralinos may open up for slightly  heavier masses.

\begin{figure}[htb]
 \subfloat[]{\includegraphics[width=3.5in, angle=0]{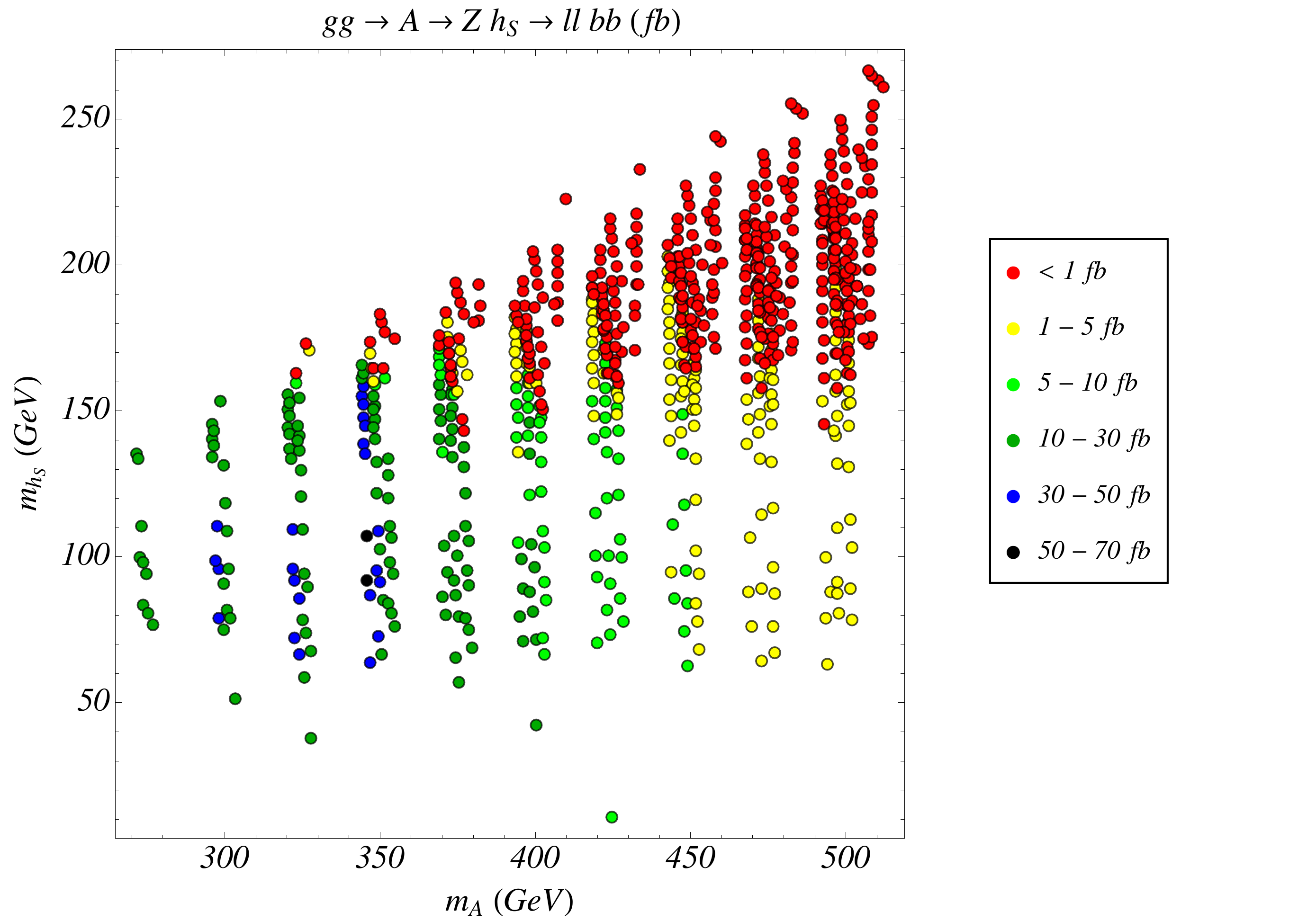}}
\subfloat[]{\includegraphics[width=3.5in, angle=0]{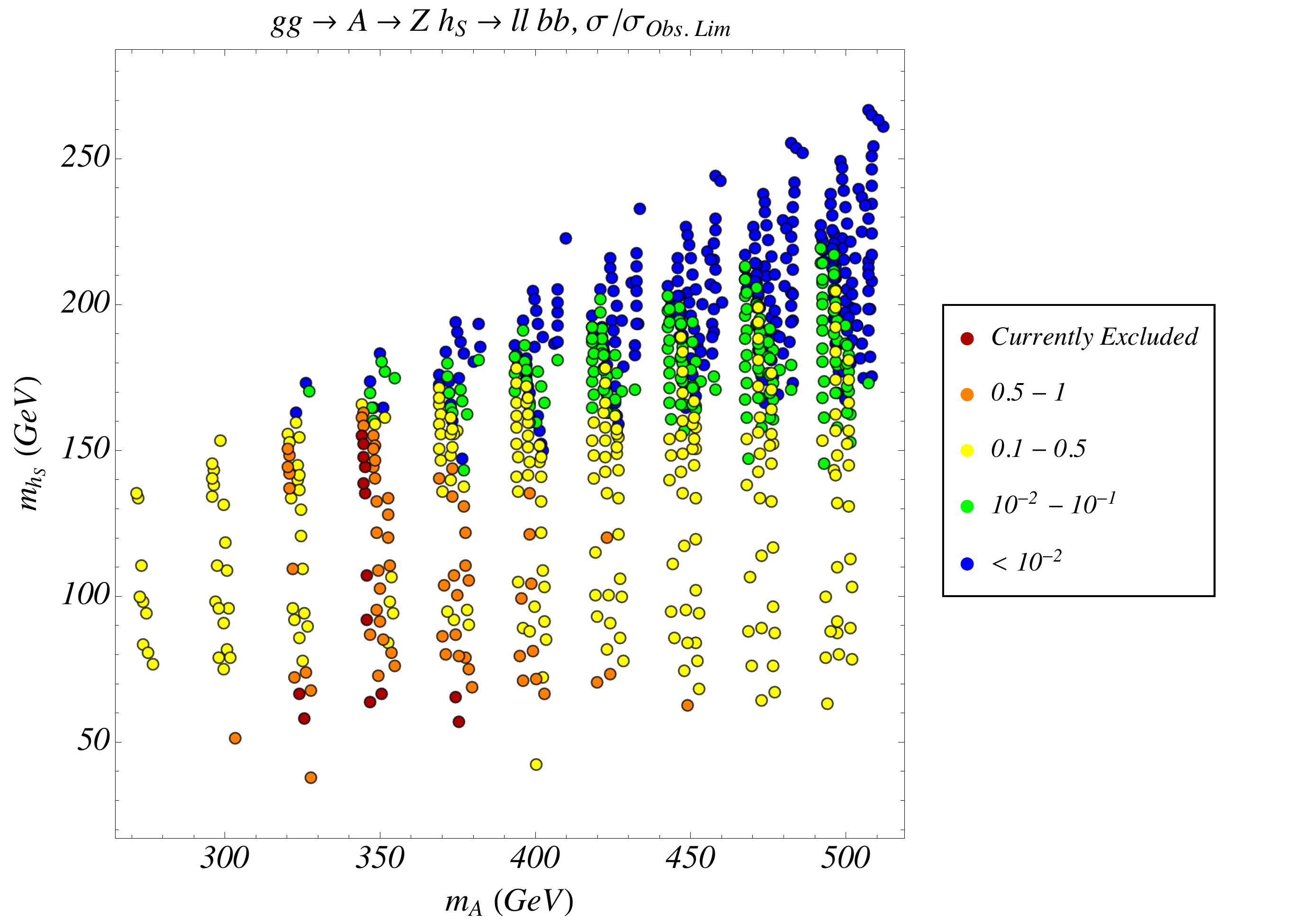}}
\caption{\label{fig:sigmaAtozh}\em  The production cross section times branching ratio (left),
and the ratio of the observed limit to the production cross section times branching ratio (right) of the decay of the heaviest CP-odd Higgs boson into a $Z$ and a CP-even  Higgs boson 
as a function of the heaviest CP-odd and the singlet like CP-even Higgs boson masses.
The cross sections are computed for LHC processes with $\sqrt{s}$ = 8 TeV, and the branching ratio includes the 
subsequent decay of the Z boson into di-leptons and $h_S$  into a bottom quark pair.}
\end{figure}

Based on the study of the non-SM-like Higgs boson branching ratios presented above we will now discuss the main search channels which may lead to discovery of the additional scalar states at the LHC.
  In  Fig.~\ref{fig:sigmaAtozh} we present the 8 TeV production cross sections of the heaviest CP-odd scalar $A$,
 decaying into a $Z$ and a $h_S$ in the $m_A$ -- $m_{h_S}$ plane.  The cross sections
 presented in the left panel of Fig.~\ref{fig:sigmaAtozh}  take into account the decay branching ratios of $Z \to \ell\ell$ and $h_S \to b\bar{b}$,
 since these final states provide excellent search modes at the LHC. 
The CMS experiment has already 
 performed searches for  scalar resonances decaying into a $Z$ and lighter scalar resonance
 using 8 TeV data \cite{CMS:2015mba}. In  the right panel of  Fig.~\ref{fig:sigmaAtozh}
  we have used the CMS {\tt ROOT} files\footnote{These have been obtained from \texttt{https://twiki.cern.ch/twiki/bin/view/CMSPublic/Hig15001TWiki}.}
to compare the limits extracted from these
 searches with the predictions of the scenario considered here. 
 
We observe that although this mode fails at present to probe
 a large fraction of the NMSSM Higgs parameter space, the current limit is close to the expected cross section for values of  $m_{h_S} \simlt 130$ GeV. Hence,
 $A\to Zh_S\to (\ell\ell)(b\bar{b})$ provides a very promising channel for non-SM-like Higgs boson searches in the next run of the LHC. 
 It is also clear  from Fig.~\ref{fig:sigmaAtozh}
that for values of the $h_S$ mass above 130 GeV, where its decay branching ratio into bottom quarks 
 becomes small, the $A\to Zh_S$ search channel becomes less efficient.  However, in this case the decay modes into weak gauge bosons may become relevant,  and searches for $h_S \to WW^{(*)}/ZZ^{(*)}$ may provide an excellent complementary probe.

\begin{figure}[t!]
\subfloat[]{\includegraphics[width=4.5in, angle=0]{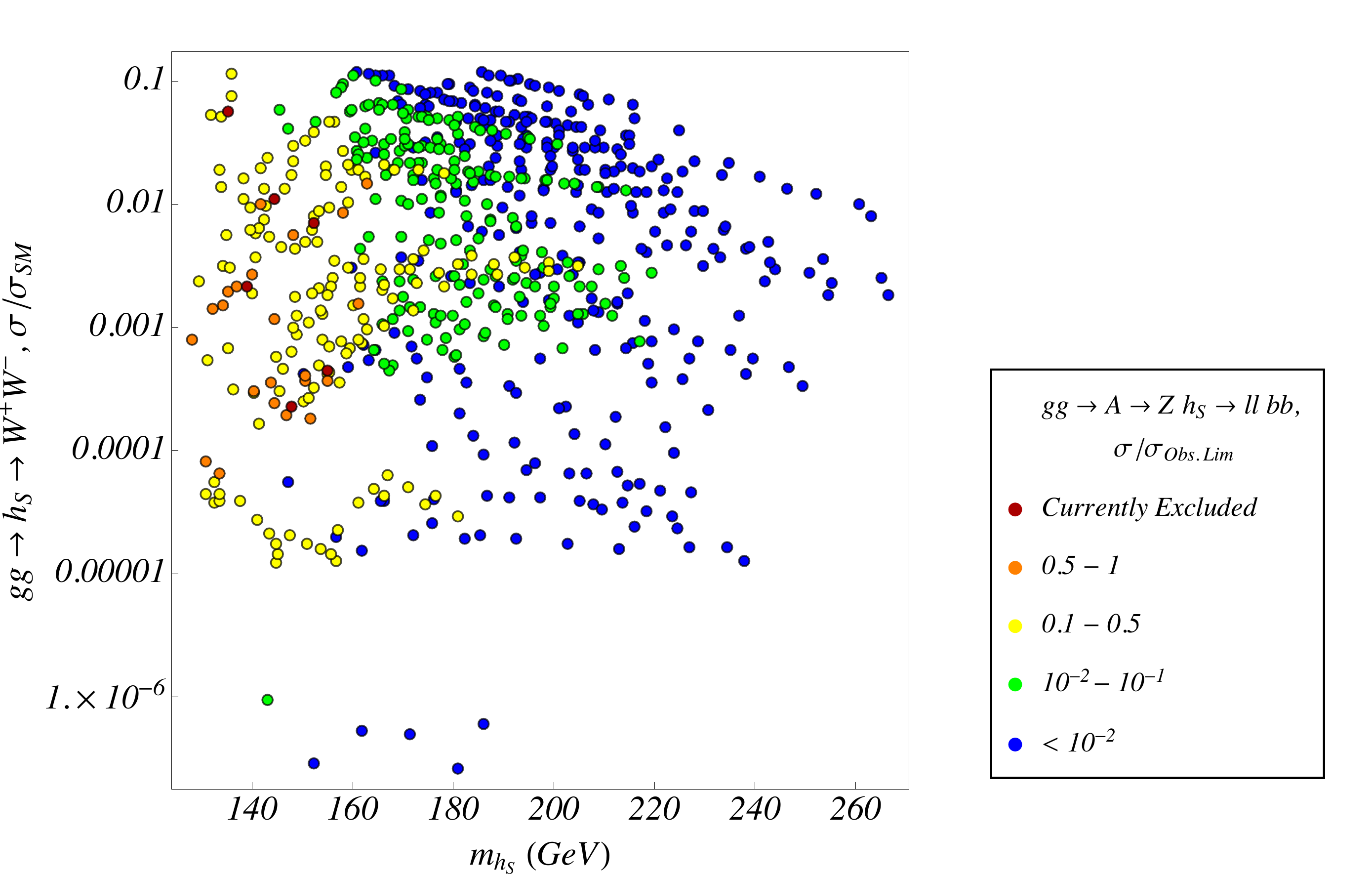}}
\caption{\label{fig:sigmaAtozh3}\em The production cross section times branching ratio of the decay of the second heaviest CP-even Higgs into pairs of W, showing the ratio of the observed limit for the heaviest CP-odd Higgs boson into a $Z$ and a CP-even Higgs bosons.\\[10pt]}
\end{figure}
 \begin{figure}[h!]
\subfloat[]{\includegraphics[width=4.5in, angle=0]{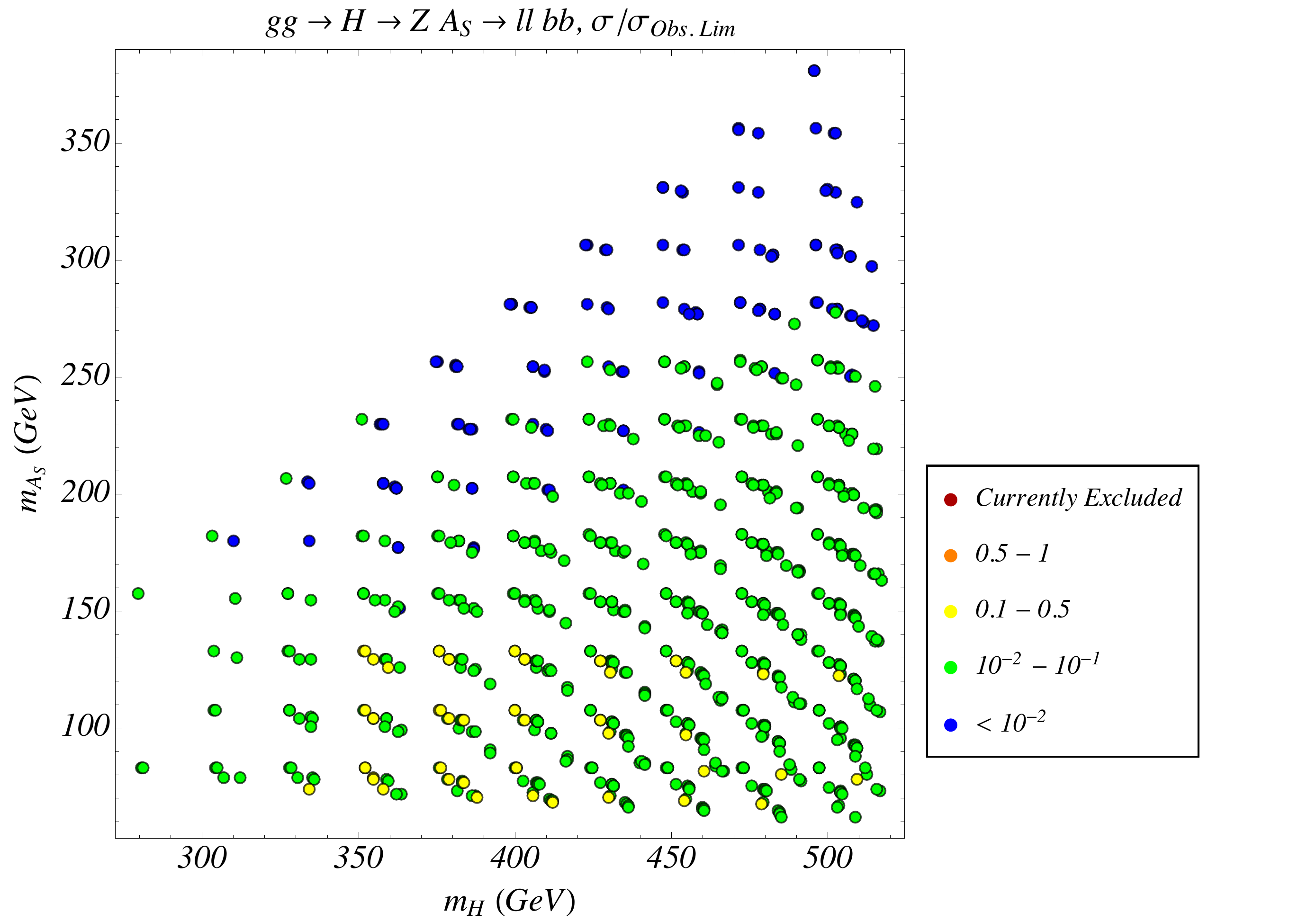}}
\caption{\label{fig:sigmaHtoza}\em Ratio of the observed limit to the production cross section times Branching ratio of the decay of the heaviest CP-even Higgs boson into a $Z$ and the lightest CP-odd 
Higgs bosons.}
\end{figure}

As discussed in Section \ref{alignment}, searches for heavy scalar resonances decaying to $W W^{(*)}$ have been performed at the LHC
and already constrain  the signal strength in the channel $ gg \rightarrow h_S \to WW^{(*)}$ to be less than 10\% of the signal strength from a SM Higgs boson of 
the same mass. 
Since the suppression of the decay branching ratio of $h_S$ into bottom quarks is in part caused by the increase
of the branching ratio into $W$ pairs, it is interesting to investigate the correlation between the search for
heavy CP-odd Higgs bosons decaying into $h_S Z$ in the $(b\bar{b})(\ell\ell)$  channel and the search for the
mainly singlet CP-even Higgs $h_S$ decaying into $WW^{(*)}$. To exhibit the complementarity between the two channels, 
we also show in Fig.~\ref{fig:sigmaAtozh3} the ratios of the event rates for the heavy 
CP-odd scalar decaying to $h_S Z$, with the same colors used in the right panel of Fig.~\ref{fig:sigmaAtozh}.  We observe
that a large fraction of the parameter space that is difficult to probe in the $A \to Z h_S\to (\ell\ell)(b\bar{b})$ channel becomes viable
in the search for $gg\to h_S \to WW^{(*)}$.  There is a small region where searches in both channels become
difficult. This is the region where $h_S$ has a small coupling  to the top quark, thereby suppressing its production
cross section, or where the singlet CP-odd scalar mass $m_{A_S}$ is small and the decay $h_S \to A_S A_S$ may be allowed.  In the latter case, we may use the decay channel $H \to Z A_S$ instead.

\begin{figure}[t!]
\subfloat[]{\includegraphics[width=3.2in, angle=0]{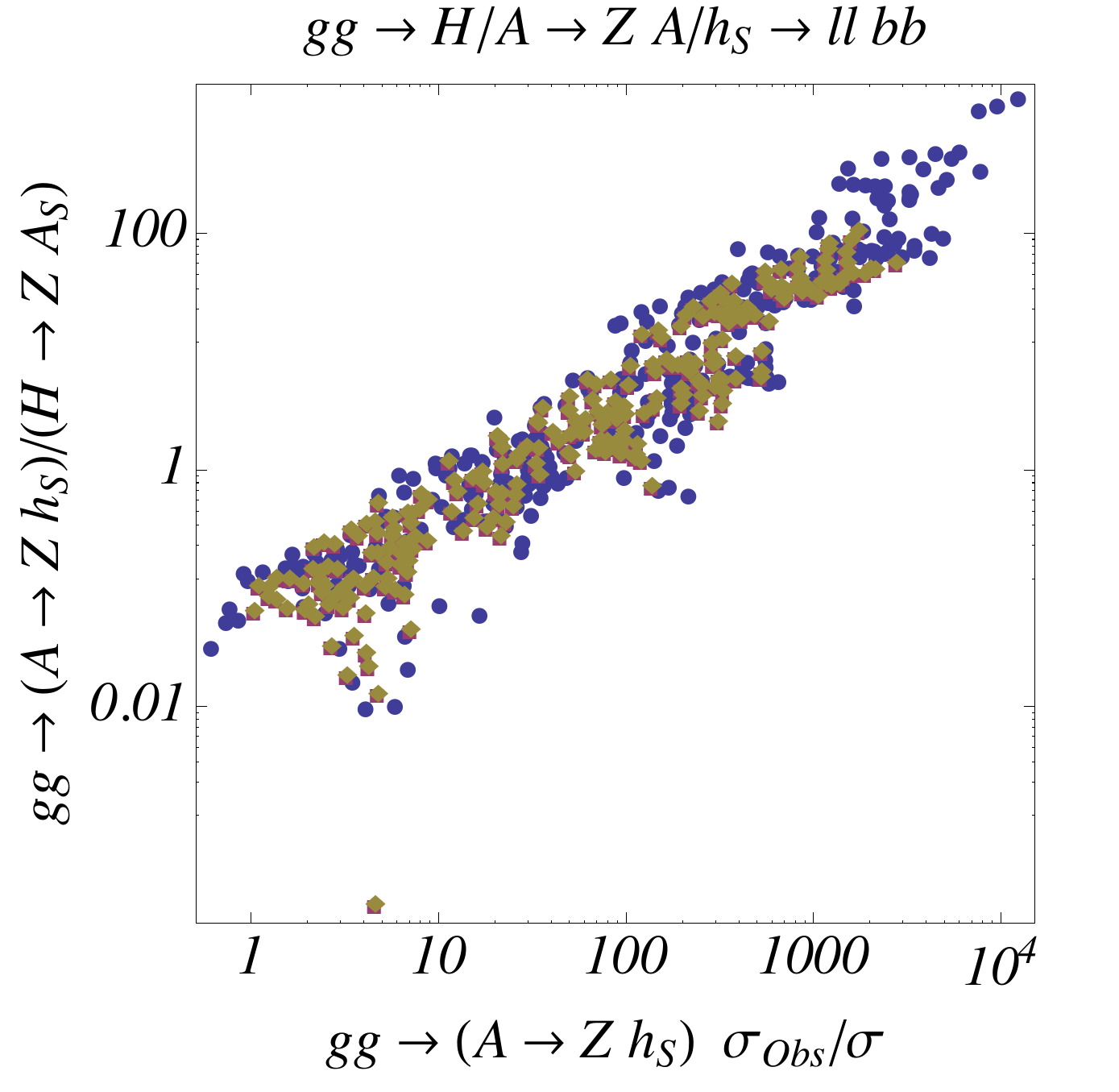}}
\subfloat[]{\includegraphics[width=3.2in, angle=0]{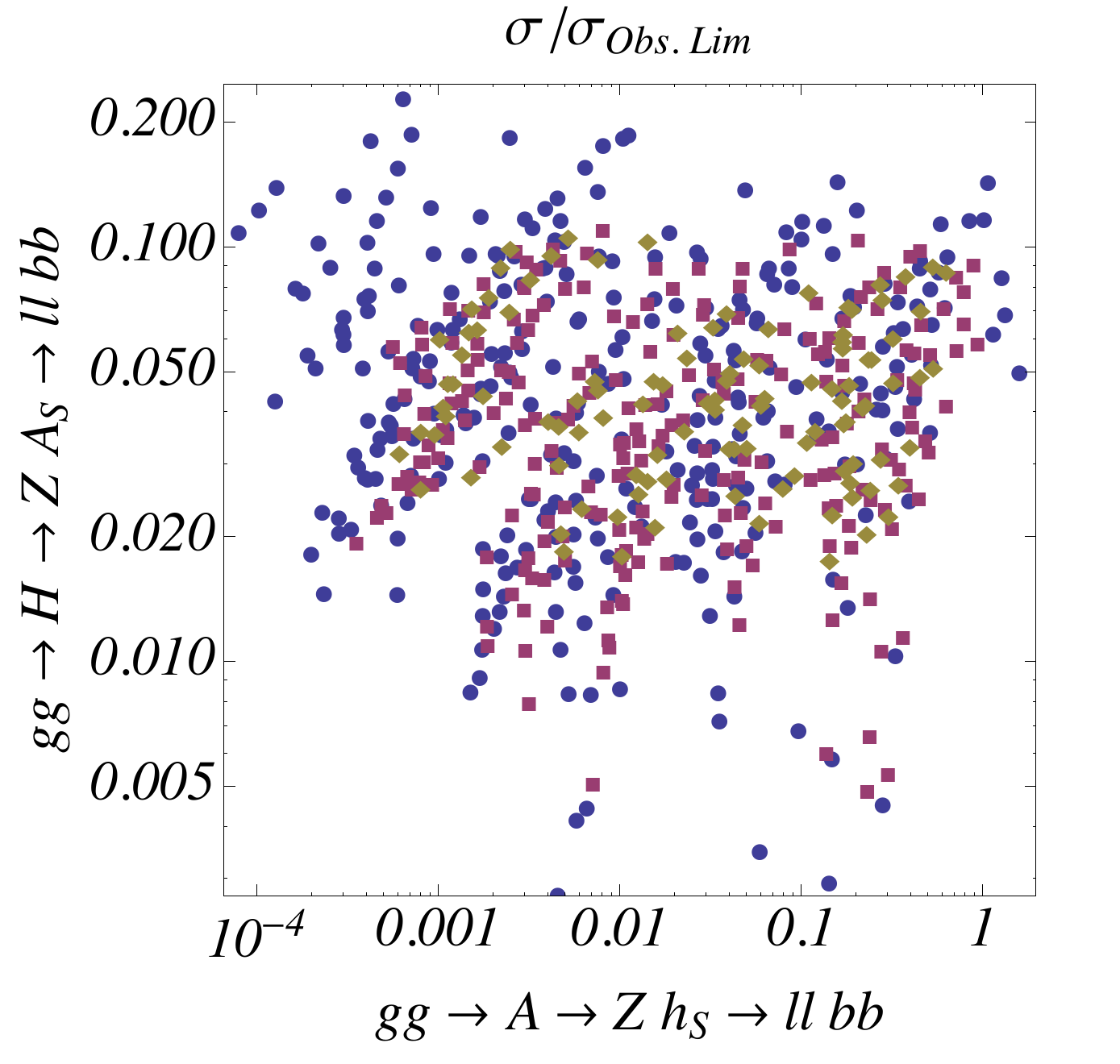}}
\caption{\label{fig:HtoZavsAtozh}{ \em Correlation between the ratio of the observed limit to the production cross sections of $A \to Z h_S$ and $H \to Z A_S$.}}
\end{figure}

In Fig.~\ref{fig:sigmaHtoza} we display the ratio of the observed limit to the production cross section of a heavy CP-even Higgs
boson $H$ decaying into $H \to Z A_S$, with $Z \to \ell\ell$ and $A_S \to b\bar{b}$. Due to the somewhat smaller production of $H$ as compared to $A$, there is no point in the NMSSM Higgs parameter space that can be probed at the 8~TeV run of the LHC in this channel. However, for low values of the $A_S$ mass, the LHC will become increasingly sensitive to searches in this channel. Moreover,
in Fig.~\ref{fig:HtoZavsAtozh} we observe the correlation between this ratio and the same ratio for the $A \to Zh_S \to (\ell\ell)(b\bar{b})$ channel.
The left panel of this figure shows that there is a complementarity in  the LHC sensitivity in these two search channels. 
The right panel shows that an increase of the sensitivity in these two channels by two orders of magnitude would serve to test
the full parameter space. 

\begin{figure}[t!]
\subfloat[]{\includegraphics[width=4.5in, angle=0]{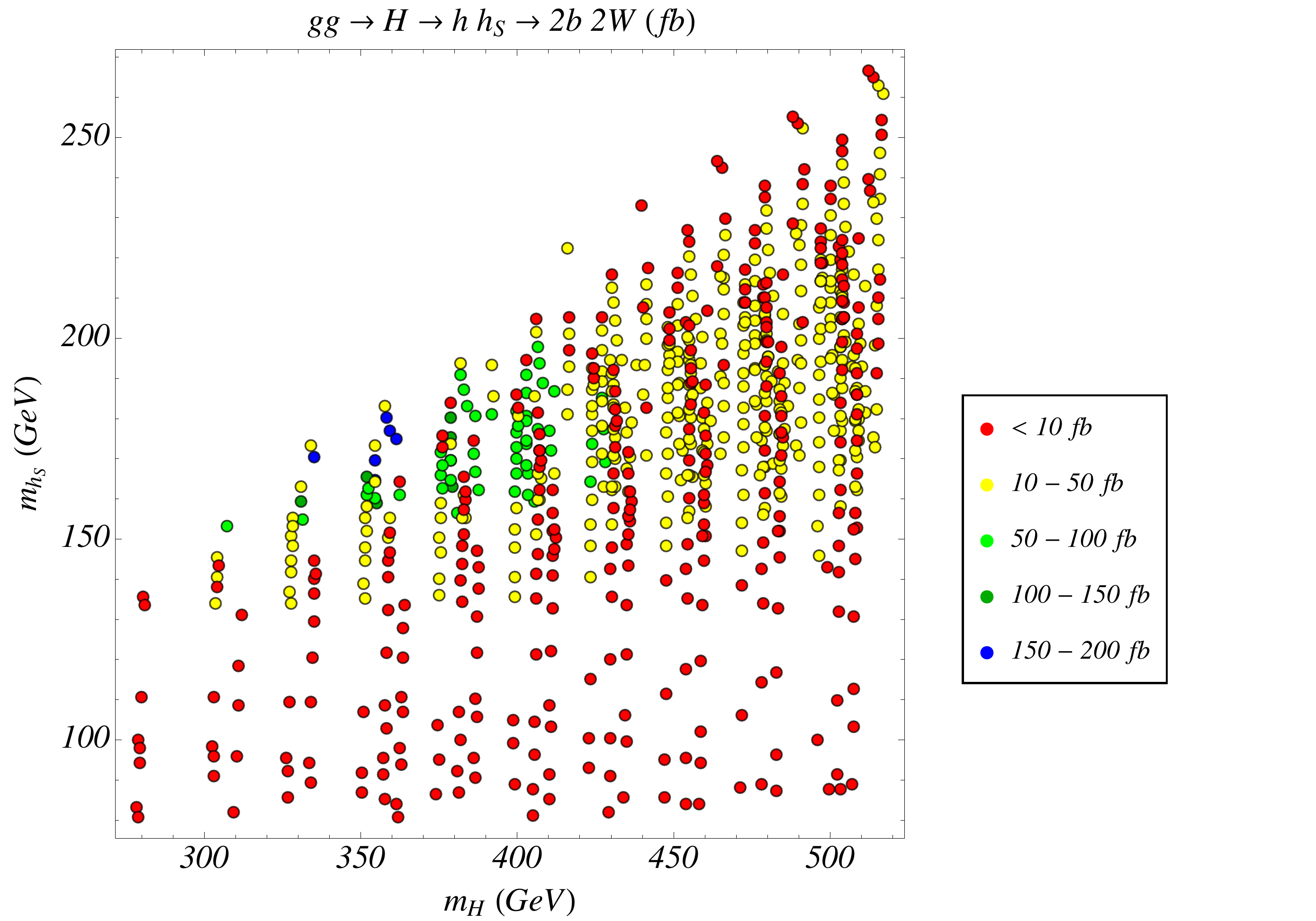}}
\caption{\label{fig:sigmaHtohhs} \em Production cross section times branching ratio of the decay of the heaviest CP-even Higgs boson into
$h$ and $h_S$, with $h$ decaying into $b\bar{b}$ and $h_S$ decaying into $WW$.\\[10pt] }
\end{figure}
\begin{figure}[h!]
\subfloat[]{\includegraphics[width=4.5in, angle=0]{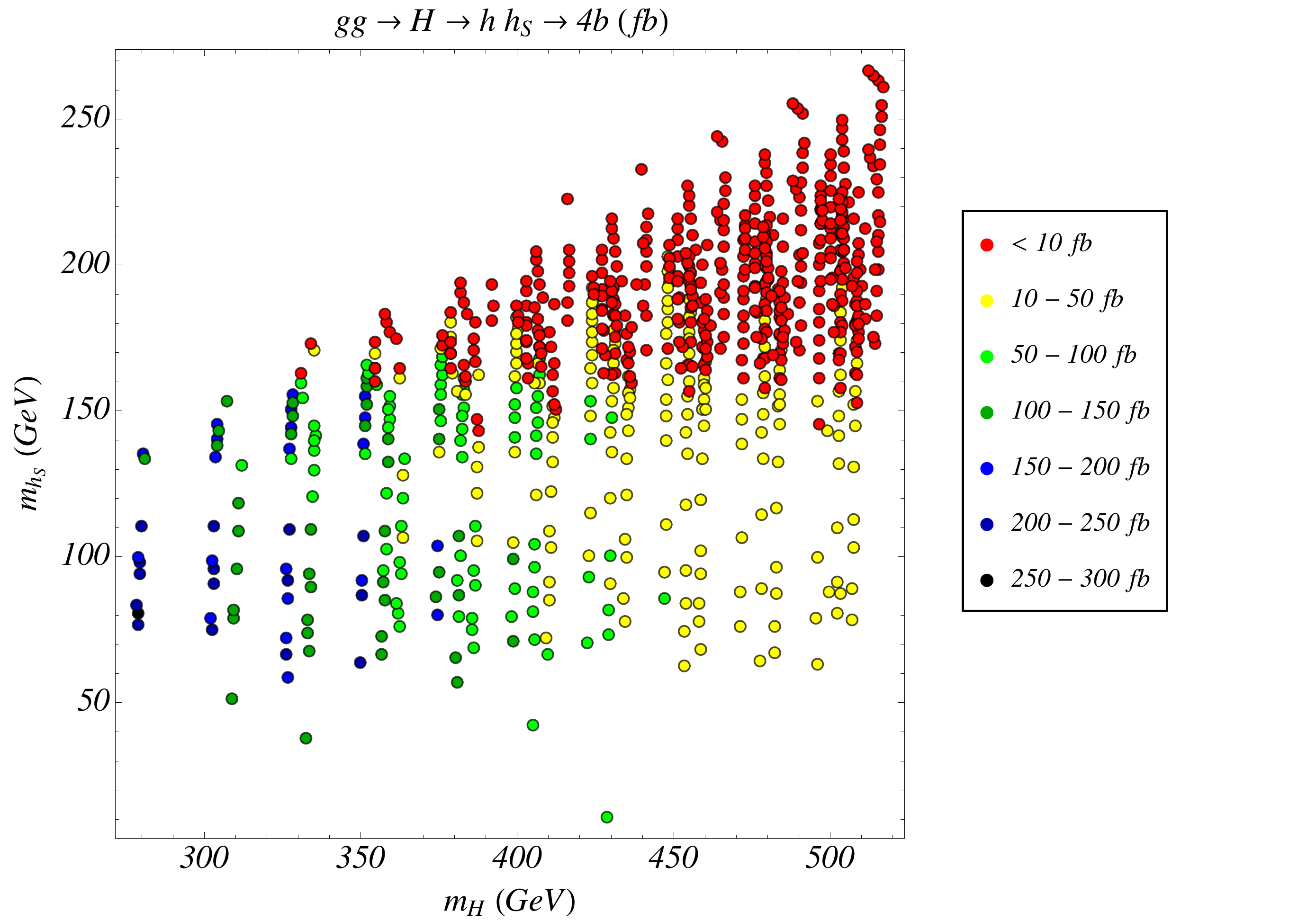} }
\caption{\label{fig:sigmaHtohhs2}\em Production cross section times Branching ratio of the decay of the heaviest CP-even Higgs boson into
$h$ and $h_S$, with both $h$ and $h_S$ decaying into $b\bar{b}$. }
\end{figure}

Finally, we consider the decays of the heavy CP-even Higgs bosons into  two lighter CP-even scalars, which as shown in
Figs.~\ref{fig:toh1h1h2h2} and~\ref{fig:toh12h2a1z} become prominent in a large region of parameter space.  Due to the large
size of the branching ratio, it is instructive to focus on the decays of the heavy
Higgs bosons into $h h_S$.  This is shown in Figs.~\ref{fig:sigmaHtohhs}  and \ref{fig:sigmaHtohhs2}, where we 
display the 8 TeV LHC cross section of these channels assuming that the SM-like Higgs decays into a pair of bottom quarks
and $h_S$ decays into $WW^{(*)}$ and bottom pairs, respectively. We see that the cross sections are sizable, of orders of tens
or hundreds of fb, and there is a large complementarity between the $bbWW$ and 4$b$  search channels, 
associated with the significant size of the corresponding $h_S$ decay branching ratios.

Most aspects of the NMSSM Higgs phenomenology outlined above can be illustrated by choosing specific benchmarks points in the NMSSM Higgs parameter space.
In Appendix~\ref{Benchmarks}, we  present three particular NMSSM benchmarks that illustrate the most important features of the Higgs phenomenology considered in this Section.

\section{Conclusions}
\label{Conclusions}

In this paper, we have studied the conditions for the presence of a SM-like Higgs boson in models containing two Higgs doublets and an additional complex singlet scalar.  In this so-called alignment limit, one of the neutral Higgs fields approximately points in the same direction in field space as the doublet scalar vacuum expectation value.
The main focus of this work is the
$\boldsymbol{\mathbb{Z}_3}$ invariant NMSSM, which provides a predictive framework in which
the interactions of scalars and fermions are well defined. Moreover, in this model the SM-like Higgs mass receives additional 
tree-level contributions with respect to the MSSM and the Higgsino mass parameter $\mu$ arises from the
vacuum expectation value of the singlet field. 

The condition of alignment is naturally obtained for the same values of the singlet-doublet coupling, $\lambda \simeq 0.65$, that leads
to a relevant contribution to the SM-like Higgs mass at low values of $\tanb$, while maintaining the perturbative consistency
of the theory up to the Planck scale.  Consequently, the stops can be light, inducing only a moderate contribution to the SM-like Higgs mass via radiative loop corrections. 

Moreover, the condition of perturbative consistency of the theory up to the Planck scale implies small values of the singlet self-coupling $\kappa$. 
The mixing of the SM-like Higgs boson with the singlet is reduced and  alignment is obtained for values of the mass parameter $M_A $ not far from $2|\mu|/s_{2\beta}$. For these values of $M_A$, $\kappa$  and $\mu$, the constraints coming from current Higgs boson measurements are satisfied, and the  spectrum of
the theory in the Higgs sector may be obtained as a function of $A_\kappa$, which 
controls the masses of the CP-even and CP-odd singlet components. 

We have shown that for values of $M_A \alt  500$ GeV,  the entire Higgs and Higgsino spectra is accessible at the LHC.
Two of the most important probes of this scenario are the searches for heavy scalar resonances, decaying into lighter scalar resonances and a $Z$,
as well as the searches for resonances in the $WW$ and $ZZ$ channels. Moreover, the search for scalar resonances decaying into two
lighter scalars is also important (with the exception of the decay into $hh$ which tends to be suppressed). Thus it is very important to expand these searches into final states in which at least one of the two light scalars has a mass different from 
$m_h=125$~GeV.  We have presented detailed studies of the Higgs phenomenology and considered three benchmarks that capture the dominant features discussed. A comprehensive study of the discovery prospects of these benchmark points at the upcoming LHC Run 2 will be treated in future work.\\

\acknowledgements

Fermilab is operated by Fermi Research Alliance, LLC under Contract No. DE-AC02-07CH11359 with the U.S. Department of Energy. Work at University of Chicago is supported in part by U.S. Department of Energy grant number DE-FG02-13ER41958.  H.E.H. is supported in part by U.S. Department of Energy grant number DE-FG02-04ER41286. I.L. is supported in part by the U.S. Department of Energy under Contract No. DE-SC0010143.  Work at ANL is supported in part by the U.S. Department of Energy under Contract No. DE-AC02-06CH11357. N.R.S is supported in part by the U.S. Department of Energy grant No. DE-SC0007859, by the Michigan Center for Theoretical Physics and the Wayne State University Start-up package. M.C, H.E.H., N.R.S. and C.W thank the hospitality of the Aspen Center for Physics, which is supported by the National Science Foundation under Grant No. PHYS-1066293. M.C., I.L. and C.W. also thank the hospitality of MIAPP Program ``LHC 14," which was supported by the Munich Institute for Astro- and Particle Physics (MIAPP) of the DFG cluster of excellence ``Origin and Structure of the Universe".\\

\newpage
\appendix
\section{The Higgs scalar potential in the Higgs basis}

\enlargethispage{\baselineskip}
It is convenient to rewrite the NMSSM Higgs potential [cf.~\eqst{soft}{vhiggs}] in terms of the Higgs basis fields $H_1$ and $H_2$ [defined in \eq{cphiggsbasisfields}] and the singlet field 
$S$,\footnote{A linear term in $S$ can always be omitted by a linear shift in the definition of $S$.
We have also omitted a possible term, $Y_5 (S^2+{\rm h.c.})$, in $\mathcal{V}$, as it is absent from the $\mathbb{Z}_3$-invariant NMSSM Higgs potential.}
\bea
\mathcal{V}&=& Y_1 H_1^\dagger H_1+ Y_2 H_2^\dagger H_2 +[Y_3
H_1^\dagger H_2+{\rm h.c.}]+Y_4 S^\dagger S
\nn \\
&&
+\bigl[C_1 H_1^\dagger H_1S+C_2 H_2^\dagger H_2S+C_3 H_1^\dagger H_2S+C_4 H_2^\dagger H_1 S+C_5(S^\dagger S)S+C_6 S^3+{\rm h.c.}\bigr]
\nn\\
&&
+\half Z_1(H_1^\dagger H_1)^2+\half Z_2(H_2^\dagger H_2)^2
+Z_3(H_1^\dagger H_1)(H_2^\dagger H_2)
+Z_4( H_1^\dagger H_2)(H_2^\dagger H_1)\nn\\
&&\qquad\qquad +\left\{\half Z_5 (H_1^\dagger H_2)^2 +\big[Z_6 (H_1^\dagger
H_1) +Z_7 (H_2^\dagger H_2)\big] H_1^\dagger H_2+{\rm
h.c.}\right\} \label{hbasispot}\\
&&
+S^\dagger S\bigl[Z_{s1}H_1^\dagger H_1+Z_{s2}H_2^\dagger H_2+
(Z_{s3}H_1^\dagger H_2+{\rm h.c.})+Z_{s4}S^\dagger S\bigr]
\nn\\
&&
+\left\{Z_{s5}H_1^\dagger H_1 S^2+Z_{s6}H_2^\dagger H_2  S^2+Z_{s7}H_1^\dagger H_2 S^2+Z_{s8}H_2^\dagger H_1 S^2+Z_{s9}S^\dagger S\,S^2+Z_{s10}S^4+{\rm h.c.}\right\}\!.
 \nn
\eea
Assuming a CP-invariant Higgs potential and vacuum,
all scalar potential coefficients can be taken real after an appropriate rephasing of $H_2$. 
At the minimum of the Higgs potential, $\langle H_1^0\rangle=v/\sqrt{2}$ and $\langle S\rangle=v_s$
(with all other vevs equal to zero), and
\bea
Y_1&=&-\half Z_1 v^2-2C_1 v_s-(Z_{s1}+2Z_{s5})v_s^2\,,\label{y1}\\
Y_3&=&-\half Z_6 v^2-(C_3+C_4)v_s-(Z_{s3}+Z_{s7}+Z_{s8})v_s^2\,,\label{y3}\\
Y_4&=&-\half C_1\frac{v^2}{v_s}-3(C_5+C_6) v_s-\half(Z_{s1}+2Z_{s5})v^2-2Z_{s4}v_s^2
-4(Z_{s9}+Z_{s10})v_s^2\label{y4}\,.
\eea

The charged Higgs mass is given by
\be
m_{H^\pm}^2=M_A^2-\half(Z_4-Z_5)v^2\,,
\ee
where the squared-mass parameter $M_A^2$ is defined by:
\be \label{MA}
M_A^2=Y_2+\half(Z_3+Z_4-Z_5)v^2+2C_2 v_s+(Z_{s2}+2Z_{s6})v_s^2\,.
\ee
The CP-even squared-mass matrix is obtained from \eq{hbasispot} by eliminating $Y_1$, $Y_3$ and $Y_4$, 
\be
\hspace{-0.05in}
\mathcal{M}^2_S=\begin{pmatrix} Z_1 v^2 & \quad Z_6 v^2 & \quad \sqrt{2}\,v\bigl[C_1+(Z_{s1}+2Z_{s5})v_s\bigr] \\[10pt]
& \quad M_A^2+Z_5 v^2 & \quad \displaystyle\frac{v}{\sqrt{2}}\biggl[C_3+C_4+2(Z_{s3}+Z_{s7}+Z_{s8})v_s\biggr] \\[10pt]
&  & \quad 
-C_1\displaystyle\frac{v^2}{2v_s}+3(C_5+C_6) v_s+4 (Z_{s4}+2Z_{s9}+2Z_{s10}) v_s^2\end{pmatrix},
\ee
where the omitted elements below the diagonal are fixed since $\mathcal{M}_S^2$ is a symmetric matrix.
\clearpage
\enlargethispage{20pt}
\noindent
Likewise, we can compute the CP-odd squared-mass matrix,
\be
\mathcal{M}^2_P=\begin{pmatrix} M_A^2 & \quad -\displaystyle\frac{v}{\sqrt{2}}\bigl[C_3-C_4+2(Z_{s7}-Z_{s8})v_s\bigr]
\\[10pt]
& \qquad\quad
-C_1\displaystyle\frac{v^2}{2v_s}-(C_5+9 C_6) v_s-2Z_{s5}v^2-4(Z_{s9}+4Z_{s10})v_s^2\end{pmatrix},
\ee
where the omitted matrix element is fixed since $\mathcal{M}_P^2$ is a symmetric matrix.
Comparing the scalar potential $\mathcal{V}$ with \eqst{soft}{vhiggs}, we obtain 
the coefficients of the quadratic terms,
\bea
Y_1&=& m^2_{H_d}c_\beta^2+m^2_{H_u}s_\beta^2\,,\label{why1}\\
Y_2&=& m^2_{H_d}s_\beta^2+m^2_{H_u}c_\beta^2\,,\label{why2}\\
Y_3&=& \half(m^2_{H_u}-m^2_{H_d})s_{2\beta}\,,\label{why3}\\
Y_4&=&m_S^2\,,\label{why4}
\eea
the coefficients of the cubic terms [after employing \eqs{mub}{masq}],
\bea
C_1&=& -C_2=\lambda c_{\beta } s_{\beta }  \left(\frac{\kappa  \mu }{\lambda }-\frac{M_A^2
   }{\mu } c_{\beta } s_{\beta }\right) \,, \label{seeone}\\
C_3&=& \lambda  c_{\beta }^2 \left(\frac{\kappa  \mu }{\lambda }-\frac{M_A^2
   }{\mu } c_{\beta } s_{\beta }\right) \,,\\
C_4&=& -\lambda  s_{\beta }^2  \left(\frac{\kappa  \mu }{\lambda }-\frac{M_A^2
   }{\mu } c_{\beta } s_{\beta }\right) \,,\\
C_5&=&0\,,\\
C_6&=&\tfrac{1}{3}\kappa A_\kappa\,,
\eea
and the coefficients of the quartic terms,
\bea
Z_1&=&Z_2=-\half\bigl[\lambda^2-\tfrac{1}{2}(g^2+g^{\prime\,2})\bigr]c^2_{2\beta}+\tfrac{1}{2}\lambda^2\,,\\
Z_3&=&-\half\bigl[\lambda^2-\tfrac{1}{2}(g^2+g^{\prime\,2})\bigr]s^2_{2\beta}+\tfrac{1}{4}(g^2-g^{\prime\,2})\,,\\
Z_4&=&-\half\bigl[\lambda^2-\tfrac{1}{2}(g^2+g^{\prime\,2})\bigr]s^2_{2\beta}-\half g^2+\lambda^2\,,\\
Z_5&=&-\half\bigl[\lambda^2-\tfrac{1}{2}(g^2+g^{\prime\,2})\bigr]s^2_{2\beta}\,,\\
Z_6&=&-Z_7=\half\bigl[\lambda^2-\tfrac{1}{2}(g^2+g^{\prime\,2})\bigr]s_{2\beta}c_{2\beta}\,,\\
Z_{s1}&=&Z_{s2}=\lambda^2\,,\\
Z_{s4}&=&\kappa^2\,,\\
Z_{s5}&=&-Z_{s6}=-\half\kappa\lambda s_{2\beta}\,,\\
Z_{s7}&=&\kappa\lambda s^2_{\beta}\,,\\
Z_{s8}&=&-\kappa\lambda c^2_{\beta}\,,\\
Z_{s3}&=&Z_{s9}=Z_{s10}=0\,.\label{zsten}
\eea

Note that whereas $Y_1$, $Y_3$ and $Y_4$ are determined from the Higgs potential 
minimum conditions [\eqst{y1}{y4}], $Y_2$ in generic two-doublet/one-singlet models is a free
parameter.  However, in the $\mathbb{Z}_3$-symmetric NMSSM Higgs sector, there is no bare
$H_u\cdot H_d$ term in \eq{vhiggs}.  Consequently, $Y_2$ is no longer an independent parameter.  Indeed, \eqs{why1}{why3} yield
\be
Y_2=Y_1+\frac{2c_{2\beta}}{s_{2\beta}}\,Y_3\,.
\ee
Inserting the results of \eqs{y1}{y3} then yields
\be
Y_2=-\half Z_1 v^2-2C_1 v_s-(Z_{s1}+2Z_{s5})v_s^2-\frac{2c_{2\beta}}{s_{2\beta}}\biggl[
\half Z_6 v^2+(C_3+C_4)v_s+(Z_{s3}+Z_{s7}+Z_{s8})v_s^2\biggr]\,.
\ee
Inserting this result into \eq{MA} ,
\bea
M_A^2&=&-\half\left(Z_1-Z_3-Z_4+Z_5+\frac{2c_{2\beta}}{s_{2\beta}}Z_6\right)v^2+2\left[C_2-C_1-
\frac{c_{2\beta}}{s_{2\beta}}(C_3+C_4)\right]v_s \nonumber \\
&&\qquad -\left[Z_{s1}-Z_{s2}+2\biggl(Z_{s5}-Z_{s6}+\frac{c_{2\beta}}{s_{2\beta}}(Z_{s3}+Z_{s7}+Z_{s8})\biggr)\right]v_s^2\,.\label{MAtwo}
\eea
Using the results of this Appendix, one can check that \eq{MAtwo} then reduces to the simple expression given in \eq{masq}.

All the results above correspond to tree-level results.  Including the leading $\mathcal{O}(h_t^4)$ loop corrections, the $Z_i$ are modified as follows:
\bea
\!\!\!\!\!\!\!\!\!\!\!
Z_1 v^2 &=& (m_Z^2-\half\lambda^2 v^2) c^2_{2\beta}+\half\lambda^2 v^2+\frac{3v^2 s_\beta^4 h_t^4}{8\pi^2}\left[\ln\left(\frac{M_S^2}{m_t^2}\right)+\frac{X_t^2}{M_S^2}\left(1-\frac{X_t^2}{12M_S^2}\right)\right],\label{mhmax}\\[8pt]
\!\!\!\!\!\!\!\!\!\!\!
Z_2 v^2 &=&  (m_Z^2-\half\lambda^2 v^2) c^2_{2\beta}+\half\lambda^2 v^2+\frac{3v^2 c_\beta^4 h_t^4}{8\pi^2}\left[\ln\left(\frac{M_S^2}{m_t^2}\right)+\frac{Y_t^2}{M_S^2}\left(1-\frac{Y_t^2}{12M_S^2}\right)\right],\label{zeetwocorr}\\[8pt]
\!\!\!\!\!\!\!\!\!\!\!
Z_3 v^2 &=&\tfrac{1}{4}(g^2-g^{\prime\,2})v^2+ s_{2\beta}^2\biggl\{m_Z^2-\half\lambda^2 v^2+\frac{3v^2 h_t^4}{32\pi^2}\left[\ln\left(\frac{M_S^2}{m_t^2}\right)+\frac{(X_t+Y_t)^2}{4M_S^2}-\frac{X_t^2 Y_t^2}{12M_S^4}\right]\biggr\}, \nonumber\\[8pt]
&&\label{zeethreecorr}\\
Z_4 v^2 &=&(\lambda^2-\half g^2)v^2+ s_{2\beta}^2\left\{m_Z^2-\half\lambda^2 v^2+\frac{3v^2 h_t^4}{32\pi^2}\left[\ln\left(\frac{M_S^2}{m_t^2}\right)+\frac{(X_t+Y_t)^2}{4M_S^2}-\frac{X_t^2 Y_t^2}{12M_S^4}\right]\right\}, \nonumber\\[8pt]
&& \label{zeefourcorr} \\
Z_5 v^2  & = & s_{2\beta}^2\left\{m_Z^2-\half\lambda^2 v^2+\frac{3v^2 h_t^4}{32\pi^2}\left[\ln\left(\frac{M_S^2}{m_t^2}\right)+\frac{X_t Y_t}{M_S^2}\left(1-\frac{X_t Y_t}{12M_S^2}\right)\right]\right\},
\label{mHcorr2}\\[8pt]
\!\!\!\!\!\!\!\!\!\!\!
Z_6 v^2&=& -s_{2\beta}\left\{(m_Z^2-\half\lambda^2v^2) c_{2\beta}-\frac{3v^2 s_\beta^2  h_t^4}{16\pi^2}\biggl[\ln\left(\frac{M_S^2}{m_t^2}\right)+\frac{X_t(X_t+Y_t)}{2M_S^2}-\frac{X_t^3 Y_t}{12 M_S^4}\biggr]\right\},
\label{zeesixcorr}\\[8pt] 
\!\!\!\!\!\!\!\!\!\!\!
Z_7 v^2&=& s_{2\beta}\left\{(m_Z^2-\half\lambda^2 v^2) c_{2\beta}+\frac{3v^2 c_\beta^2  h_t^4}{16\pi^2}\biggl[\ln\left(\frac{M_S^2}{m_t^2}\right)+\frac{Y_t(X_t+Y_t)}{2M_S^2}-\frac{X_t Y_t^3}{12 M_S^4}\biggr]\right\}.
\label{zeesevencorr} 
\eea
\section{Components of the Mass Eigenstates}
\label{app:eigenstates}

We present here generic expressions for the components of the CP-even Higgs mass eigenstates in terms of the mass eigenvalues
and the elements of the CP-even Higgs squared-mass matrix.  The interaction eigenstates and the mass eigenstates are related by
\be 
\left(\begin{array}{c}
h \\
H \\
h_S
\end{array}\right) = 
\left(\begin{array}{ccc}
\kappa^h_{\rm SM} & \kappa^h_{\rm NSM} & \kappa^h_{\rm S} \\
\kappa^H_{\rm SM} & \kappa^H_{\rm NSM} & \kappa^H_{\rm S} \\
\kappa^{h_S}_{\rm SM} & \kappa^{h_S}_{\rm NSM} & \kappa^{h_S}_{\rm S}
\end{array}\right)  
\left(\begin{array}{c}
H^{\rm SM} \\
H^{\rm NSM}\\
H^{\rm S}
\end{array}\right) \ .
\label{eq:mixingmatrix}
\ee

For the Higgs mass eigenstate $h$, 
\bea
\frac{\kappa_{\rm NSM}^h}{\kappa_{\rm SM}^h} & = & -\frac{\mathcal{M}^2_{12}(m_h^2 -\mathcal{M}^2_{33})
+ \mathcal{M}^2_{13} \mathcal{M}^2_{23}}{\mathcal{M}_{23}^4 + (\mathcal{M}^2_{22} - m_h^2)(m_h^2 - \mathcal{M}^2_{33})} \,,
\label{h1}\\
\frac{\kappa_{\rm S}^h}{\kappa_{\rm SM}^h} & = & \frac{\mathcal{M}^2_{13}(\mathcal{M}^2_{22} - m_h^2) 
- \mathcal{M}^2_{12} \mathcal{M}^2_{23}}{\mathcal{M}^4_{23} + (\mathcal{M}^2_{22} - m_h^2)(m_h^2 - \mathcal{M}^2_{33})}\,. 
\label{h2}
\eea

For the Higgs mass eigenstate $H$, 
\bea
\frac{\kappa_{SM}^H}{\kappa_{\rm NSM}^H} & = & - 
 \frac{\mathcal{M}^2_{12} (m_H^2 - \mathcal{M}^2_{33}) +
\mathcal{M}^2_{13} \mathcal{M}^2_{23}}{M_{13}^4 + (\mathcal{M}^2_{11} - m_H^2)(m_H^2 - \mathcal{M}^2_{33})} \,,
\label{H1}\\
\frac{\kappa_{S}^H}{\kappa_{\rm NSM}^H} & = & 
\frac{\mathcal{M}^2_{23} (\mathcal{M}^2_{11} - m_H^2) -
\mathcal{M}^2_{12} \mathcal{M}^2_{13}}{\mathcal{M}^4_{13} + (\mathcal{M}^2_{11} - m_H^2)(m_H^2 - \mathcal{M}^2_{33})} \,.
\label{H2}
\eea

For the Higgs mass eigenstate $h_S$, 
\bea
\frac{\kappa_{\rm SM}^{h_S}}{\kappa_{\rm S}^{h_S}} & = & - \frac{\mathcal{M}^2_{13}( m_{h_S}^2 -\mathcal{M}^2_{22} )
+ \mathcal{M}^2_{12} \mathcal{M}^2_{23}}{M_{12}^4 + (\mathcal{M}^2_{22} - m_{h_S}^2)(m_{h_S}^2 - \mathcal{M}^2_{11})} \,,
\label{hs1}\\
\frac{\kappa_{\rm NSM}^{h_S}}{\kappa_{\rm S}^{h_S}} & = & \frac{\mathcal{M}^2_{23} ( \mathcal{M}^2_{11} - m_{h_S}^2 ) 
- \mathcal{M}^2_{12} \mathcal{M}^2_{13}}{M_{12}^4 + (\mathcal{M}^2_{22} - m_{h_S}^2)(m_{h_S}^2 - \mathcal{M}^2_{11})}\,.
\label{hs2}
\eea 

\bigskip\bigskip

\section{Analytic Expressions for the Higgs Couplings}
\label{couplings}
\subsection{Trilinear Higgs Couplings}

In Table~\ref{processes} we present the tree-level Higgs trilinear couplings in terms of Higgs-basis scalar fields.  The second column of Table~\ref{processes} displays the corresponding coefficients derived 
from the Higgs potential, Eq.~(\ref{hbasispot}), and the third column of Table~\ref{processes} 
evaluates these coefficients in the $\mathbb{Z}_3$-invariant NMSSM using the results of \eqst{seeone}{zsten}.
The corresponding Feynman rules are obtained by multiplication by a symmetry factor $-i\, n!$, where $n$ is the number of identical bosons 
that are associated with the trilinear coupling. 
From Table~\ref{processes} we see that the coefficient of the Higgs trilinear couplings $H^{\rm NSM} H^{\rm SM} H^{\rm SM}$ and $H^{\rm S} H^{\rm SM} H^{\rm SM}$ are proportional
to $\mathcal{M}^2_{12}$ and $\mathcal{M}^2_{13}$, Eq.~(\ref{eq:CPMM}), respectively,  and approach zero in the alignment limit.  
\begin{table}[t!]
\centering    
\begin{tabular}{|c|c|c|c|}
\hline
& coefficient in $\mathcal{V}$ [\eq{hbasispot}]&    $\mathbb{Z}_3$-invariant NMSSM \\
\hline
$H^{\rm SM} H^{\rm SM} H^{\rm SM}$  & $\half vZ_1$ &  $\tfrac{1}{8}v\bigl[2\lambda^2 s^2_{2\beta}+(g^2+g^{\prime\,2})c^2_{2\beta}\bigr]$\\
\hline
$H^{\rm SM} H^{\rm SM} H^{\rm NSM}$  &  $\tfrac{3}{2}v Z_6$ &  $\tfrac{3}{4}v(\lambda^2-g^2-g^{\prime\,2})s_{2\beta}c_{2\beta}$  \\
\hline
$H^{\rm SM} H^{\rm NSM} H^{\rm NSM}$  &  $\half v(Z_3+Z_4+Z_5)$ & $\tfrac{1}{8}\bigl[2\lambda^2+(2\lambda^2-g^2-g^{\prime\,2})(1-3s^2_{2\beta})\bigr]$\\
\hline
$H^{\rm NSM} H^{\rm NSM} H^{\rm NSM}$ &  $\half vZ_7$ & $-\tfrac{1}{4}v(\lambda^2-g^2-g^{\prime\,2})s_{2\beta}c_{2\beta}$   \\
\hline
$H^{\rm SM} H^{\rm SM} H^{\rm S}$  & $\bigl[C_1+v_s(Z_{s1}+2Z_{s5})\bigr]/\sqrt{2}$  &  $\frac{\lambda \mu}{\sqrt{2}}\left[ 1- \half s_{2 \beta}\left(\frac{\kappa}{\lambda}+\frac{M_A^2}{2 \mu^2} s_{2\beta}\right)\right]$\\
\hline
$H^{\rm SM} H^{\rm NSM} H^{\rm S}$  &  $\bigl[C_3+C_4+2v_s(Z_{s3}+Z_{s7}+Z_{s8})\bigr]/\sqrt{2}$ & $-\frac{\lambda \mu c_{2\beta}}{\sqrt{2}}\left(\frac{\kappa}{\lambda}+\frac{M_A^2}{2\mu^2}s_{2\beta}\right) $ \\
\hline
$H^{\rm NSM} H^{\rm NSM} H^{\rm S}$  & $\bigl[C_2+v_s(Z_{s2}+2Z_{s6})\bigr]/\sqrt{2}$  &
 $\frac{\lambda \mu}{\sqrt{2}}\left[ 1+\half s_{2\beta}\left(\frac{\kappa}{\lambda}+\frac{M_A^2}{2\mu^2}s_{2\beta}\right)\right]$ \\
\hline
$H^{\rm SM} H^{\rm S} H^{\rm S}$  &  $\half v(Z_{s1}+2Z_{s5})$ & $\half v\lambda(\lambda-\kappa s_{2\beta})$ \\
\hline
$H^{\rm NSM} H^{\rm S} H^{\rm S}$  &  $\half v(Z_{s3}+Z_{s7}+Z_{s8})$ & $-\half v\kappa\lambda c_{2\beta}$\\
\hline
$H^{\rm S} H^{\rm S} H^{\rm S}$  &  $\bigl[C_5+C_6+2v_s (Z_{s4}+2Z_{s9}+2Z_{s10})\bigr]/\sqrt{2}$ &  $\tfrac{\kappa}{3 \sqrt{2}}(A_\kappa+6\frac{\kappa \mu}{\lambda})$\\
\hline
$H^{\rm SM} A^{\rm NSM} A^{\rm NSM}$  & $\half v(Z_3+Z_4-Z_5)$  &  $\tfrac{1}{4}v\bigl[\lambda^2+\bigl(\lambda^2-\half(g^2+g^{\prime\,2})\bigr)c_{2\beta}^2\bigr]$ \\
\hline
$H^{\rm NSM} A^{\rm NSM} A^{\rm NSM}$  &  $\half vZ_7$ & $-\tfrac{1}{4}v\bigl[\lambda^2-\half(g^2+g^{\prime\,2})\bigr]s_{2\beta}c_{2\beta}$ \\
\hline
$H^{\rm S} A^{\rm NSM} A^{\rm NSM}$  &  $\bigl[C_2+v_s(Z_{s2}+2Z_{s6})\bigr]/\sqrt{2}$ &
$\frac{\lambda \mu}{\sqrt{2}}\left[ 1+ \half s_{2 \beta} \left(\frac{\kappa}{\lambda}+\frac{M_A^2}{2 \mu^2} s_{2 \beta}\right)\right]$  \\
\hline
$H^{\rm SM} A^{\rm NSM} A^{\rm S}$  & $\bigl[C_4-C_3-2v_s(Z_{s7}-Z_{s8})\bigr]$   & 
$\frac{\lambda\mu}{ \sqrt{2}}\left(\frac{M_A^2}{2 \mu^2} s_{2 \beta}-3\frac{\kappa}{\lambda}\right)$\\
\hline
$H^{\rm NSM} A^{\rm NSM} A^{\rm S}$  &  0 & 0 \\
\hline
$H^{\rm S} A^{\rm NSM} A^{\rm S}$  &  $v(Z_{s8}-Z_{s7})$ & $-\kappa\lambda v$ \\
\hline
$H^{\rm SM} A^{\rm S} A^{\rm S}$  &  $\half v(Z_{s1}-2Z_{s5})$ &  $\half v\lambda(\lambda+\kappa s_{2\beta})$\\
\hline
$H^{\rm NSM} A^{\rm S} A^{\rm S}$  & $\half v(Z_{s3}-Z_{s7}-Z_{s8})$  &$\half v\kappa\lambda c_{2\beta}$  \\
\hline
$H^{\rm S} A^{\rm S} A^{\rm S}$  &  $\bigl[C_5-3C_6+2v_s(Z_{s4}+6Z_{s10})\bigr]/\sqrt{2}$ & $-\kappa(A_\kappa-2\frac{\kappa\mu}{\lambda})/\sqrt{2}$\\
\hline
$H^{\rm SM}H^+H^-$ & $vZ_3$ & $-\half\bigl[\lambda^2-\half(g^2+g^{\prime\,2})\bigr]s^2_{2\beta}+\tfrac{1}{4}(g^2-g^{\prime\,2})$\\
\hline
$H^{\rm NSM}H^+H^-$ & $vZ_7$ & $-\half\bigl[\lambda^2-\half(g^2+g^{\prime\,2})\bigr]s_{2\beta}c_{2\beta}$\\
\hline
$H^{\rm S}H^+H^-$ & $\sqrt{2}\bigl[C_2+v_s(Z_{s2}+2Z_{s6})\bigr]$ & $\sqrt{2}\,\lambda\mu\left[ 1+\half  s_{2\beta}\left(\frac{\kappa}{\lambda}+\frac{M_A^2}{2\mu^2}s_{2\beta}\right)\right]$ \\
\hline
\end{tabular}
\caption{\label{processes} \em Tree-level trilinear scalar interactions.}
\end{table}

We can also include the effects of the dominant contributions to the
one-loop radiative corrections to the trilinear scalar interactions by
employing the leading $\mathcal{O}(h_t^4)$ corrections given in
\eqst{mhmax}{zeesevencorr}.   These corrections modify the trilinear
Higgs couplings shown in Table~\ref{processesrad}.   The results
presented in Table~\ref{processesrad} have been obtained as follows.
First, we work in the approximation that $m_h^2\simeq Z_1 v^2$.  We
then use \eq{mhmax} to solve for $\ln(M_S^2/m_t^2)$ in terms of
$m_h^2$, $m_Z^2$, $\lambda$, and $X_t$.  The resulting expression is
then used in \eqst{zeetwocorr}{zeesevencorr} to eliminate the
logarithmic terms.  Using the resulting expressions for the $Z_i$ to
evaluate the trilinear couplings in Table~\ref{processes}, we obtain
the results shown in Table~\ref{processesrad} after dropping the
additional corrections proportional to $(X_t-Y_t)/M_S$.\footnote{Although it is straightforward to keep track of
  the terms proportional to $(X_t-Y_t)/M_S$, in practice these
  terms provide only a small correction to the results shown in
  Table~\ref{processesrad}.}  Note in particular that $m_h$, which
appears in the trilinear Higgs couplings shown in
Table~\ref{processesrad}, is the radiatively-corrected Higgs mass in
the NMSSM, which we set equal to 125 GeV.   That is, the leading
radiative corrections to the Higgs trilinear couplings have been
absorbed in the definition of $m_h$.

\begin{table}[t!]
\centering
\begin{tabular}{|c|c|c|}
\hline
&  $2v\,\times$ trilinear Higgs coupling of the $\mathbb{Z}_3$-invariant NMSSM\\
\hline
$H^{\rm SM} H^{\rm SM} H^{\rm SM}$  & $m_h^2$\\
\hline
$H^{\rm SM} H^{\rm SM} H^{\rm NSM}$ & $3s^{-1}_\beta
\left(m_h^2 c_\beta-m_Z^2c_{2\beta}c_\beta-\half\lambda^2 v^2s_{2\beta}s_\beta\right)$  \\
\hline
$H^{\rm SM} H^{\rm NSM} H^{\rm NSM}$ & $3s^{-2}_\beta\left[m_h^2 c^2_\beta-m_Z^2(c^2_{2\beta}-\tfrac{2}{3} s^2_\beta)-\lambda^2 v^2 s^2_\beta(c_{2\beta}+\tfrac{2}{3})\right]$\\
\hline
$H^{\rm NSM} H^{\rm NSM} H^{\rm NSM}$ & $s_\beta^{-3}\left[m_h^2 c_\beta^3+m_Z^2 c_{2\beta} c_\beta(2s^2_\beta-c^2_\beta)-\half\lambda^2 v^2 s_{2\beta} s_\beta(2c^2_\beta-s^2_\beta)\right]$ \\
\hline
$H^{\rm SM} A^{\rm NSM} A^{\rm NSM}$  &  $s^{-2}_{\beta}\left(m_h^2 c_\beta^2-m_Z^2 c^2_{2\beta}-\lambda^2 v^2 c_{2\beta}s_\beta^2 \right)$ \\
\hline
$H^{\rm NSM} A^{\rm NSM} A^{\rm NSM}$  &$s_\beta^{-3}\left[m_h^2 c_\beta^3+m_Z^2 c_{2\beta} c_\beta(2s^2_\beta-c^2_\beta)-\half\lambda^2 v^2 s_{2\beta} s_\beta(2c^2_\beta-s^2_\beta)\right]$ \\
\hline
$H^+ H^- H^{\rm SM}$ & $4m_W^2+2s^{-2}_\beta\left(m_h^2 c_\beta^2-m_Z^2 c^2_{2\beta}-\half\lambda^2 v^2s^2_{2\beta}\right)$
\\
\hline
$H^+ H^- H^{\rm NSM}$ &$2s_\beta^{-3}\left[m_h^2 c_\beta^3+m_Z^2 c_{2\beta} c_\beta(2s^2_\beta-c^2_\beta)-\half\lambda^2 v^2 s_{2\beta} s_\beta(2c^2_\beta-s^2_\beta)\right]$
\\
\hline
\end{tabular}
\caption{\label{processesrad} \em Approximate one-loop corrected trilinear scalar interactions.}
\end{table}

\subsection{Coupling of neutral Higgs bosons to neutral gauge bosons}

In contrast to the coupling of a CP-even Higgs boson to pairs of gauge
bosons, which is present only for $H^{\rm SM}$, the derivative
couplings of pairs of neutral scalars to the neutral gauge boson are governed by the gauge interactions of the
non-SM Higgs doublet.  That is,
\begin{equation}
g_{H^{\rm NSM} A^{\rm NSM} Z} =  \half i \sqrt{g_1^2 + g_2^2} \left( p - p' \right)^{\mu} 
\end{equation}
where $p$ and $p'$ are the incoming momentum of $H^{\rm NSM}$ and $A^{\rm NSM}$, respectively and $ig_{H^{\rm NSM} A^{\rm NSM} Z} $ is the corresponding Feynman rule for the $H^{\rm NSM} A^{\rm NSM} Z$ vertex.

\subsection{Couplings of the mass eigenstate Higgs fields}

It is instructive to derive the expressions for the couplings among the different mass eigenstate
Higgs bosons in the exact alignment limit.
In light of \eq{hierarchy},
we shall also assume that the mixing between the doublet and singlet CP-even scalar fields are small. In the notation introduced in Section~\ref{nmssm}, we take $\epsilon_1=\epsilon_2=\eta^\prime=0$ (corresponding to the exact alignment limit) and $|\eta|\ll 1$, in which case $m_h^2=Z_1 v^2$ and \eq{eq:kappamatrix} reduces to
\be
\left(\begin{array}{c}
h \\
H \\
h_S
\end{array}\right)  \simeq 
\left(\begin{array}{ccc}
1 & \phm 0 \ & \phm 0 \\
0 & -1 &  -\eta \\
0 & \ -\eta &  \phm 1
\end{array}\right)  
\left(\begin{array}{c}
H^{\rm SM} \\
H^{\rm NSM}\\
H^{\rm S}
\end{array}\right) \,.
\label{app:kappamatrix}
\ee 
Similarly, we shall assume that the mixing between the doublet and singlet CP-odd scalar fields are small.  In this approximation,
\be
\left(\begin{array}{c}
A \\
A_S
\end{array}\right)  \simeq 
\left(\begin{array}{ccc}
\phm 1 & \phm \xi  \\
-\xi & \phm 1
\end{array}\right)  
\left(\begin{array}{c}
A^{\rm NSM}\\
A^{\rm S}
\end{array}\right) \,,
\ee
where $|\xi|\ll 1$.\footnote{In our numerical scans, we find that typical
  values of $\sin\xi$ lie in a range between about 0.1 and 0.3.
Thus, the results of Table~\ref{masseigenstatescalars} provide a
useful first approximation to the effects of the mixing between the
doublet and singlet CP-odd scalar fields.} 
The interactions of the scalar mass-eigenstates are given in
Table~\ref{masseigenstatescalars}, where terms quadratic (and higher order) in $\eta$ and $\xi$
have been neglected.  The trilinear Higgs interactions are expressed in terms of the coefficients, $C_{ijk}$
that appear in Table~\ref{processes},
where the subscripts $i$, $j$, $k$ label the Higgs basis scalar
fields.  In particular, $C_{H^{\rm NSM}A^{\rm NSM}A^{\rm S}}=0$ for
the scalar potential given in \eq{hbasispot}, and $C_{H^{\rm SM}H^{\rm
    SM}H^{NSM}}=C_{H^{\rm SM}H^{\rm SM}H^{S}}=0$ in the exact alignment
limit.  These relations have been implemented in obtaining Table~\ref{masseigenstatescalars}.

The Higgs interactions with a single $Z$ boson are expressed in terms of the
$H^{\rm NSM} A^{\rm NSM}Z$ interaction, denoted by $G$ in Table~\ref{masseigenstatescalars}.
The corresponding Feynman rules, denoted by $-ig_{abc}$ (where $a$,
$b$ and $c$ label the Higgs mass eigenstate fields),
are obtained by multiplying the entries of the second column of Table~\ref{masseigenstatescalars} by $-i\, n!$, where $n$ is the number of identical boson fields appearing in the
interaction term.

\begin{table}[ht!]
\centering
\begin{tabular}{|c|c|c|}
\hline
vertex &  term in the interaction Lagrangian \\
\hline
$hhh$ &  $ C_{H^{\rm SM} H^{\rm SM} H^{\rm SM} }$\\
\hline
$h h H$ & $0$\\
\hline
$h H H$&   $C_{H^{\rm SM} H^{\rm NSM} H^{\rm NSM} } + 2 \eta C_{H^{\rm SM} H^{\rm NSM} H^S}$\\
\hline
$H H H$& $- C_{H^{\rm NSM} H^{\rm NSM} H^{\rm NSM}} - 3 \eta C_{H^{\rm NSM} H^{\rm NSM} H^S}$\\
\hline
$h h h_S$ & $0$ \\
\hline
$h h_S H$ & $- C_{H^{\rm SM} H^{\rm NSM} H^S}  -  \eta C_{H^{\rm SM} H^S H^S} +  \eta C_{H^{\rm SM} H^{\rm NSM} H^{\rm NSM}}$\\ 
\hline
$HHh_S$ & $\phantom{xxxx} C_{H^{\rm NSM} H^{\rm NSM} H^{\rm S}} -  \eta C_{H^{\rm NSM} H^{\rm NSM} H^{\rm NSM}} 
 + 2 \eta C_{H^{\rm NSM} H^{\rm S} H^{\rm S}}\phantom{xxxx}$\\
\hline
$h h_S h_S$ &  $C_{H^{\rm SM} H^{\rm S} H^{\rm S}}
-2 \eta C_{H^{\rm SM} H^{\rm NSM} H^{\rm S}} $\\ 
\hline
$H h_S h_S$ & $-C_{H^{\rm NSM} H^{\rm S} H^{\rm S} }   + 2 \ \eta \ C_{H^{\rm NSM} H^{\rm NSM} H^{\rm S}} -  \eta \ C_{H^{\rm S} H^{\rm S} H^{\rm S} } $\\ 
\hline
$h_S h_S h_S$ &  $ C_{H^{\rm S} H^{\rm S} H^{\rm S}}  + 3 \eta C_{H^{\rm NSM} H^{\rm S} H^{\rm S}}$\\
\hline
$hAA$ & $C_{H^{\rm SM} A^{\rm NSM} A^{\rm NSM}} + 2\xi C_{H^{\rm SM} A^{\rm NSM} A^{\rm S}}$\\
\hline
$H A A$ & $- C_{H^{\rm NSM} A^{\rm NSM} A^{\rm NSM}} -  \eta C_{H^{\rm S} A^{\rm NSM} A^{\rm NSM}}$\\
\hline
$h_S A A$ & $C_{H^{\rm S} A^{\rm NSM} A^{\rm NSM}} - \eta C_{H^{\rm NSM} A^{\rm NSM} A^{\rm NSM}}
+ 2 \xi C_{H^{\rm S} A^{\rm NSM} A^{\rm S}}$\\
\hline
$h A A_S$ & $C_{H^{\rm SM} A^{\rm NSM} A^{\rm S}}  -  \xi C_{H^{\rm SM} A^{\rm NSM} A^{\rm NSM}}
+  \xi C_{H^{\rm SM} A^{\rm S} A^{\rm S}}$ \\
\hline
$HAA_S$ & $-\eta C_{H^{\rm S} A^{\rm NSM} A^{\rm S}} 
- \xi C_{H^{\rm NSM} A^{\rm S} A^{\rm S}} + \xi C_{H^{\rm NSM} A^{\rm NSM} A^{\rm NSM}}$\\
\hline
$h_S A A_S$& $C_{H^{\rm S} A^{\rm NSM} A^{\rm S}}  
+  \xi C_{H^{\rm S} A^{\rm S} A^{\rm S}}  -  \xi C_{H^{\rm S} A^{\rm NSM} A^{\rm NSM}}$\\
\hline
$h A_S A_S$ & $C_{H^{\rm SM} A^{\rm S} A^{\rm S}} - 2 \xi C_{H^{\rm SM} A^{\rm NSM} A^{\rm S}}$\\
\hline
$H A_S A_S$ & $-C_{H^{\rm NSM} A^{\rm S} A^{\rm S}} - \eta C_{H^{\rm S} A^{\rm S} A^{\rm S}}$\\ 
\hline
$h_S A_S A_S$ & $ C_{H^{\rm S} A^{\rm S} A^{\rm S}} - \eta C_{H^{\rm NSM} A^{\rm S} A^{\rm S}} 
-2 \xi g_{H^{\rm S} A^{\rm NSM} A^{\rm S}}$\\
\hline
$hH^+H^-$ & $C_{H^{\rm SM} H^+ H^-} $\\
\hline
$H H^+H^-$ & $-C_{H^{\rm NSM} H^+ H^-} -\eta C_{H^{\rm S} H^+ H^-} $\\
\hline
$h_S H^+H^-$ & $C_{H^{\rm S} H^+ H^-} -\eta C_{H^{\rm NSM} H^+ H^-} $\\
\hline
$A h Z$ & $0$\\
\hline
$AHZ$ & $G$\\
\hline
$A h_S Z$  & $-\eta G$\\
\hline
$A_S h Z$ & $0$\\  
\hline
$A_S HZ$ & $-\xi G$\\
\hline
$A_S h_S Z$ & $0$\\  
\hline
\end{tabular}
\caption{\label{masseigenstatescalars} \em Interactions of the mass-eigenstate scalars in the alignment limit.  The coefficients $C$ are given in Table~\ref{processes} and $G$ is the $A^{\rm NSM}H^{\rm NSM}Z$ interaction coefficient.}
\end{table}
\clearpage

\clearpage

\section{Benchmarks}
\label{Benchmarks}

In this section we present three benchmarks that illustrate the most important features of the Higgs phenomenology considered in Section~\ref{HiggsDecays}.
 Except for the third generation squarks, the gluinos, the sleptons and the squarks are all kept at the TeV scale and  decouple from the low energy phenomenology
at the electroweak scale.   The value of $\lambda = 0.65$ is chosen to
obtain alignment and preserve the perturbativity up to the Planck
scale. The relevant parameters are given in Table~\ref{parameters}.
\begin{table}[h!]
\centering
\begin{tabular}{|c|c|c|c|c|}
\hline
& Benchmark 1a &  Benchmark 1b & Benchmark 2 & Benchmark 3 \\
\hline
$\tan\beta$ & 2.1 & 2.1 & 2.5 &  2.5 \\
\hline
$M_1$(GeV)  & 122 & 200 & 135 & -400 \\
\hline
$M_2$(GeV)  & -500 & 600 & $- 300$ & $-800$ \\
\hline
$A_t$(GeV)  & $-650$ &    -750 &$-900$ & $-1400$\\
\hline
$m_{Q_3}$(GeV) &  700 & 700  & 700 & 800 \\
\hline 
$m_{U_3}$(GeV)  & 340 & 340 & 700 & 800 \\
\hline
$\kappa$ & 0.3 &  0.3 & 0.3  & 0.3 \\
\hline 
$A_\lambda$(GeV)  & 210 & 210 & 350 & 350   \\
\hline 
$A_\kappa$(GeV) & $-90$ & -75 & $-270. $ & -100 \\
\hline 
$\mu$(GeV) &  122  &  120 & 174.  & 200. \\  
\hline
\end{tabular}
\caption{\label{parameters} \em Parameters for the three different benchmarks.}
\end{table}

The Higgs and stop spectra obtained by {\tt NMSSMTools} using  these input parameters are
displayed in Table~\ref{Higgsandstopmasses}, and the chargino and neutralino masses are given in Table~\ref{EWinomasses}.  
The production cross sections for the  neutral
Higgs scalars at the LHC are presented in Table~\ref{XSs}, while some relevant processes, including the Higgs decay branching ratios are
summarized in Table~\ref{processes1}.  In what follows we will focus on the low-energy phenomenology and discuss the salient features of each benchmark scenario.

\begin{table}[t!]
\centering
\begin{tabular}{|c|c|c|c|c|}
\hline
& Benchmark 1a &  Benchmark 1b & Benchmark 2 & Benchmark 3 \\
\hline
$m_h$(GeV)   & 124.5  &     125.3      &125.4 & 124.5 \\
\hline
$m_{h_S}$(GeV)  &  93.4 &        94.5  & 72.54 & 160.3\\
\hline
$m_H$(GeV) & 301.0  &      293.0       & 470.37 & 513.1\\
\hline
$m_{A_S}$(GeV)  & 175.4 &   167.7       & 280.16 & 208.4 \\
\hline
$m_A$(GeV)  & 295.3 &     286.4  & 466.26 & 507.6\\
\hline
$m_{H^+}$(GeV) & 280.6 &         272.0  & 456.5  & 500.0 \\
\hline
$m_{\tilde{t}_1}$(GeV) &  272.7 &         255.3      & 625.77  & 693.6    \\
\hline
$m_{\tilde{t}_2}$(GeV) &  722.3 &           726.7       & 826.26   & 966.6  \\
\hline
\end{tabular}
\caption{\label{Higgsandstopmasses} \em Higgs and stop masses in the three benchmarks.}
\end{table}

\begin{table}[h!]
\centering
\begin{tabular}{|c|c|c|c|c|}
\hline
& Benchmark 1a &  Benchmark 1b & Benchmark 2 & Benchmark 3\\
\hline
$m_{\chi^0_1}$(GeV)   &  77.0  &  77.7  & 106.6 & 170.7 \\
\hline
$m_{\chi^0_2}$(GeV)   &   145.5  &164.4   & 171.3 & 226.9\\
\hline
$m_{\chi^0_3}$(GeV)   & 164.0  & 169.2   & 200.1 & 255.1\\
\hline
$m_{\chi^0_4}$(GeV)   &  187.8  &  216.9  & 237.1 & 401.4\\
\hline
$m_{\chi^0_5}$(GeV)   & 519.5   & 619.5   &  327.4  & 812.4 \\
\hline
$m_{\chi^\pm_1}$(GeV)   &  130.4  & 110.9  & 179.9 & 207.2 \\
\hline
$m_{\chi^\pm_1}$(GeV)   &   519.5 & 619.4 & 327.3 & 812.4 \\
\hline
\end{tabular}
\caption{\label{EWinomasses} \em Electroweakino masses in the three benchmarks.}
\end{table}

\begin{table}[h!]
\centering
\begin{tabular}{|c|c|c|c|c|}
\hline
 & Benchmark 1a & Benchmark 1b & Benchmark 2 & Benchmark 3 \\
\hline
$\sigma(gg \to h \to b\bar{b})/\sigma_{SM}$  & 0.85 & 0.93  & 1.00  & 0.80\\
\hline
$\sigma(gg \to h \to VV)/\sigma_{SM}$ & 1.28 & 1.16 &  1.01 &  1.12  \\
\hline
$\sigma(gg \to h_S \to VV)/\sigma_{SM}$   & $1.1 \times 10^{-3}$  & 8.1 $10^{-4}$ & -- &  0.05 \\
\hline 
$\sigma(VV \to h_S \to b\bar{b})/\sigma_{SM}$ &  0.054 &  0.036 & 6.2 $10^{-4}$ & 0.8  \\
\hline
$\sigma(gg \to H )$(pb) ( 8 TeV)  & 1.20  & 1.28 & 0.31 & 0.21\\ 
\hline
$\sigma(gg \to H)$(pb) (14 TeV) & 3.83 & 4.14 & 1.28 & 0.89 \\
\hline
$\sigma(gg \to A)$(pb) (8 TeV) & 2.18 & 2.21 & 0.57 & 0.35\\
 \hline
$\sigma(gg \to A)$(pb) ( 14 TeV) & 7.10 & 7.11 & 2.28 & 1.48 \\
\hline
\end{tabular}
\caption{\label{XSs} \em Relevant production cross sections for the three benchmarks.}
\end{table}

\clearpage
\noindent
{\bf \phantom{x}\\[15pt] Benchmark scenarios 1a and 1b}
\bigskip

\begin{table}[t!]
\centering
\begin{tabular}{|c|c|c|c|c|}
\hline
& Benchmark 1a &  Benchmark 1b & Benchmark 2 & Benchmark 3 \\ 
\hline
BR($b \to s \gamma$)  $\times$ $10^4$  &  3.76  & 3.57 & $3.68$ & $3.59$ \\
\hline
 $\Omega h^2$  &   0.119  &      0.013  &  0.128 & 0.011\\
\hline
$\sigma_{SI}$(pb) $\times$ $10^{10}$  & 2.41 $10^{-2}$ &       3.17  & 11.0 & 0.02\\
\hline
BR($h_S \to b \bar{b}$)   & 0.91   &          0.91   &  0.91 & 0.57\\
\hline
BR($h_S \to W^+ W^-$) &  7.5 $\times 10^{-5}$  &     8 $10^{-5}$  &    $--$  & 0.23 \\
\hline
BR($H \to t \bar{t}$)   &  $-$  &    $--$  &        0.39 & 0.52 \\
\hline
BR($H \to h h_S$)   & 0.47   &   0.39   &        0.24 &  0.16\\
\hline
BR($H \to  \chi_i^{0} \chi_j^{0})$ &  0.33  &    0.31  &    0.26 & 0.20  \\
\hline
BR($H \to \chi_1^+ \chi_1^-$)  & 0.009 &        0.14 &     0.008 & 0.001 \\
\hline
BR($A \to t \bar{t}$)  &   $-$ &  $--$ &     0.53 & 0.59\\
\hline 
BR($A \to Z h_S$)  &  0.36   &   0.21   &      0.16 &  0.14\\ 
\hline
BR($A \to \chi_i^{0} \chi_j^{0}$)  & 0.51 &    0.47 &        0.31 & 0.18 \\
\hline
BR($A \to \chi_1^+ \chi_1^-$) &  0.001  &  0.19  &           0.01  & 0.0005 \\
\hline
BR($A_S \to b \bar{b}$)  &0.01   &       0.005   &      0.007  & 0.87 \\
\hline
BR($A_S \to \chi_1^0 \chi_1^0$) &   0.99 & 0.99 & 0.96 & $-$  \\
\hline
BR($H^+ \to t \bar{b}$) & 0.73  &   0.73  &        0.55 & 0.62 \\
\hline
BR($H^+ \to W^+ h_S$)  &  0.15 &     0.15 &     0.18 & 0.15 \\
\hline
BR($H^+ \to \chi_1^+ \chi_i^0$) & 0.10 &  0.11 &        0.24 & 0.18 \\
\hline
\end{tabular}
\caption{\label{processes1} \em Relevant processes in the three benchmarks.}
\end{table}

The first two Benchmarks, 1a and 1b have similar spectra but differ slightly in the degree
of alignment of the SM-like Higgs with the singlet state and the value of the electroweak gaugino masses.  Benchmark  1a  has a dark matter relic density consistent with the observed one and a spin independent direct detection scattering cross section significantly below the current experimental bound, while
Benchmark 1b has a heavier gaugino spectrum and a relic density an order of magnitude below the observed one. 
In both of these benchmarks the stop spectrum has been
 fixed to obtain the observed 125 GeV Higgs mass and the $b \to s \gamma$ rate, keeping the non-SM Higgs bosons light.  In addition, Benchmark 1a and 1b have the following properties:
 
 \underline{\sl Higgs Searches}: The second lightest Higgs boson $h$ behaves like the observed (SM-like) Higgs boson with mass 125 GeV due to alignment at low $\tanb$. The mostly doublet non-SM Higgs boson masses $m_A$ and $m_H$  are around 300 GeV, and hence the neutral Higgs boson decays into top quark pairs are forbidden, while the charged Higgs boson decays mostly into top and bottom quarks, with BR$(H^\pm \rightarrow tb) ~\approx 0.7$.   Both neutral  Higgs bosons decay  into
electroweakinos with combined branching ratios of  about 40\% or larger, whereas the charged Higgs decay into electroweakinos is  only at the 10\% level. The other relevant decays of the neutral CP-even Higgs boson are  into $h h_S$ (40\%) and $h_S h_S$ (10\%).

The CP-odd scalar  $A$ has a sizable decay with a branching ratio of about  36\%  (20\%)  into $h_S Z$ 
in scenario 1a (1b) and the charged Higgs decays into $ W h_S$  with a 15\% branching ratio. The increase in the  branching ratio of the decay of $A\to h_S Z$ in scenario 1a compared to 1b
is due to the decrease of the decay into charginos. Such an increase makes the $A$ signatures
compatible with an excess observed by CMS in the $bb \ell\ell$ channel~\cite{CMS:2015mba} for masses of the heavier
and lighter Higgs states consistent with the one assumed in Benchmark 1a. On the other
hand, the decay of $H$ and $A$ into charginos in Benchmark 1b leads to a chargino production cross section of the same order as the one coming from Drell Yan processes and makes it possible to test this scenario in the search for charginos at the Run 2 of the LHC.

The mainly singlet CP-even Higgs boson $h_S$ decays dominantly into bottom quark pairs, while  the mainly singlet CP-odd Higgs bosons $A_S$ decays overwhelmingly into a pair of the lightest neutralinos. The small increase of the misalignment in Benchmark 1a compared to 1b  implies
a possible contribution to the LEP $e^+ e^- \to Z^* \to Z h_S$ cross section of the order of 
5.5\%, consistent with a small excess observed at LEP in this channel for this range of masses. 

It therefore follows that the most promising discovery modes  for these two benchmarks at the LHC are  in the topologies $2\ell 2b$, 4$b$ or  $2b2 W$ arising mainly from the gluon fusion production of $H$ and $A$ with subsequent decays $A \rightarrow Z h_S$ and $H \rightarrow h h_S$, respectively,  as discussed in section IV, as well as in the search for chargino pair production.

\underline{\sl Stop searches}: In both Benchmark 1a and 1b, the mass of one of the stops is approximately equal
to the sum of the mass of the top and the lightest neutralino.  This motivates the search for stops at the LHC in this challenging 
region of parameters.  
The other stop, mainly $\widetilde t_L$, is about 725 GeV in mass and can be searched for in decays into top or  bottom quarks and electroweakinos.   The lightest sbottom is also about 700 GeV in mass, and can be searched for in several channels at the LHC.

\underline{\sl Electroweakino searches}: The lightest  neutralinos are singlino-Higgsino admixtures, with an additional bino component in Benchmark 1a. Both the second and third
lightest neutralinos have a mass gap with respect to the lightest neutralino which is less than $m_Z$, and therefore will decay into $Z^* \chi^0_1$.
The lightest chargino has a mass of  about 110 GeV, and is Higgsino-like.  The small mass difference between the lightest chargino and neutralino makes the leptons 
coming from the chargino decays soft and difficult to detect. 

\bigskip
\noindent
{\bf Benchmark scenario 2}
\bigskip

Benchmark 2  is more traditional in the sense that the relic density is consistent with the observed one.  The
spin independent direct detection cross section is below but close to the 
LUX~\cite{Akerib:2013tjd} experimental bound, and therefore
can be soon tested by the next generation of Xenon experiments. In addition, it has the following properties

\underline{\sl Higgs searches}: The second lightest Higgs boson $h$  behaves like the observed (SM-like) Higgs boson with mass 125~GeV due to alignment at low $\tanb$.
The lightest, mostly singlet CP-even Higgs boson $h_S$ is lighter than the $Z$ boson and decays predominantly to bottom quark pairs. The lightest  CP-odd Higgs boson $A_S$ has a mass of about 300 GeV and decays predominantly into neutralinos,  with a 4\% branching ratio into $Zh_S$.
The non-SM doublet Higgs boson masses $m_H$ and $m_A$ are about 470 GeV.  Consequently, both neutral Higgs bosons have relevant decays into top quark pairs. Given that the charged Higgs boson mass is about 460 GeV,  the $b \to  s \gamma$ rate is
consistent with observations without the need of a light stop.  
The dominant decays for the heavy neutral CP-even Higgs boson $H$ are:  40\% into  $t\bar{t}$, 25\% into $h h_S$ and about 30\% into electroweakinos.  Similarly, $A$ decays 55\% of the time into $t \bar{t}$, 16\% 
into $Zh_S$ and about  30\% into electroweakinos.   The charged Higgs boson decays 55\% of the time into $t \bar{b}$, 20\% into $W h_S$, and 25\% into electroweakinos.
Similar to Benchmarks 1a and 1b, the most promising discovery modes  for $A$ and $H$ in this benchmark scenario at the LHC are via the topologies $2\ell 2b$, $4b$ or  $2b2 W$. However the fact that they are heavier and both have significant decays into top quark pairs makes detection more challenging.

\underline{\sl Stop searches}: Both stops are in the $625$ -- $825$ GeV range and decay into many different channels, including
bottom--chargino and top--neutralino final states. Their masses can be raised somewhat by lowering the stop mixing, without spoiling the consistency with the observed Higgs mass. The left-handed sbottom masses are of the same order.  

\underline{\sl Electroweakino searches}: Many different electroweakinos are present in the mass range of 100 -- 350 GeV. The lightest neutralino mass is
about 110 GeV. The second and third lightest neutralinos, as well as the charginos,  are about 180 GeV and hence can be looked for
in trilepton searches. Since the lightest electroweakinos are Higgsino and singlino like, the cross sections are smaller than for winos~\cite{Beenakker:1996ed, Ellwanger:2013rsa}. In particular,  observe that $\chi^0_2$  is in the region marginally excluded by CMS for winos, but since it is mostly an admixture of Higgsino and singlino, its production cross section 
is suppressed with respect to the wino one. Hence there are good prospects to search for some of the electroweakinos  efficiently at Run 2 of the LHC.  

\bigskip

\noindent
{\bf Benchmark scenario 3}
\bigskip

Benchmark 3 presents a scenario where $h_S\to WW$ is a relevant search channel at the LHC.  The thermal relic density contribution is small, demanding the presence of non-thermal production of the lightest neutralino.  The spin independent cross section is an order of magnitude smaller than the current LUX bound.  In addition, this scenario has the following properties:

\underline{\sl Higgs searches}: The lightest Higgs boson $h$ is the observed (SM-like) Higgs  boson with mass 125 GeV, while the second lightest CP-even Higgs $h_S$ is mostly singlet and has a mass close to the $WW$ threshold, hence decays dominantly into $W^{\pm}$ pairs. The  gluon fusion production cross section of  $h_S$  times its branching ratio  into $W^{\pm}$ pairs is about 4$\%$ of the SM cross section for a Higgs boson of the same mass. Hence $h_S$ can be efficiently searched for at the current run of the LHC.  The main difference between the Higgs phenomenology for Benchmarks 2 and 3 is the exchange of  roles between the two  lightest mainly singlet Higgs bosons, $h_S$ and $A_S$,  since now $A_S$ has a mass of about 130 GeV and decays predominantly in bottom quark pairs. The
heavy Higgs boson  $H$ decays prominently into top pairs (45$\%$), into two different lightest Higgs bosons, $h h_S$ (21$\%$) and into electroweakinos (32$\%$).  The CP-odd Higgs boson $A$ has also prominent decays into top pairs (54$\%$), into $Z \; h_S$ ($15\%$) and into electroweakinos ($21\%$). Therefore, this benchmark may be tested efficiently at the LHC  in the topologies $2\ell 2W$,  $2b 2 W$ or $4 W$ through the gluon production of $A$ and $H$ and their subsequent decays into $Zh_S$ and $h h_S$, respectively.

\underline{\sl Stop Searches}: The stop and sbottom spectra are similar to the ones in Benchmark 2. Charginos and neutralinos are heavier, but the the third generation squarks may decay 
into multiple channels and may be searched for efficiently at the Run 2 of LHC.

\underline{\sl Electroweakino searches}: The lightest electroweakinos are admixtures of singlinos and Higgsinos, with mass gaps that are smaller than the weak gauge boson masses. Hence, searches
at the LHC are difficult and will demand high luminosity.

\end{document}